# Suppressing Acoustomigration and Temperature Rise for High-power Robust Acoustics

Fangsheng Qian, Shuhan Chen, Wei Wei, Jiashuai Xu, Kai Yang, Junyan Zheng, Zijun Ren, Xingyu Liu, and Yansong Yang*

Department of Electronic and Computer Engineering, The Hong Kong University of Science and Technology, Hong Kong, China.

Author to whom correspondence should be addressed. Electronic mail: * Yansong Yang: eeyyang@ust.hk

## Abstract

High-frequency acoustic wave transducers, vibrating at gigahertz (GHz), favored for their compact size, are not only dominating the front-end of mobile handsets but are also expanding into various interdisciplinary fields, including quantum acoustics, acoustic-optics, acoustic-fluids, acoustoelectric, and sustainable power conversion systems. However, like strong vibration can "shake off" substances and produce heat, a long-standing bottleneck has been the ability to harness acoustics under high-power vibration loads, while simultaneously suppressing temperature rise, especially for IDT-based surface acoustic wave (SAW) systems. Here, we proposed a layered acoustic wave (LAW) platform, utilizing a quasi-infinite multifunctional top layer, that redefines mechanical and thermal boundary conditions to overcome three fundamental challenges in high-power acoustic wave vibration: self-heating, thermal instability, and acoustomigration. By simply leveraging a simplified, thick single-material overlayer to achieve electro-thermo-mechanical co-design, this acoustic platform moves beyond prior substrate-focused thermal management in SAW technology. It demonstrates, for the first time from the top boundary, simultaneous redistribution of the von Mises stress field and the creation of an efficient vertical thermal dissipation path. The LAW transducer, vibrating at over 2 GHz, achieves a 70% reduction in temperature rise under identical power loads, a first-order temperature coefficient of frequency (TCF) of −13 ppm/°C with minimal dispersion, and an unprecedented threshold power density of 45.61 dBm/mm$^2$ - over one order-of-magnitude higher than that of state-of-the-art thin-film surface acoustic wave (TF-SAW) counterparts at the same wavelength λ. This architecture enables scalable deployment of high-power acoustic wave components in space-constrained hybrid platforms and opens the functional diversification of acoustic wave transducers.

## 1. Introduction

Acoustics has emerged as a transformative platform across quantum engineering[1,2], sustainable power transfer infrastructure[3–5], next-generation wireless communication networks[6–8], and other interdisciplinary fields[9,10] (Fig. 1a). For acoustic transducers, its unique ability to confine GHz-frequency mechanical energy into chip-scale dimensions underpins critical advances: enabling coherent phonon-mediated qubit coupling for high-fidelity quantum information processing[11,12]; powering 5G/6G RF signal processing in billions of smartphones and micro base stations; and driving high-power satellite links - exemplified by SpaceX's 2025 acquisition of Akoustis Technologies to develop high-performance acoustic filters for direct-to-cell networks, where power handling defines orbital link viability. Beyond communications, the inherently high operating frequency and power density of acoustic wave transducers also make them promising candidates for acoustic wave actuators and non-magnetic energy reservoirs in smart energy transfer systems, offering new pathways for more compact and sustainable power conversion infrastructure. In the emerging interdisciplinary fields, strong acoustoelectric interactions between mobile carriers and propagating phonons could produce non-reciprocal amplification and allow all-acoustic RF signal processing, including circulators and switches[9,13]. In addition, acoustic wave transducers feature orders-of-magnitude smaller footprints than circuit quantum electrodynamics (cQED) devices at similar frequencies and can also support quantum control of mechanical motion in the strong-coupling regime for probing quantum foundations in complex systems[14].

Despite enabling versatile acoustoelectric functionalities of broad relevance to the electronics community, these acoustic transducers confront a fundamental trade-off: while engineered as efficient energy reservoirs, their extreme energy density triggers nonlinear thermal instabilities, causing catastrophic thermal runaway and irreversible device failure. Among them, bulk acoustic wave (BAW) transducers, utilizing a thickness-extensional (TE) mode, can achieve higher power density owing to favorable stress profiles and lower electrode resistance. However, the very nature of this vertically confined, standing-wave mode limits its use in broader applications that rely on propagating acoustic waves for in-plane coupling, such as integrated sensing, microfluidics[15], and hybrid acoustic-optics systems[16]. In contrast, surface acoustic wave (SAW) transducers offer distinct advantages in planar integrability, fabrication simplicity, and natural coupling with other physical fields. Therefore, enhancing the power density of the more versatile, interdigital based (IDT) SAW platform is a critical step to unlock acoustic transducers' full potential across these diverse domains. Fundamentally, four physical phenomena prevent power-scaling into large-signal regimes: 1) acoustomigration, driven by mechanical stress and self-heating effects, degrades electromechanical coupling by transporting metallic clusters; 2) thermoelastic instability, manifesting as frequency drift from temperature-dependent acoustic velocity and thermal expansion mismatch; 3) mechanical stress accumulation, inducing thin-film cracking or delamination; and 4) electromigration, where high current densities in the IDT electrodes induce atomic diffusion via electron momentum transfer, leading to void formation and eventual open-circuit failure. Additionally, arc discharging, another parasitic effect arising from pyroelectric charging, can be significantly mitigated by material engineering[17]. However, the first three phenomena have persisted as critical challenges for decades, constrained by the intrinsic material and architectural limitations of existing acoustic transducers. As illustrated in Fig. 1b, high-power vibration-induced mechanical stress on state-of-the-art thin-film surface acoustic wave (TF-SAW) transducers leads to visible structural failure, including severe cracking of the piezoelectric layer, delamination of metal electrodes, and widespread hillock formation in the IDT region. Early microscopy studies showed the hillock elevation of several micrometers under overloaded stress conditions, compared to a baseline root-mean-square roughness of only 2.5 nm in pristine devices[18]. Therefore, efficiently suppressing hillocks and voids formation ask for a novel design paradigm. Additionally, thermal management of current TF-SAW transducers relies on a high-thermal-conductivity substrate positioned beneath the active region. However, another key limitation of traditional configuration lies in the inefficient heat transfer from localized hot spots, primarily due to the piezoelectric layer's inherently low thermal conductivity, hindering heat flow from the active area to the substrate. Therefore, the contribution of high-thermal-conductivity substrates to heat dissipation is minimal.

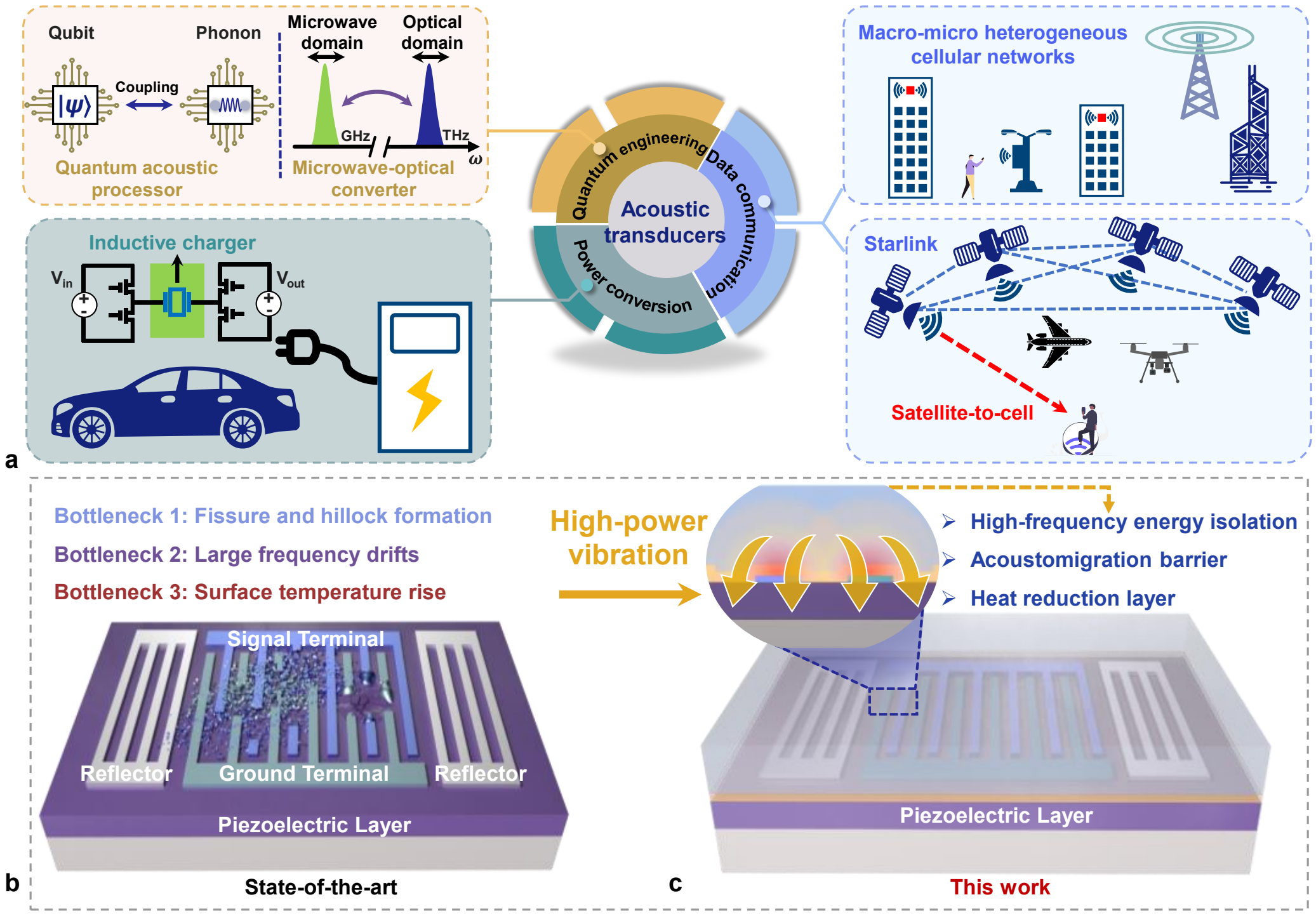

**Fig. 1 | Conceptualization of LAW transducers for ultra-high power capacity**. **a**, Acoustic transducers are ubiquitous, leveraging their compact size and strong multi-domain coupling for applications in quantum engineering, power conversion, and next-generation wireless communication. However, this miniaturization is a double-edged sword: it intensifies power density and increases susceptibility to device instabilities arising from nonlinear effects and self-heating. **b**, Schematic illustrating TF-SAW transducer failure mechanisms under high-power loads: surface strain within the piezoelectric cavity generates excessive mechanical stress in IDTs, exceeding material-dependent thresholds and inducing metallic cluster migration/diffusion. Thermal management in state-of-the-art piezoelectric-on-insulator (POI) architecture is constrained by the poor thermal conduction at the top boundaries (piezoelectric layer/air interface). Furthermore, elasticity and acoustic velocity dispersion caused by thermal expansion lead to significant frequency drifts. **c**, Conceptual perspective-view schematic of a LAW transducer on a POI platform. Mitigating device instabilities in robust-critical scenarios, such as quantum acoustodynamics and infrastructure for electrified transport and wireless communication networks, requires a key innovation: rethinking top thermal-mechanical boundary conditions to channel heat away and increase the acoustomigration barrier in a concise solution.

To march acoustic wave components toward high-power operations scenarios, overcoming the aforementioned limitations requires a concerted set of design strategies. These include: a) Disrupting grain boundary continuity or introducing additive atoms to suppress fissure and hillock formation during the vibration; b) Implementing multidirectional heat reduction routes (top, sides, and bottom) to dissipate heat from the active regions; c) Incorporating additional temperature compensating layer to stabilize high-frequency vibration across ultra-wide temperature ranges.

In this article, we implement the new design strategies by introducing a layered acoustic wave (LAW) architecture, which comprises a piezoelectric cavity and an isotropic $SiO_2$ isolation layer with thickness $h$, sandwiched between a high-velocity substrate and a quasi-infinite multifunctional overlayer (Fig. 1c). This architecture prioritizes acoustic energy confinement and mechanical stress redistribution through engineered acoustic boundaries, with constituent materials selected through a holistic electrical–mechanical–thermal co-design process. The key innovation is the decoupled yet synergistic function of each layer: the thin $SiO_2$ layer provides electrical isolation and mitigates interfacial stress, while the quasi-infinite thick silicon (Si) film, far thicker than typical acoustic wavelengths, serves as a triple role. It acts as 1) a mechanical confiner that reshapes the von Mises stress distribution, reducing peak stress at the IDT/piezoelectrics interface; 2) an integrated heat-spreader through the top and side surfaces of the transducers with the thermal conductivity exceeding that of air by approximately two orders of magnitude[19], creating vertical dissipation pathways that drastically lower operating temperatures under high power; and 3) a thermal expansion compensator that improves temperature stability without sacrificing effective electromechanical coupling ($k_t^2$) compared to traditional temperature-compensated SAW (TC-SAW) transducers. By co-designing the electrical, thermal, and mechanical boundary conditions, we achieve simultaneous enhancement in power density, thermal management, and temperature stability—without introducing complex fabrication steps or the typical performance trade-offs associated with multi-layer stacks. This integrated approach enables a fundamental shift from merely mitigating power-induced failure to actively engineering the transducer's intrinsic power-handling ceiling.

To directly evaluate the fundamental power-handling enhancement enabled by our architectural innovation, we conducted standardized high-power reliability experiments at the transducer level. This approach isolates the performance of the transducer design itself from the confounding variables inherent to system-level measurements, such as layout-dependent power flow distribution, electromagnetic parasitic effects, and potential current/voltage division effects[20], thereby providing a direct benchmark of the intrinsic advance in boundary condition engineering. At this component level, the LAW transducers exhibited a 70% reduction in steady-state temperature rise compared to their TF-SAW counterparts, highlighting significantly improved thermal management. Furthermore, the LAW architecture achieved a threshold injected power density of 45.61 dBm/mm$^2$, representing a 12.73-fold enhancement over conventional TF-SAW transducers (34.56 dBm/mm$^2$). This advancement extends to cryogenic environments, with the LAW platform reaching a threshold of 49.45 dBm/mm$^2$ (88.11 W/mm$^2$) at -85 °C, demonstrating exceptional high-power robustness in low-temperature conditions. Our work addresses these aforementioned challenges head-on, demonstrating an unprecedented, one order-of-magnitude improvement in power capacity and thus enabling the transition of acoustics from small-signal to large-signal regimes, for the first time. Critically, this architecture and design paradigm can be broadly extensible to other material systems, opening a broad pathway for performance-driven diversification in next-generation acoustic technologies.

## 2. Results

### 2.1 Boundary redefining

Following the long-standing consensus of the required top air boundary, state-of-the-art acoustic transducers can be classified into two types: i) “Free-Free” boundary conditions (Fig. 2a) and ii) “Free-Fixed” boundary conditions (Fig. 2b). As the most straightforward strategy, Type I devices, free-free platforms, such as thin-film bulk acoustic resonators (FBARs) and Lamb wave devices[21–28], achieve strong energy confinement but suffer from mechanical fragility and electrical instability due to their suspended structure. Based on the dispersion relationships of waves, acoustics wave in each medium features a specific cut-off frequency below which waves propagate through the medium rather than reflect. Therefore, Type II free-fixed platforms, such as solidly mounted FBARs (SMRs) and TC-SAW devices[29–36], mitigate some mechanical challenges by anchoring to high-velocity substrates or employing Bragg reflectors, yet they remain hampered by insufficient heat dissipation and persistent parasitic effects. Notably, TC-SAW architectures attempt to stabilize frequency response via a thin $SiO_2$ overlay. This approach, however, introduces inherent trade-offs: the nanoscale thickness of the $SiO_2$ layer results in poor thermal conductivity (~0.1 W m$^{-1}$K$^{-1}$), limiting effective heat removal from the active region[37]. Furthermore, the acoustic properties of $SiO_2$ introduce an inherent trade-off as the layer thickens for compensation: its low acoustic phase velocity undermines energy confinement and sacrifices $k_t^2$. Consequently, achieving temperature stability in this approach necessitates a fundamental compromise with the transducer's electromechanical conversion efficiency. While high-thermal-conductivity substrates mitigate certain thermal challenges, persistent acoustomigration and significant thermal instability persist and do not favor Type II devices as a suitable candidate for achieving the ideal performance metrics required in Fig. 1a. As a result, these conventional approaches fall short of the ideal combination of high-power handling, thermal stability, and minimal acoustic loss required for next-generation applications. While the conceptual use of an upper cladding to create the "fixed-fixed" boundary condition has been explored[38–40], primarily in BAW resonators, for purposes such as enhancing quality factor ($Q$) or $k_t^2$. These prior implementations often rely on complex multi-layer Bragg reflectors (Fig. 2c), which require precise thickness control and, critically, introduce multiple thermal interfaces that impede vertical heat flow, exacerbating thermal management challenges[41]. Unlike them, the proposed LAW architecture leverages a simplified, quasi-infinite medium that firstly engineers the von Mises stress distribution and creates an efficient thermal dissipation path to revolutionize power handling in IDT-based SAW platforms, directly addressing the root causes of power-induced failure without the thermal penalty of multi-interface stacks (Fig. 2c). This advancement offers inherent advantages in planar integrability and coupling to other physical fields (e.g., optics or fluids) compared to vertically confined BAW modes, making its power resilience a key enabler for broader applications.

Achieving high-quality electromechanical resonance under Fixed-Fixed boundary conditions requires new thinking about both the lower (substrate) and upper (cladding) boundaries. Key criteria for new boundary design include: (i) effective acoustic energy confinement within the piezoelectric layer; (ii) strong drift barriers to suppress acoustomigration; (iii) sufficient electrical insulation to minimize dielectric loss; (iv) multidirectional heat dissipation exceeding air convection; (v) negligible mechanical damping to preserve high quality factors; and (vi) field concentration using low-permittivity adjacent layers. While these guidelines are qualitative, they provide a systematic foundation for electrical–mechanical–thermal co-design. Compared to other common alternatives like lithium tantalate ($LiTaO_3$) and aluminum nitride (AlN), lithium niobate ($LiNbO_3$) exhibits a larger electromechanical coupling ($K^2$) behavior. This property makes $LiNbO_3$ an optimal choice for the piezoelectric layer in our LAW transducer implementation, enabling enhanced performance in wideband signal processing and high-frequency non-magnetic power conversion applications. For the foreign substrate, sapphire is selected for the bottom boundary due to its low cost and substantial acoustic velocity mismatch with the $LiNbO_3$ layer for shear horizontal (SH) modes, enabling efficient energy confinement and simplifying the heterostructure to a three-layer configuration. A thin $SiO_2$ layer is uniformly deposited atop $LiNbO_3$ to provide electrical isolation and residual stress relief for the following top quasi-infinite layer.

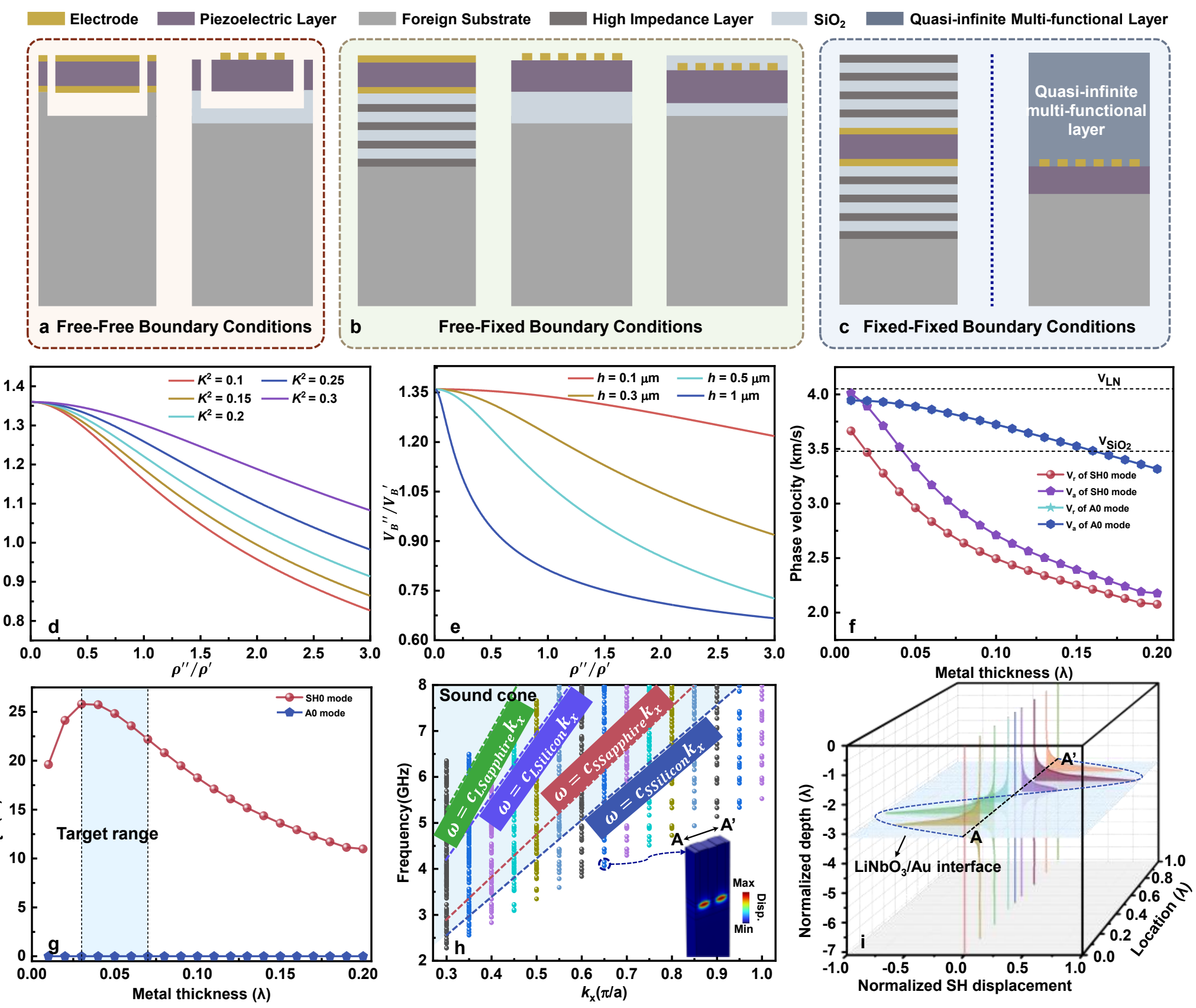

**Fig. 2 | Theoretical analysis and calculation for LAW transducers design**. **a**, Representative "free-free" acoustic boundary platforms (e.g., thin-film bulk acoustic resonators [FBARs] and antisymmetric Lamb wave resonators), requiring a cavity release process to isolate the transducer from the substrate, forming an air-transducer-air configuration. **b**, "Free-fixed" boundary platforms (e.g., solidly mounted FBAR [SMR], TF-SAW, and temperature-compensated SAW [TC-SAW]), where the transducers are anchored on substrates to create an air-transducer-substrate architecture. **c**, Two realizations of "fixed-fixed" boundaries for acoustic confinement: Dual-Bragg reflectors, requiring precise thickness control of alternating layers; Our layered acoustic wave architecture, employing a simplified, quasi-infinite single overlayer. **d**, Acoustic velocity ratio $V_B''/V_B'$ versus mass density ratio $\rho''/\rho'$ in the LAW transducer with varying $K^2$ in $LiNbO_3$. **e**, Dependence of $V_B''/V_B'$ on $\rho''/\rho'$ in LAW transducers with modulated $SiO_2$ interlayer thickness (100 nm-1 μm). **f**, Eigenmode analysis of phase velocities for the SH0 and A0 Lamb wave modes as functions of normalized Au thickness. **g**, Calculated $k_t^2$ dependence on normalized Au thickness. **h**, Dispersion relations of LAW transducers showing sub-6 GHz frequency scalability. Inset: Targeted SH mode with stress concentration at the $LiNbO_3$/Au interface; allowed acoustic waves in the sound cone form a continuum $\omega > c \cdot k_x$, which can freely radiate into the bulk. **i**, Cross-sectional SH-mode displacement profiles at 8 equally spaced positions (λ/8 intervals) along A-A', illustrating standing-wave formation in the cavity and > 90% energy attenuation within 1λ depth.

To analyze and strengthen the SH wave vibration in the LAW architecture, governing equations for the three-layer stack are derived using recursive relations[42,43], assuming ideal interfacial adhesion. The detailed derivation process is presented in Supplementary Information, Section 1. Fig. 2d illustrates the dependence of the slow shear acoustic velocity ratio $V_B''/V_B'$ ($SiO_2$ to top quasi-infinite layer) on the corresponding mass density ratio of $\rho''/\rho'$, across varying $K^2$ in $LiNbO_3$. Non-leaky SH mode is confined below the plotted curve in Fig. 2d, translating that there is a lower limit of $V_B'$ for effective top boundary design. And enhanced $K^2$ broadens the SH mode's existence range at fixed $\rho''/\rho'$. Fig. 2e demonstrates the influence of $SiO_2$ thickness on the existing range of SH-LAW. The threshold for $V_B''/V_B'$ decreases with increasing $h$, reflecting most acoustic fields concentrate near the $LiNbO_3$ surface. It can be concluded that the lower limits of $V_B''/V_B'$ obeys a negative relationship with $h$. For $h$ = 1 μm, the calculated $V_B''/V_B'$ curve decreases sharply as $\rho''/\rho'$ increases, constraining viable top-layer materials. Material candidates meeting these criteria include α-Si, $Si_xN_y$, AlN, $Al_2O_3$ for their mechanical properties. And the metals can be Cu, Au, Pt, Ag for making the slow shear acoustic velocity ($V_B$) smaller than 3500 m/s. Au is chosen for IDTs due to its chemical inertness during $SiO_2$ deposition. Beyond stability, the choice of dense electrode material such as Au provides additional benefits for the transducer's $k_t^2$ performance. The IDT structure inherently shifts the mechanical stress field from the piezoelectric layer towards the electrode layer. A high-density electrode like Au enhances mass loading and couples more effectively with this stress field and the applied electric field, leading to a higher achievable $k_t^2$. Additionally, α-Si serves as the top quasi-infinite layer for its fast shear velocity, positive temperature coefficient of elastic stiffness, and low dielectric constant[44]. Additionally, the combination of the sandwiched $SiO_2$ and upper α-Si layer contributes to relatively large coupling thanks to the negative $\Delta(V)$. Simulated performance comparisons of LAW transducers with and without optimized boundaries are detailed in Supplementary Information, Section 2.

To further refine the structural design of the LAW transducers, quasi-3D eigenmode simulations were performed in COMSOL Multiphysics by systematically varying the Au electrode thickness. This analysis evaluates its impact on phase velocity and $K^2$ for the fundamental SH and antisymmetric (A0) modes, as presented in Fig. 2f and 2g. The slow shear bulk wave velocities of $LiNbO_3$ and amorphous $SiO_2$ are indicated in Fig. 2f. To maximize the coupling and prevent bulk wave leakage, the phase velocity of SH mode is engineered to remain below the $V_{SiO_2}$ by slightly leveraging the mass loading effect of Au[45,46]. This translates to an increased acoustic impedance mismatch between $LiNbO_3$ and $SiO_2$ at optimal thickness, contributing to strong SH-wave reflection, which corresponds to a mode conversion from a leaky Love wave to a nonleaky SH wave[47]. However, thicker Au electrodes shift the SH stress field upward, indicating a moderate coupling, as confirmed by Fig. 2g. A targeted Au thickness range of 0.04λ to 0.07λ is then identified.

Fig. 2h presents the dispersion relation for the acoustic modes in the LAW configuration. Owing to the structure's periodicity in the $x$-direction, only the $k_x$ component of the wave vector $k$ is conserved throughout the entire system. Each point in Fig. 2h represents an eigenmode of the LAW device. The asymmetric boundary conditions partition the dispersion diagram into five regions bounded by sound lines $\omega = c \cdot k$, where $c$ denotes the longitudinal or shear acoustic velocity in either the sapphire substrate or quasi-infinite Si layer. Below the lowest shear sound line, elastic waves are prevented from radiating into the quasi-infinite α-Si layer or sapphire substrate. Such localized modes exhibit evanescent exponential decay along the depth direction. Above the highest sound line[48,49], elastic waves propagate freely in the bulk substrate and quasi-infinite top layer, defining this region as the "sound cone". Within the three intermediate regions, elastic waves of distinct modes exhibit unidirectional leakage into silicon.

Overall, the stacked multilayer acoustic waveguides enable dispersion-engineered SH-LAW transducers to operate in the sub-6 GHz frequency range. This complete vertical boundary architecture, optimized for actual material properties and adjacent layer thicknesses, maximizes acoustic energy confinement within the piezoelectric cavity and ensures rapid exponential decay of elastic energy away from the active region, as shown in Fig. 2i.

## 2.2 Multifunctional implementation and characterization of LAW transducers

The performance of LAW transducers is governed by their 3D geometry and constituent material properties. In particular, the large $K^2$ in $LiNbO_3$-based acoustic transducers, while enabling wideband operation, also increases their sensitivity to temperature fluctuations. The temperature coefficient of frequency (TCF), a critical parameter for spectral stability in practical environments, can be expressed as: TCF = TCV - CTE, where TCV is the temperature coefficient of acoustic velocity and CTE represents the coefficient of thermal expansion along wave propagating directions. Minimizing TCF requires both suppression of material deformation and reduction of TCV across operational temperatures.

Although $SiO_2$ is exploited in TC-SAW devices for temperature compensation, its efficacy depends critically on the Si-O-Si bond angle, which is highly sensitive to deposition conditions. This sensitivity complicates precise modeling of $SiO_2$'s temperature-dependent elasticity, necessitating experimental isolation of material-specific TCF contributions in defining geometric specifics of LAW transducers (see Supplementary Information, Section 3). Furthermore, increasing $SiO_2$ thickness narrows the LAW existence range, as demonstrated by Fig. 2e. After comprehensive consideration, a 270 nm amorphous $SiO_2$ layer was selected to balance the frequency stability, energy confinement, and electrical isolation.

Thermal management also plays a crucial role in high-power operation. Considering the thermal boundary conditions, heat-spreading in amorphous materials exhibits thickness-dependent behavior: Below 100 nm, α-Si thermal conductivity is dominated by diffusion (non-propagating modes), transitioning to propagating-dominated transport above 100 nm[50]. Previous experimental investigations confirm that both in-plane thermal conductivity $k_{||}$ and cross-plane thermal conductivity $k_{\perp}$ increase rapidly as film thickness approaches 2 μm[51].

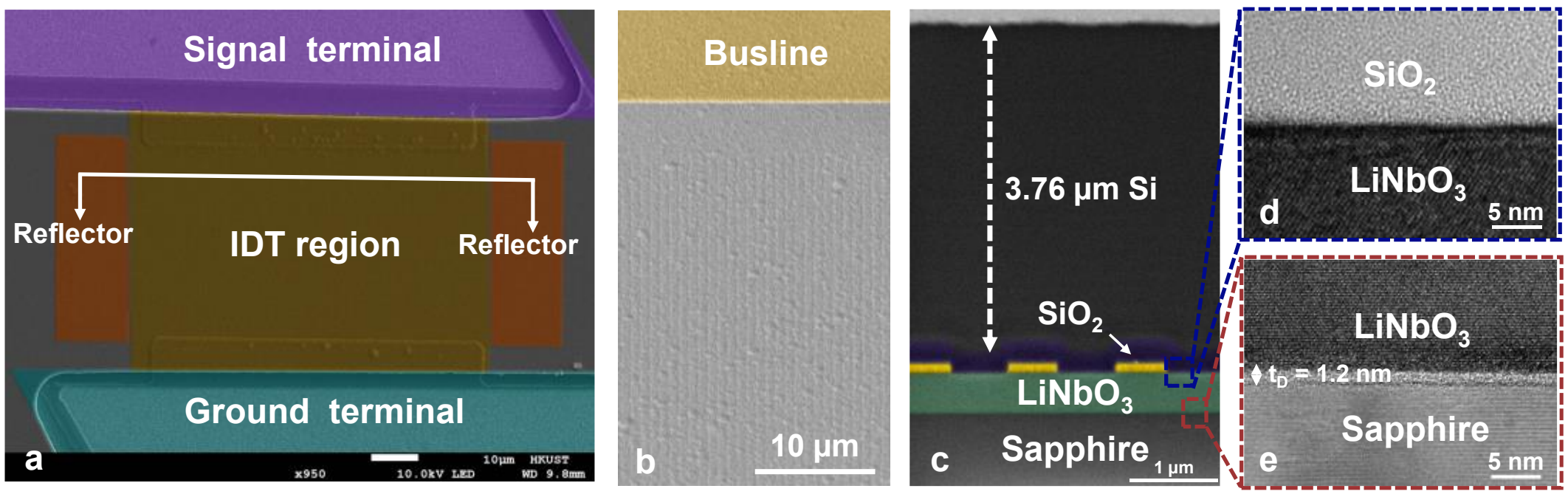

**Fig. 3 | Boundary discussion and implementation for LAW transducers**. **a**, False-color top-view SEM image of a fabricated LAW transducer. **b**, Enlarged SEM image of the active region, showing a void-free surface. **c**, Cross-sectional SEM image of the IDT region, showing layers: 3.76 μm α-Si, 270 nm $SiO_2$ (purple), 85 nm Au (yellow), 500 nm $LiNbO_3$ (green), and sapphire substrate. Cross-sectional scanning transmission electron microscopy (STEM) images of the **d**, $SiO_2/LiNbO_3$ interface in the blue rectangle in **c** and **e**, $LiNbO_3$/sapphire interface in the red rectangle in **c**.

Besides, beyond a thickness of ~ 0.6 μm, the α-Si cladding layer acts as a semi-infinite medium from the perspective of the guided mode, making the frequency independent of further thickness increases (Supplementary Information, Section 4). From the standpoint of thermal boundary, the quasi-infinite layer was therefore designed to exceed 2 μm in thickness to provide efficient thermal dissipation.

The structural integrity and interface quality of the fabricated LAW transducers were confirmed through a series of microscopy analyses. Figs. 3a and 3b display top-view scanning electron microscope (SEM) images of the fabricated LAW transducers, with Fig. 3b detailing the IDT active region. Cross-sectional SEM analysis (Fig. 3c) confirms a void-free 3.76 μm-thick α-Si layer deposited via conformal plasma-enhanced chemical vapor deposition (PECVD). This process ensures uniform $SiO_2$ insulation and robust interfacial adhesion to the $LiNbO_3$ substrate, enhancing thermal dissipation through maximized interfacial contacts. Transmission electron microscopy (TEM) image in Fig. 3d reveals a sharp $SiO_2/LiNbO_3$ interface with a dense, defect-free metal migration barrier. Further, the $LiNbO_3$/sapphire bonded interface (Fig. 3e) exhibits minimal interfacial damage (1.2 nm), with both the $LiNbO_3$ thin film and sapphire substrate retaining high crystallinity post-fabrication. Energy-dispersive spectroscopy (EDS) elemental mapping of the scanning transmission electron microscopy (STEM) image confirms uniform distribution of Si, O, Au, Nb, and Al across the transducer structure, as detailed in Supplementary Information, Section 5.

To isolate the mechanical damping effects of α-Si and $SiO_2$, conventional and $SiO_2$-overcoated TF-SAW transducers were measured after the fabrication of each transducer configuration. A standardized wafer cleaning protocol was performed prior to each subsequent fabrication step to remove potential contaminants introduced during measurements. Befitting from the design of acoustic boundary conditions, the strong plate modes (in-band A0 mode, S0, and SH1 mode) inherent to the baseline TF-SAW and $SiO_2$-overcoated TF-SAW structures are suppressed simultaneously after adding the α-Si multi-functional layer. These results can be directly attributed to the tailored sound cone and stress field, providing more freedom in acoustic boundary designs on the LAW platform (Fig. 4a). The in-band spurious modes on three platforms are transversal modes arising from the waveguiding effect in the aperture direction. The suppression strategies are discussed in Supplementary Information, Section 6. Furthermore, the identical admittance observed across all three transducer configurations at frequencies distant from $f_r$, indicates unaffected static capacitance by the top boundary, confirming effective electric field confinement inside the piezoelectric cavity. Compared to the TF-SAW transducer, measured admittance curves of $SiO_2$-overcoated TF-SAW transducers exhibit an upward frequency shift attributed to velocity compensation. However, the additional $SiO_2$ overlayer presents degraded device performances, including the TCF, $k_t^2$ and $Q$. Subsequent deposition of the high-velocity Si cladding layer continuously increases the resonant frequency ($f_r$) from 2.205 GHz ($SiO_2$-overcoated configuration) to 2.458 GHz and boosts Bode-$Q_{max}$ from 234 ($SiO_2$-overcoated TF-SAW configuration) to 445, as illustrated in Fig. 4b. The $Q$-factor enhancement stems from stress field manipulation enabled by the high-velocity α-Si cladding layer. Detailed performance merits of these three types of transducers are summarized in Supplementary Information, Section 7 and Section 8. Following that, LAW transducers with a range of lateral wavelengths (λ ranging from 1.0 μm to 1.8 μm) were tested, as presented in Fig. 4c. The measured admittance ratio (AR), defined as the difference between admittance magnitudes at $f_r$ and the anti-resonant frequency ($f_a$) of each transducer, demonstrating that the LAW architecture achieves AR values comparable to or exceeding those of state-of-the-art TF-SAW devices with the same wavelength on $LiNbO_3$/SiC platform[17], as shown in Fig. 4d. It can be observed that other spurious modes occur beyond the targeted SH0 mode, especially for wavelengths > 1.2 μm. These are higher-order bulk waves scattered at vertical acoustic boundaries, which can be easily suppressed via further mechanical bandgap engineering, as detailed in Supplementary Information, Section 6. Additional key performance metrics, including quality factor (Bode-$Q_{max}$), $k_t^2$, and figure of merit (Bode-$Q_{max}$ × $k_t^2$), are also summarized in Figs. 4d-4e. Notably, the highest measured Bode-$Q_{max}$ inside the passband is 559 at $\lambda = 1.8$ μm. While the moderate $k_t^2$ arises from dispersive stress field in the upper layer, the $LiNbO_3$ layer itself maintains strong acoustic field confinement, with most LAW transducers exhibiting $k_t^2$>17%, confirming sufficient electromechanical coupling for wideband applications. The optimized boundaries yield a FoM of 94 at $\lambda = 1.8$ μm, with only 7.9% degradation relative to TF-SAW counterparts. Device performance remains stable for $\lambda \geq 1.2$ μm, and $Q_{3dB}$ at $f_r$ for all devices shows little variation, indicating minimal electrical loss as λ decreases from 1.8 μm to 1.0 μm. Below 1.2 μm, reduced coupling arises from lateral electric field dispersion, suggesting future optimization via $LiNbO_3$ thickness adjustment.

Temperature stability was systematically characterized using a vacuum cryogenic probe station over an ultra-wide range (-150 °C to 350 °C). Fig. 4f shows the admittance response of a LAW device (λ = 1.2 μm), presenting a 12 dB increase in AR as temperature decreases from 350 °C to −150 °C, which arises from reduced mechanical damping and ohmic loss. Note that $f_r$ and $f_a$ increase linearly with cooling, yielding fractional frequency variations proportional to temperature (Fig. 4g). Neglecting third and higher-order terms, the TCF can be determined based on the following formula: $(f_T - f_0)/f_0 = A_1 * (T - T_0) + A_2 * (T - T_0)^2$, where $A_1$ and $A_2$ refer to first- and second-order TCF, and room temperature $T_0$ is denoted as the reference temperature. Fitted TCF for $f_r$ are $A_1$ = -21.8 ppm/°C and $A_2$ = -3.387 ppb/°C$^2$, while those for TCF at $f_a$ ($TCF_a$) are $A_1$ = -13 ppm/°C and $A_2$ = -0.948 ppb/°C$^2$. Enhanced TCF in the LAW transducer arises from two mechanisms: suppression of thermal expansion provided by the extra upper interlayer force and modulation of TCV, which is highly dependent on the temperature coefficient of piezoelectric and elastic constants. More importantly, the phase velocity of SH mode exhibits dispersion with $\lambda$, that is, variation in $\lambda$ caused by thermal expansion or contraction inherently alter the TCV term[52]. In this case, TCF can be captured by: $\mathrm{TCF} \equiv \frac{1}{f}\frac{df(T)}{dT} = \frac{\lambda}{v_p}\frac{d}{dT}\frac{\lambda(T)}{v_p(T)}$, suggesting a coupling between TEC and TCV. As shown in Fig. 4(h), LAW transducers achieve superior temperature stability and minimal dispersion compared to state-of-the-art SH-SAW devices[31–34,52–56], marking the lowest reported TCF values for $LiNbO_3$-based transducers. The architecture decouples TEC and TCV, mitigating dispersion-induced TCF instability. Comprehensive TCF at $f_r$ ($TCF_r$) comparison between this work and state-of-the-art works are presented in Supplementary Information, Section 9. Compared to $SiO_2$-overcoated TF-SAW devices in Supplementary Information, Table S1, our LAW architecture improves the first- and second-order $TCF_r$ performance from -117.51 ppm/°C to -21.8 ppm/°C, stemming from α-Si's positive TCV and efficient interlayer stress compensation.

### 2.3 Power handling capability evaluation

To rigorously assess power capacity across different acoustic architectures using the setup described in Supplementary Information, Section 10, several aspects in the evaluation process should be taken into consideration to allow for accelerated testing and ensure a fair comparison. First, a fundamental prerequisite for a meaningful comparison is a benchmarking method that isolates the intrinsic impact from architectural innovation. Power handling measured at the filter level, however, reflects a system-level property influenced by numerous design-specific factors, such as cascading resonators, power-routing layout, matching networks, and test conditions. These variables can obscure the intrinsic improvement attributable solely to the core transducer technology, as comprehensively discussed in Supplementary Information, Section 11. We thus conducted high-power testing at the individual transducer level to eliminate confounding effects from multi-transducer interactions within filters and enable a direct comparison of the fundamental design. Second, a direct comparison of power-handling performance requires TF-SAW and LAW transducers fabricated from an identical layout. This control is essential because the static capacitance (C0) governs the frequency-dependent von Mises stress profile, which in turn dictates the intrinsic, architecture-specific weakest point for failure in each platform, as analyzed in Supplementary Information, Section 12. See Supplementary Information, Section 13, for the measured $S$-parameters responses of these two transducers, highlighting a minimal device-to-device variation and no degradation due to boundary-engineered energy confinement. Third, their impedance mismatch with 50 Ω systems demands precise driving frequency selection. See Supplementary Information, Section 14 for detailed power dissipation distribution curves within a representative LAW transducer, showing their dependence on driving frequencies and measurement configurations. Most importantly, it must be considered that port impedance mismatch can be a significant concern when performing power tests at the individual transducer level, as it leads to reflected power that does not contribute to device stress. To ensure that the reported power metrics accurately represent the stress experienced by the device, all quoted power values, including the critical "injected power" in the following part, are rigorously defined as the delivered power, i.e., the incident power minus the reflected power. This definition explicitly excludes the reflected portion, thereby eliminating any bias introduced by impedance mismatch and ensuring a fair comparison of the intrinsic power-handling capability of each architecture. For the testing frequency selection, the preliminary criteria are based on the fact that the higher the dissipated power, the higher the temperature in the active region, and the shorter the time-to-failure. A detailed experimental validation to support this claim is presented in Supplementary Information, Section 12 and Section 15. Thus, driving frequencies favoring maximum power absorption for transducers should be obtained from the *S*-parameters measured under small-signal operational conditions. Last, the power absorption peak shifts downward by several megahertz (MHz) under elevated power due to the non-zero TCF. To compensate, a predefined frequency window was used to mitigate thermal shifts in the power absorption peak, replacing single-frequency continuous-wave testing to maintain accuracy. Guided by these principles, standardized reliability measurements were performed to study acoustomigration dynamics and quantify the threshold power durability of each architecture.

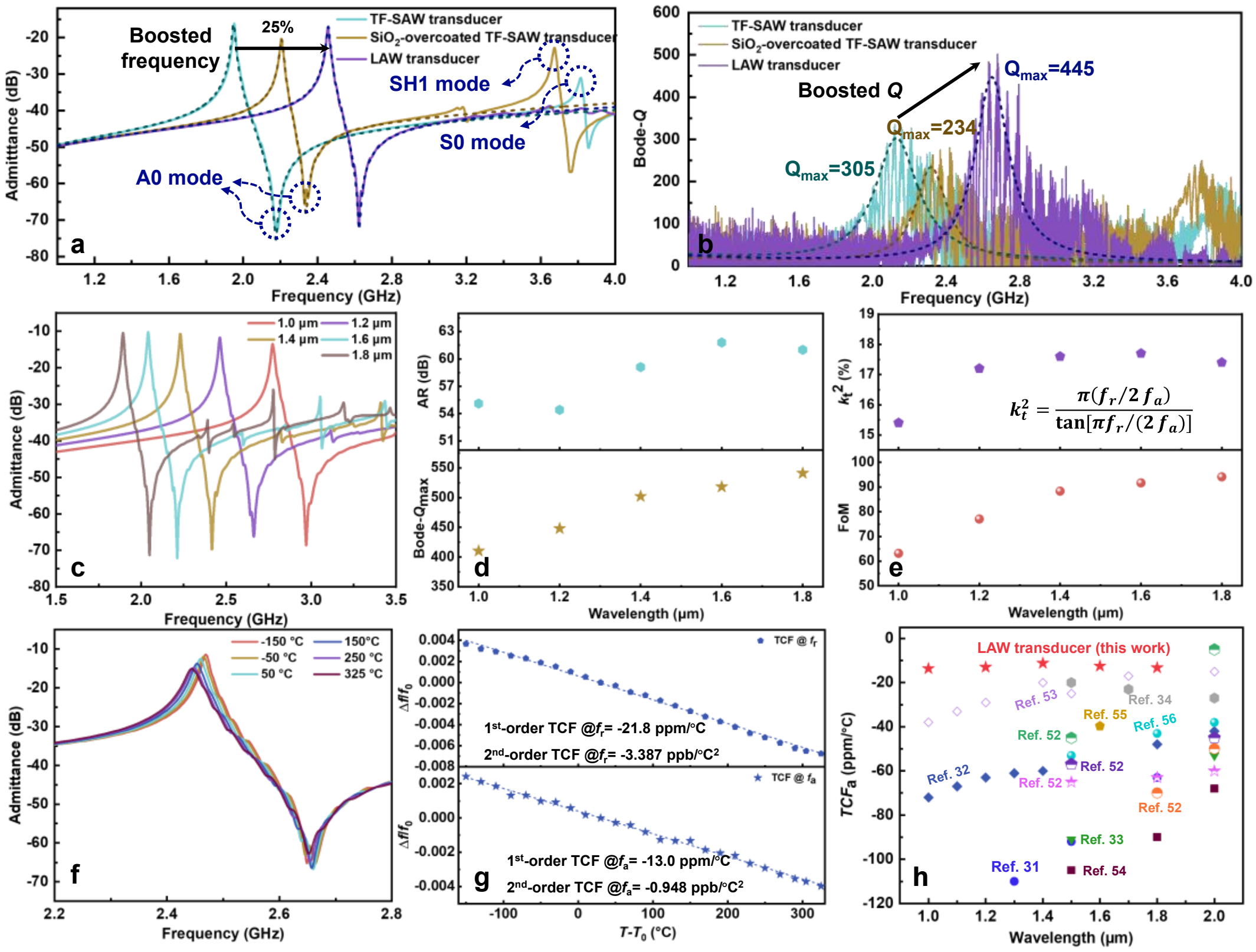

**Fig. 4 | Measurement results from the LAW transducers**. Measured **a**, admittance and **b**, corresponding Bode-$Q$ curves of transducers in conventional TF-SAW, $SiO_2$-overcoated TF-SAW, and LAW configurations. **c**, Admittance curves of LAW transducers as a function of wavelength. Extracted metrics include **d**, admittance ratio (AR) and Bode-$Q_{max}$ in the passband, **e**, $k_t^2$ and figure of merit (FoM) for LAW transducers across varying wavelengths. **f**, Temperature-dependent admittance responses of a LAW transducer (λ = 1.2 μm) measured over an ultra-wide temperature range from -150 °C to 325 °C. **g**, Extracted relative frequency drifts under temperature variation, with the first-order and second-order $TCF$ at $f_r$ and $f_a$. **h**, Scatter plot comparison of $TCF_a$ for the LAW transducers in this work and for advanced TF-SAW transducers reported in the literature. Data points are compiled from the following references: $LiNbO_3/SiO_2$/Quartz[31]; $LiNbO_3$/SiC[32]; $LiNbO_3/SiO_2$/Si[33]; $SiO_2/LiNbO_3$[34]; $LiNbO_3$/69°Y90°X-Quartz, $LiNbO_3$/60°Y90°X-Quartz, $LiNbO_3$/AT-Quartz and $LiNbO_3$/YZ-Quartz[52]; $LiNbO_3/SiO_2$/SiC[53, 54]; $LiNbO_3$/SiC[55]; and $LiNbO_3/SiO_2$/p-Si/Si[56].

Fig. 5a and 5b present the frequency-dependent dissipation, reflection, and transmission coefficients of both TF-SAW and LAW transducers at −15 dBm load power. While nonlinear distortions emerge at elevated powers, their effect on overall absorption efficiency in the target frequency window remains marginal and is assumed equivalent between device types[57]. Outside the defined frequency window, most transmitted power passes through the two-port configuration with minimal energy loss. Since electromigration and acoustomigration are thermally facilitated, in situ monitoring of active-region temperature is critical for understanding the failure modes. For conventional contact-method surface temperature measurements, a light-absorbing black paint, infrared (IR)-opaque and highly emissive, is typically coated on the device under test (DUT) surface to ensure uniform emissivity and consistent thermal boundary conditions. However, applying this IR-opaque coating would interfere with the operation of TF-SAW transducers. Therefore, a non-contact method, with a pre-calibrated IR camera and macro lens, was utilized for real-time temperature mapping of acoustic wave transducers without coating the IR opaque black paint. The emissivity correction protocol for quantitative extraction of temperature rise in both architectures is presented in the

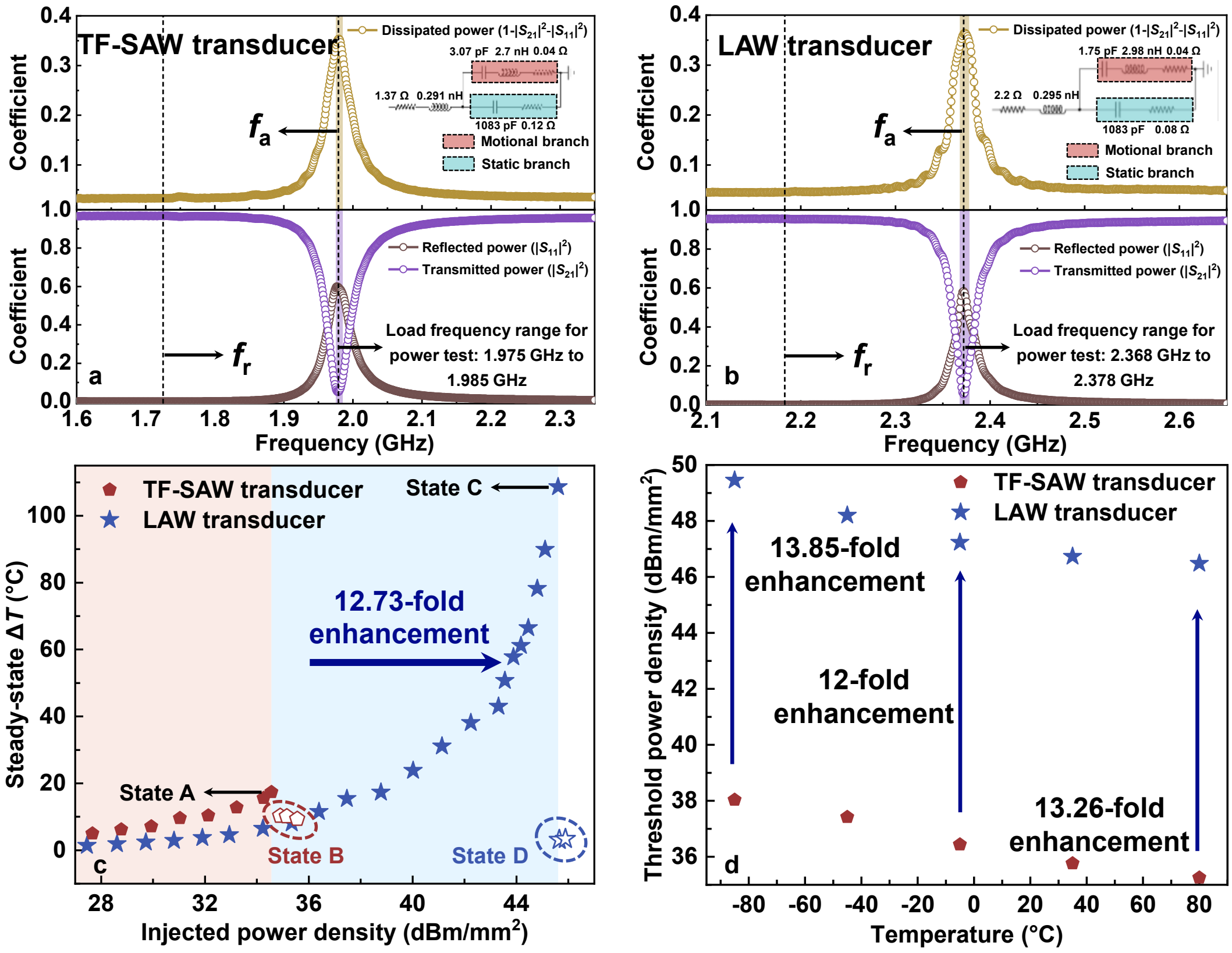

**Fig. 5 | Power test design and measurement for TF-SAW and LAW transducers**. **a**, Measured dissipated, reflected, and transmitted power coefficients as a function of driving frequency for a 2-port TF-SAW transducer operating at a load power of $P$ = -15 dBm. **b**, Measured dissipated, reflected, and transmitted power coefficients as a function of driving frequencies for a 2-port LAW transducer operating at a load power of $P$ = -15 dBm. Both transducers share an identical layout design to ensure equitable high-power comparison. $f_r$ and $f_a$ are marked by the dashed lines in each plot. The loading frequency ranges for the two transducers are also indicated. Fitted parameters from the modified Butterworth-Van Dyke (mBVD) model for each configuration are summarized in the respective insets. **c**, Steady-state temperature rises extracted from thermal mapping data of both transducers are plotted against the injected power density within the specified load frequency range. Injected power density is defined as the actual power delivered to the DUT (incident power minus reflected power) divided by the whole rectangular transduction area, including the IDT area, reflectors, and bus lines. States A to D denote operating conditions of the transducers under progressively increasing power load levels. **d**, Comparisons of injected power density threshold for LAW and TF-SAW transducers are presented across testing temperatures ranging from −85 °C to 80 °C, where the threshold is defined as the maximum power density supporting reliable operation for 5-minute power loading under the specified conditions. Prior to reaching the power density thresholds, both DUTs withstood sequential tests at incrementally higher power levels.

Methods section.

Due to the low reverse transmission in the power amplifier and isolator, direct measurement of admittance responses during high-power testing was not possible. However, operational conditions can be continuously tracked via in-situ IR temperature imaging and $S_{21}$ scattering parameter responses. While the metal migration process in TF-SAW transducers exhibits temporal continuity, manifested by a gradual progression of voids and hillocks, both device types exhibit rapid drops in $S_{21}$ and temperature upon reaching their threshold injected power. A detailed explanation of the correction method for the scattering parameters and reflection coefficients calculation is provided in the Methods section. Fig. 5(c) summarizes the extracted temperature rises of TF-SAW and LAW transducers under varying injected power levels. The injected power threshold is defined as the maximum power level that DUTs can withstand for 10 minutes without an $S_{21}$ drop. For a single TF-SAW transducer, the measured temperature experiences a nonlinear rise as the injected power increases from 15.75 dBm (37 mW) to 22.65 dBm (184 mW), driven by escalating electrical/mechanical losses and increased power dissipation coefficients. A sudden temperature drop occurs at an injected power of 22.99 dBm (199 mW). After this point, increasing the input power further results in a relatively constant temperature, indicating consistent dissipated power despite higher output power from the isolator. Given the relatively large area (0.0644 mm$^2$) of the active area, this corresponds to a threshold power density of 34.56 dBm/mm$^2$. For the LAW configuration, the measured surface temperature rises in the injected power range below 22.65 dBm are lower than those of the TF-SAW counterparts. The maximum temperature rise is only 5.2 °C, while the TF-SAW transducer exhibits a temperature rise of 17.4 °C, corresponding to a 70% reduction. This improvement is attributed to the extra heat dissipation routes provided by the top quasi-infinite layer. The single LAW transducer sustains stable operation up to an injected power of 33.70 dBm (2.34 W), corresponding to an ultrahigh power density threshold of 45.61 dBm/mm$^2$. As detailed in Supplementary Section 12, we performed comprehensive high-power characterizations on TF-SAW and LAW devices with varying C0 values at multiple frequency points across the resonator band. The results consistently demonstrate that the proposed LAW architecture achieves approximately an order of magnitude improvement in power handling capability across the entire operating band, not merely at a single frequency point. Supplementary Information, Section 16, provides further examples of in-situ temperature measurement and analysis for LAW transducers driven by various frequencies with high power loads.

Temperature-dependent robustness was further investigated using a cryogenic probe station, systematically evaluating devices from −85 °C to 80 °C (Fig. 5d). To avoid the destructive impact from a single test, multiple devices of each type, matched for power-dissipation profiles (selected using the criteria in Figs. 5a and 5b), were evaluated across the temperature range. A consistent increase in threshold injected power density is observed in the LAW transducers as ambient temperatures decrease, with a nonlinear enhancement becoming pronounced below -45 °C. The LAW transducer achieves a injected power density threshold of 49.45 dBm/mm$^2$ (88.11 W/mm$^2$) at -85 °C, demonstrating unprecedented high-power resilience in low-temperature environments - a 13.85-fold improvement over TF-SAW counterparts. This enhancement is attributed to accelerated thermal dissipation and reduced energy loss in LAW platforms under cryogenic conditions. Above -45 °C, LAW devices exhibit a gradual decline in threshold power density from 48.20 dBm/mm$^2$ at -45 °C to 46.48 dBm/mm$^2$ at 80 °C. In contrast, TF-SAW devices exhibit a sharper reduction in power capacity, dropping from 37.42 dBm/mm$^2$ to 35.26 dBm/mm$^2$ over the same temperature range. The narrower variation of power threshold in LAW transducers across the entire temperature range reflects that their power durability is less prone to degradation under thermal stress, demonstrating a superior robust and temperature-stable performance compared to TF-SAW devices. See Supplementary Information, Section 17 and 18, for detailed performance comparison of the tested transducers before and after the high-power loads.

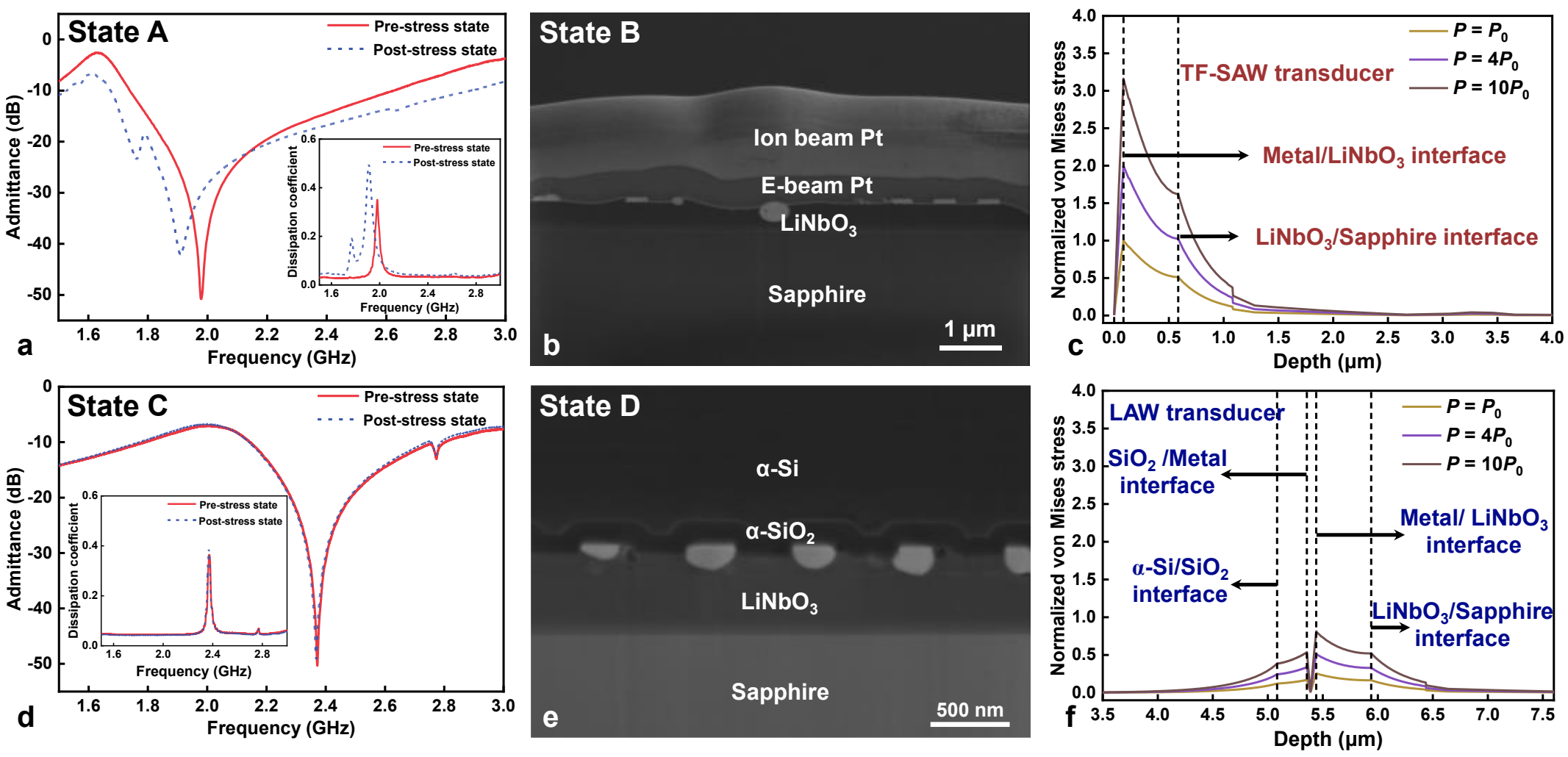

**Fig. 6 | Electrical, structural, and mechanistic analysis of TF-SAW and LAW transducers. a**, Measured admittance curve and power dissipation characteristics of the TF-SAW transducer at state A, showing spurious mode, degraded $k_t^2$, and $Q$. **b**, Cross-sectional view SEM image of the failed TF-SAW transducer at state B, revealing severe electrode migration (acoustomigration) and material deformation. **c**, Depth-resolved von Mises stress profiles for the TF-SAW transducer under increasing injected power ($P_0$, $4P_0$, $10P_0$), all normalized to the maximum stress in the TF-SAW device at $P_0$. **d**, Measured admittance curve and power dissipation profiles of the LAW transducer at state C, exhibiting preserved frequency response. **e**, Cross-sectional view SEM image of the failed LAW transducer at state D, showing localized damage without catastrophic electrode migration. **e**, Measured admittance curve and power dissipation profiles of the LAW transducer at state C. **f**, Corresponding von Mises stress profiles for the LAW device under the same power levels ($P_0$, $4P_0$, $10P_0$), normalized to the TF-SAW-at-$P_0$ reference. The peak stress in the LAW transducer is only ~1/4 of the baseline TF-SAW's value at $P_0$ through stress redistribution via upper-boundary engineering—a mechanism not addressed by conventional substrate engineering.

To elucidate post-stress failure mechanical and electrical conditions, two states were defined for each transducer configuration: pre- and post-threshold power exposure (Fig. 5c, States A and C for TF-SAW and LAW devices, respectively), followed by further testing above threshold (States B and D). After 10 minutes at threshold power, TF-SAW devices showed reduced $k_t^2$, an in-band spurious mode, degraded $Q$, and smaller static capacitance (Fig. 6a). The inset of Fig. 6a reveals a downward shift in the dissipation coefficient, corroborating temperature measurements in Fig. 5c. Cross-sectional SEM and EDS mapping of failed TF-SAW transducers (State B, Fig. 6b) revealed severe deformation, gold migration, and hillock formation, confirming acoustomigration-induced failure. A bright line is visible along the $LiNbO_3$ surface between failed electrodes (see Supplementary Information, Section 19, for comprehensive acoustomigration investigation and EDS mapping). The EDS analysis shows the bright line is comprised of Pt and Au, indicating that acoustic-induced migration under high-power conditions has caused gold to accumulate in these regions. Besides, the EDS mapping results confirm the severely deformed $LiNbO_3$ layer and IDTs under high-power loads.

In contrast, the LAW transducer in State C maintained electrical and mechanical properties nearly identical to their pre-stress state (Fig. 6c), even after withstanding over 12 times higher power density. Cross-sectional SEM of stressed LAW devices (State D, Fig. 6d) shows no evidence of electrode migration, due to the suppressed grain boundary continuity provided by the upper α-Si cladding. However, the deformed electrodes are diffused into the $LiNbO_3$ layer, and an electrically short connection between two metal strips is formed in the left region highlighted by a rectangular frame (see Supplementary Information, Section 20, for exhaustive EDS mapping and discussion on the cross-section). Some voids between electrodes and the $SiO_2$ isolation layer were observed, attributed to mechanical stress under extreme loading, but without catastrophic device failure.

The fundamental mechanism behind this dramatic improvement in power durability is revealed by the von Mises stress distribution at the critical interface. As shown in the depth-resolved profiles (Fig. 6c, f), normalized to the maximum stress in the baseline TF-SAW device at a reference power $P_0$, the maximum stress in the LAW platform is only about 1/4 of that in the SAW device under the same injected power ($P_0$). Even as the power increases to $10P_0$, the relative stress level in the LAW device rises only marginally to approximately 4/5 of the baseline TF-SAW's stress at $P_0$. This substantial and persistent stress suppression demonstrates that the LAW architecture enhances power capacity primarily through the redistribution and mitigation of mechanical stress at the fundamental level, thereby directly inhibiting the driving force for acoustomigration. This result underscores a key advantage of upper-boundary engineering over conventional substrate engineering. As reported by previous work, merely improving substrate thermal conductivity does not effectively reshape this interfacial stress distribution; the failure bottleneck remains[58]. In contrast, the co-designed top cladding in the LAW platform simultaneously manages heat and actively tailors the acoustic boundary condition, achieving a fundamental improvement in power handling that overcomes the inherent bottlenecks of traditional approaches.

## 3. Discussion

We have presented new findings in acoustics and proposed a new layered acoustic wave architecture that fundamentally redefines the electrical, mechanical, and thermal boundary conditions of high-frequency acoustic transducers for vibrating at large power density. Through the synergistic integration of a piezoelectric cavity and a quasi-infinite multifunctional cladding, the LAW platform addresses three persistent challenges in high-power acoustic wave vibration with reduced loss: acoustomigration suppression, efficient heat dissipation, and enhanced thermal stability. Our experimental results demonstrate that LAW transducers achieve a 70% reduction in temperature rise and a 12.7-fold increase in threshold power density compared to state-of-the-art TF-SAW devices, while maintaining robust frequency stability with a first-order TCF of −13 ppm/°C across a wide operational temperature range of 475 °C. Notably, the architecture enables reliable GHz acoustics at ultra-high power densities. By decoupling the constraints imposed by conventional acoustic transducer architectures, this work establishes a generalizable design paradigm for next-generation acoustic technologies. The principles demonstrated here are broadly applicable to a wide range of IDT-based acoustics systems and open pathways for the scalable deployment of ultracompact, high-power acoustic transducers in emerging applications - from 5G/6G communications to miniaturized, non-magnetic energy reservoirs.

## Methods

### Device fabrication

The devices were fabricated on a 500 nm-thick X-cut single-crystalline $LiNbO_3$ thin film bonded to a c-axis sapphire substrate using surface-activated bonding (SAB) and grinding technology. IDTs were first patterned by electron-beam lithography, followed by deposition of a 5 nm Cr adhesion layer and 80 nm Au via electron-beam evaporation and lift-off (metal ratio = 0.5). After that, photolithography and an additional 100 nm Au layer enhanced electrical contact and transversal energy confinement. IDTs exhibited uniform sidewall profiles. To avoid accumulated thermal stress and prevent surface damage, a 270 nm $SiO_2$ layer was deposited by plasma-enhanced chemical vapor deposition (PECVD) at a low deposition rate, leveraging its low thermal budget. After that, a 3.76 μm-thick α-Si layer was deposited by PECVD. Residual stress introduced during the α-Si deposition process leads to buckling, delamination, fatigue, and other failure modes in the thick film. Large residual stress also facilitates device failure and thin-film cracking under high-power operation, degrading device performance by altering the mechanical properties of the LAW transducer[59]. Intrinsic stress in α-Si arises from systematic changes in the position of Si atoms that happen after a slip-free adhesion layer on the underlying substrate. These atomic arrangements can be controlled by adjusting the deposition conditions and processing time. To mitigate accumulated stress, the deposition process was divided into 18 cycles, each lasting 5 minutes. This strategy prevents the combination of both intrinsic stress and thermal stress, which could otherwise lead to the formation of cracks and hillocks in thick films. The residual stress in the α-Si thick film can be extracted via curvature measurement on a 4-inch standardized wafer. Detailed extraction method and influence of residual stress in the α-Si thick film on LAW device performance are discussed in Supplementary Information, Section 21. Note that the optimized residual stress in α-Si thick film is 4.38 MPa. In the following steps, photoresist (HPR504) was employed as an etching mask, followed by α-Si etching using $SF_6$/Ar hybrid gas in an inductively coupled plasma reactive ion etching (ICP-RIE) system. Subsequently, $SiO_2$ was etched using $CHF_3/O_2$ hybrid gas in a reactive ion etching (RIE) system. The etching procedures were controlled by pre-calibrated etching time and depth, and the etching depth was examined via profilometry. Then, the etching mask was stripped in acetone, followed by a standard cleaning process.

**Electrical characterization**

Following off-wafer short-load-open calibration, one-port measurement configuration is utilized to measure the scattering parameters of transducers under room temperature and atmospheric pressure, using a Keysight P5028A Vector Network Analyzer with a load power of -15 dBm. The admittance responses ($Y$-parameters) were subsequently calculated from the measured $S$-parameters without applying any extra de-embedding step. The performance merits of each device are evaluated by extracting the Bode-$Q$ using established methods[60]. While $k_t^2$ exhibits significant variation depending on the calculation formula, making it less meaningful when compared with prior works. Here, $k_t^2$ in this work is calculated using the standard IEEE definition[61]: $k_t^2 = \frac{\pi(f_r/2f_a)}{tan[\pi f_r/(2f_a)]}$. Temperature-dependent frequency drift was characterized in a vacuum cryogenic probe station across a range of -150 °C to 325 °C, with 20 °C increments. To mitigate thermal instability from the measurement setup, the wafer was stabilized for 20 minutes at each target temperature prior to probe contact, followed by an additional 5-minute settling period post-landing to ensure thermal equilibrium at the probe tips.

**Power durability measurement**

The power handling capability of acoustic devices is conventionally evaluated on the basis of the time to failure (TTF) metric, which correlates closely with not only the input power ($P_{in}$), the device temperature ($T$), but also the driving frequency. TTF follows Eyring's model: $\tau = \alpha \cdot \exp(E/kT) \cdot P_{in}^m$, where $k$ and $E$ represent the Boltzmann constant and activation energy, and $\alpha,\ m$ are device-specific constants. In this model, excessive power and high temperature are typically applied to accelerate device failure to a realistic value due to the Arrhenius law. While accelerated failure tests for acoustic transducers under elevated power and temperature (enabled by the Arrhenius law) allow TTF prediction via numerical models built on massive experimental data, no standardized protocol exists for acoustic transducers.

To address this gap, we established a benchmarking method for power capacity by directly monitoring temperature and $S_{21}$ variations in TF-SAW and LAW transducers. Device failure is identified by an abrupt temperature drop and $S_{21}$ degradation, which indicates impedance mismatch preventing power delivery to the DUTs. We incrementally increased the isolator's output power from 18 dBm in 1 dB steps until reaching 30 dBm, followed by 0.5 dB steps up to 33 dBm, and finally 0.25 dB steps until reaching the injected power threshold, at which point device failure occurred within a 5-minute loading time. For each frequency sweep, the time of absorption is 10 s, sufficient for the device to reach thermal equilibrium at each power level[62]. Besides, a port-extension calibration was performed at each power level using a CS-5 substrate's "Thru" standard. The $S_{21}$ response measured directly at 10 dBm input power after necessary off-wafer calibration is plotted in Supplementary Information, Section 22, and aligns with the de-embedded $S_{21}$ response of the DUT by using the Thru standard as a reference. Reflection coefficients of transducers under varying power loads were then extracted via the de-embedded $S_{21}$ transmission response modelling.

Prior to in-situ temperature distribution measurements of the DUT during power load experiments, an emissivity correction is required for quantitative thermal mapping. While emissivity calibration is typically performed per constituent layer, the heterogeneous geometry with nonuniformity within the active region complicates this process. To address this, TF-SAW and LAW transducers were uniformly heated to predefined temperatures (50–130 °C, 20 °C increments) on a controlled hotplate. The surface emissivity was then calibrated by correlating IR camera radiation data with known temperatures. At each increment, samples were stabilized for 30 minutes, and measurements were averaged over five devices to minimize uncertainty.

**Nano/microstructure analysis**

Cross-section samples of the broken TF-SAW transducer, as well as pristine and damaged LAW transducers, were prepared for transmission electron microscopy (TEM) using an FEI Helios G4 UX DualBeam focused-ion-beam/SEM system. Surface and cross-sectional morphologies were examined with a JEOL-7800F SEM. High-resolution TEM (HR-TEM) and high-angle annular dark-field scanning TEM (HAADF-STEM) images were obtained using a JEOL JEM-ARM200F STEM equipped with a Cs probe corrector. This instrument, featuring dual wide-area (100 $mm^2$) silicon drift detectors, was also used for energy-dispersive X-ray spectroscopy (EDS) elemental mapping to assess compositional uniformity and interfacial quality across the device stack.

## Availability of Data and Materials

All data are fully available without restriction.

## References

1. Chou, M.-H. *et al.* Deterministic multi-phonon entanglement between two mechanical resonators on separate substrates. *Nat. Commun.* **16**, 1450 (2025).
2. Andrews, R. W. *et al.* Bidirectional and efficient conversion between microwave and optical light. *Nat. Phys.* **10**, 321–326 (2014).
3. Williamson, S. S. The success of electric mobility will depend on power electronics. *Nat. Electron.* **5**, 14–15 (2022).
4. Stolt, E. *et al.* A spurious-free piezoelectric resonator based 3.2 kW DC–DC converter for EV on-board chargers. *IEEE Trans. Power Electron.* **39**, 2478–2488 (2024).
5. Manzaneque, T., Lu, R., Yang, Y. & Gong, S. A high FoM lithium niobate resonant transformer for passive voltage amplification. In *Proc. 2017 19th International Conference on Solid-State Sensors, Actuators and Microsystems (TRANSDUCERS)* 798–801 (IEEE, 2017).
6. Tharpe, T., Hershkovitz, E., Hakim, F., Kim, H. & Tabrizian, R. Nanoelectromechanical resonators for gigahertz frequency control based on hafnia–zirconia–alumina superlattices. *Nat. Electron.* **6**, 599–609 (2023).
7. Giribaldi, G., Colombo, L., Simeoni, P. & Rinaldi, M. Compact and wideband nanoacoustic pass-band filters for future 5G and 6G cellular radios. *Nat. Commun.* **15**, 304 (2024).
8. Hakim, F., Rudawski, N. G., Tharpe, T. & Tabrizian, R. A ferroelectric-gate fin microwave acoustic spectral processor. *Nat. Electron.* **7**, 147–156 (2024).
9. Hackett, L. *et al.* Towards single-chip radiofrequency signal processing via acoustoelectric electron–phonon interactions. *Nat. Commun.* **12**, 2769 (2021).
10. Kuang, T. *et al.* Nonlinear multi-frequency phonon lasers with active levitated optomechanics. *Nat. Phys.* **19**, 414–419 (2023).
11. Lüpke, U., Rodrigues, I. C., Yang, Y., Fadel, M. & Chu, Y. Engineering multimode interactions in circuit quantum acoustodynamics. *Nat. Phys.* **20**, 564–570 (2024).
12. Chu, Y. *et al.* Quantum acoustics with superconducting qubits. *Science* **358**, 199–202 (2017).
13. Hackett, L. *et al.* Non-reciprocal acoustoelectric microwave amplifiers with net gain and low noise in continuous operation. *Nat. Electron.* **6**, 76–85 (2023).
14. Wollack, E. A. *et al.* Quantum state preparation and tomography of entangled mechanical resonators. *Nature* **604**, 463–467 (2022).
15. Devendran, C., Collins, D. J., Ai, Y. & Neild, A. Huygens-Fresnel acoustic interference and the development of robust time-averaged patterns from traveling surface acoustic waves. *Phys. Rev. Lett.* **118**, 154501 (2017).
16. Lin, Q. *et al.* Optical multi-beam steering and communication using integrated acousto-optics arrays. *Nat. Commun.* **16**, 4501 (2025).
17. Zhang, S. *et al.* Surface acoustic wave devices using lithium niobate on silicon carbide. *IEEE Trans. Microw. Theory Tech.* **68**, 3653–3666 (2020).
18. Hesjedal, T., Mohanty, J., Kubat, F., Ruile, W. & Reindl, L. M. A microscopic view on

acoustomigration. *IEEE Trans. Ultrason. Ferroelectr. Freq. Control* **52**, 1584–1593 (2005).

19. Zink, B. L., Pietri, R. & Hellman, F. Thermal conductivity and specific heat of thin-film amorphous silicon. *Phys. Rev. Lett.* **96**, 055902 (2006).
20. Wang, W. *et al.* Ultra-high power handling capability bulk acoustic wave filters using square array topology. *J. Micromechanics Microengineering* **35**, 085009 (2025).
21. Aigner, R. *et al.* BAW filters for 5G bands. In *Proc. 2018 IEEE International Electron Devices Meeting (IEDM)* 14.5.1-14.5.4 (IEEE, 2018).
22. Izhar *et al.* Periodically poled aluminum scandium nitride bulk acoustic wave resonators and filters for communications in the 6G era. *Microsyst. Nanoeng.* **11**, 1–11 (2025).
23. Schaffer, Z., Simeoni, P. & Piazza, G. 33 GHz overmoded bulk acoustic resonator. *IEEE Microw. Wirel. Compon. Lett.* **32**, 656–659 (2022).
24. Kramer, J. *et al.* 57 GHz acoustic resonator with $k^2$ of 7.3 % and $Q$ of 56 in thin-film lithium niobate. In *Proc. 2022 International Electron Devices Meeting (IEDM)* 16.4.1-16.4.4 (IEEE, 2022).
25. Yang, K. *et al.* Nanosheet lithium niobate acoustic resonator for mmWave frequencies. *IEEE Electron Device Lett.* **45**, 272–275 (2024).
26. Ouyang, P., Yi, X. & Li, G. Single-crystalline bulk acoustic wave resonators fabricated with AlN film grown by a combination of PLD and MOCVD methods. *IEEE Electron Device Lett.* **45**, 538–541 (2024).
27. Yang, Y., Lu, R., Gao, L. & Gong, S. 10–60-GHz electromechanical resonators using thin-film lithium niobate. *IEEE Trans. Microw. Theory Tech.* **68**, 5211–5220 (2020).
28. Rassay, S. *et al.* Intrinsically switchable ferroelectric scandium aluminum nitride lamb-mode resonators. *IEEE Electron Device Lett.* **42**, 1065–1068 (2021).
29. Qamar, A. & Rais-Zadeh, M. Coupled BAW/SAW resonators using AlN/Mo/Si and AlN/Mo/GaN layered structures. *IEEE Electron Device Lett.* **40**, 321–324 (2019).
30. Zhang, L. *et al.* High-performance acoustic wave devices on $LiTaO_3$/SiC hetero-substrates. *IEEE Trans. Microw. Theory Tech.* **71**, 4182–4192 (2023).
31. Liu, P. *et al.* A spurious-free SAW resonator with near-zero TCF using $LiNbO_3$/$SiO_2$/quartz. *IEEE Electron Device Lett.* **44**, 1796–1799 (2023).
32. Xu, H. *et al.* SAW filters on $LiNbO_3$/SiC heterostructure for 5G n77 and n78 band applications. *IEEE Trans. Ultrason. Ferroelectr. Freq. Control* **70**, 1157–1169 (2023).
33. Xu, H. *et al.* Large-range spurious mode elimination for wideband SAW filters on $LiNbO_3$/$SiO_2$/Si platform by $LiNbO_3$ cut angle modulation. *IEEE Trans. Ultrason. Ferroelectr. Freq. Control* **69**, 3117–3125 (2022).
34. Shen, J. *et al.* A low-loss wideband SAW filter with low drift using multilayered structure. *IEEE Electron Device Lett.* **43**, 1371–1374 (2022).
35. Zheng, P. *et al.* Miniaturized dual-mode SAW filters using 6-inch $LiNbO_3$-on-SiC for 5GNR and WiFi 6. In *Proc. 2023 International Electron Devices Meeting (IEDM)* 1–4 (IEEE, 2023).
36. Zhang, L. *et al.* High-performance surface acoustic wave filters for X-Band applications. *IEEE Trans. Microw. Theory Tech.* 1–8 (2024).

37. Larkin, J. M. & McGaughey, A. J. H. Thermal conductivity accumulation in amorphous silica and amorphous silicon. *Phys. Rev. B* **89**, 144303 (2014).
38. Segovia-Fernandez, J. & Yen, E. T.-T. Resonant confiners for acoustic loss mitigation in bulk acoustic wave resonators. In *Proc. 2023 IEEE 36th International Conference on Micro Electro-Mechanical Systems (MEMS)* 165–168 (2023).
39. Lakin, K. M. Bulk acoustic wave coupled resonator filters. In *Proc. 2002 IEEE International Frequency Control Symposium and PDA Exhibition* 8–14 (IEEE, 2002).
40. Bahr, B., Marathe, R. & Weinstein, D. Phononic crystals for acoustic confinement in CMOS-MEMS resonators. In *Proc. 2014 IEEE International Frequency Control Symposium (IFCS)* 1–4 (2014).
41. Woon, W.-Y. *et al.* Thermal management materials for 3D-stacked integrated circuits. *Nat. Rev. Electr. Eng.* **2**, 598–613 (2025).
42. Hashimoto, K. -y., Watanabe, Y., Akahane, M. & Yamaguchi, M. Analysis of acoustic properties of multi-layered structures by means of effective acoustic impedance matrix. In *Proc. 1990 IEEE Symposium on Ultrasonics* 937–942 (IEEE, 1990).
43. Wang, Y., Hashimoto, K., Omori, T. & Yamaguchi, M. Change in piezoelectric boundary acoustic wave characteristics with overlay and metal grating materials. *IEEE Trans. Ultrason. Ferroelectr. Freq. Control* **57**, 16–22 (2010).
44. Qian, F., Zheng, J., Xu, J. & Yang, Y. Heterogeneous interface-enhanced thin-film SAW devices using lithium niobate on Si. *IEEE Microw. Wirel. Technol. Lett.* 1–4 (2024).
45. Du, X. *et al.* Near 6-GHz Sezawa mode surface acoustic wave resonators using AlScN on SiC. *J. Microelectromechanical Syst.* **33**, 577–585 (2024).
46. Plessky, V., Makkonen, T. & Salomaa, M. M. Leaky SAW in an isotropic substrate with thick electrodes. In *Proc. 2001 IEEE Ultrasonics Symposium* 239–242 (IEEE, 2001).
47. Guo, Y., Kadota, M. & Tanaka, S. Hetero acoustic layer surface acoustic wave resonator composed of $LiNbO_3$ and quartz. *IEEE Trans. Ultrason. Ferroelectr. Freq. Control* **71**, 182–190 (2024).
48. He, Y., Bahr, B., Si, M., Ye, P. & Weinstein, D. A tunable ferroelectric based unreleased RF resonator. *Microsyst. Nanoeng.* **6**, 8–8 (2020).
49. Anderson, J., He, Y., Bahr, B. & Weinstein, D. Integrated acoustic resonators in commercial fin field-effect transistor technology. *Nat. Electron.* **5**, 611–619 (2022).
50. Cahill, D. G., Watson, S. K. & Pohl, R. O. Lower limit to the thermal conductivity of disordered crystals. *Phys. Rev. B* **46**, 6131–6140 (1992).
51. Braun, J. L. *et al.* Size effects on the thermal conductivity of amorphous silicon thin films. *Phys. Rev. B* **93**, 140201 (2016).
52. Xiao, B. *et al.* Anisotropy-matched LN/quartz heterostructure with inherent spurious mitigation for wideband SAW devices. *IEEE Trans. Microw. Theory Tech.* **73**, 10080–10094 (2025).
53. Shen, J. *et al.* High-performance surface acoustic wave devices using $LiNbO_3/SiO_2/SiC$ multilayered substrates. *IEEE Trans. Microw. Theory Tech.* **69**, 3693–3705 (2021).
54. Liu, P. *et al.* Monolithic 1–6-GHz multiband acoustic filters using SH-SAW and LLSAW on $LiNbO_3/SiO_2/SiC$ platform. *IEEE Trans. Microw. Theory Tech.* **72**, 5653–5666 (2024).

55. Zheng, P. *et al.* Near 5-GHz longitudinal leaky surface acoustic wave devices on $LiNbO_3$/SiC substrates. *IEEE Trans. Microw. Theory Tech.* **72**, 1480–1488 (2024).
56. Xu, H. *et al.* Higher order mode elimination for SAW resonators based on $LiNbO_3$/$SiO_2$/poly-Si/Si substrate by Si orientation optimization. *J. Microelectromechanical Syst.* **33**, 163–173 (2024).
57. Ivira, B. *et al.* Self-heating study of bulk acoustic wave resonators under high RF power. *IEEE Trans. Ultrason. Ferroelectr. Freq. Control* **55**, 139–147 (2008).
58. Shen, J. *et al.* SAW filters with excellent temperature stability and high power handling using $LiTaO_3$/SiC bonded wafers. *J. Microelectromechanical Syst.* **31**, 186–193 (2022).
59. Shashwat, B. *et al.* Impact of in-plane residual stress on the performance of the film bulk acoustic resonators. In *Proc. 2023 IEEE International Ultrasonics Symposium (IUS)* 1–4 (IEEE, 2023).
60. Feld, D. A., Parker, R., Ruby, R., Bradley, P. & Dong, S. After 60 years: a new formula for computing quality factor is warranted. In *Proc. 2008 IEEE Ultrasonics Symposium* 431–436 (IEEE, 2008).
61. Naik, R. S., Lutsky, J. J., Reif, R. & Sodini, C. G. Electromechanical coupling constant extraction of thin-film piezoelectric materials using a bulk acoustic wave resonator. *IEEE Trans. Ultrason. Ferroelectr. Freq. Control* **45**, 257–263 (1998).
62. Gamble, K. & Buettner, W. Steady-state and transient thermal modeling for a SAW duplexer. In Proc. 2016 IEEE International Ultrasonics Symposium (IUS) 1–6 (IEEE, 2016).

## Acknowledgments

This work is supported in part by the National Natural Science Foundation of China 62304193, in part by the Hong Kong Research Grants Council 26202122, in part by the Hong Kong Innovation and Technology Commission ITS/144/23, and in part by the Hong Kong RGC Strategic Topics Grant STG3/E-602/23N. We acknowledge the support by the Nanosystem Fabrication Facility (CWB), HKUST, for device fabrication. We are grateful to N. Li from the Department of ECE, HKUST, for his assistance in the SEM characterization, and Y. Cai from the Material Characterization and Preparation Facility (MCPF), HKUST, for her assistance in FIB sample preparation and TEM characterization.

## Authors Contributions

Y. Y. conceived the idea, supervised the project and provided guidance to the authors. Y. Y. and F.Q. proposed the technical approach. F.Q. designed the layout. F.Q. and S.C. performed the numerical calculations and FEA simulations related to acoustic devices design and conducted the fabrication. F.Q. W.W. and J.X. performed the device characterizations. K.Y. J.Z. participated in the material characterizations. Z.R. X.L.

processed and analyzed the data. F.Q. and Y.Y. wrote the manuscript. All authors commented on the manuscript.

## Competing Interests

The authors declare no competing interests.

# Supplementary Information for ‘Suppressing Acoustomigration and Temperature Rise for High-power Robust Acoustics’

Fangsheng Qian, Shuhan Chen, Wei Wei, Jiashuai Xu, Kai Yang, Junyan Zheng, Zijun Ren, Xingyu Liu, and Yansong Yang*

Department of Electronic and Computer Engineering, The Hong Kong University of Science and Technology, Hong Kong, China.

*Correspondence and requests for materials should be addressed to: * Yansong Yang: eeyyang@ust.hk

# Table of contents

## 1 Boundary analysis for maximizing acoustic wave confinement

We assume a three-layer heterostructure in which an isotropic and homogeneous electrical insulating $SiO_2$ layer with thickness $h$ is inserted in between the $LiNbO_3$ and the top cladding layers. The initial boundary engineering of the proposed LAW configuration aims at maximizing the acoustic wave confinement. The LAW velocity in the proposed configuration can be calculated starting from Equations (S1) and (S2)[1]:

$$\Omega(V_{\mathrm{LAW}}) + \Delta(V_{\mathrm{LAW}}) = K^2, \tag{S1}$$

$$\Delta(V) = \frac{\rho'}{\rho}\left(\frac{V_B'}{V_B}\right)^2 \Omega'(V), \tag{S2}$$

where $\Omega(V) = \sqrt{1-(V/V_B)^2}$ and $\Omega'(V) = \sqrt{1-(V/V_B')^2}$. $K^2$ refers to the electromechanical coupling coefficient for thickness-shear vibration and $V_B$ represents the SH-mode velocity of the $LiNbO_3$ layer. For the specific three-layer structure, $\Delta(V)$ can be derived from the recursive relation of acoustic impedance matrixes[2], this yields:

$$\Delta(V) = \frac{\rho''}{\rho}\left(\frac{V_B''}{V_B}\right)^2 \Omega''(V) \frac{1 + \frac{\rho''\Omega''(V)}{\rho'\Omega'(V)}\left(\frac{V_B''}{V_B'}\right)^2 \tanh\big(\beta h \Omega''(V)\big)}{\frac{\rho''\Omega''(V)}{\rho'\Omega'(V)}\left(\frac{V_B''}{V_B'}\right)^2 + \tanh\big(\beta h \Omega''(V)\big)}, \tag{S3}$$

where $\Omega''(V) = \sqrt{1-(V/V_B'')^2}$, and $\beta$ is denoted as the wavenumber of LAW along the boundary between the sandwiched $SiO_2$ layer and the $LiNbO_3$ layer. Generally, the LAW velocity $V_{\mathrm{LAW}}$ should be slower than that of slow shear bulk waves in three layers. Otherwise, the LAW exhibits a leaky nature with degraded performance. When $V_B'' < V < V_B'$, it should be noted that $\Omega'(V)$ and $\Omega''(V)$ are purely real and purely imaginary, respectively. Under this circumstance, $\Omega''(V)$ can be rewritten as: $\Omega''(V) = \mathrm{i}\zeta$, $\zeta = \mathrm{Im}[\Omega''(V)]$, $\zeta = (V/V_B'')^2 - 1$. Rearrange Equation (S3) by using the relationship of $\tanh(\mathrm{i}x) = \mathrm{i}\tan(x)$, we get

$$\Delta(V) = \frac{\rho''}{\rho}\left(\frac{V_B''}{V_B}\right)^2 \zeta \frac{1 - \frac{\rho''\zeta}{\rho'\Omega'(V)}\left(\frac{V_B''}{V_B'}\right)^2 \tan(\beta h \zeta)}{\frac{\rho''\zeta}{\rho'\Omega'(V)}\left(\frac{V_B''}{V_B'}\right)^2 + \tan(\beta h \zeta)}. \tag{S4}$$

From the perspective of material designs, it is recommended that materials for the upper and sandwiched layers should feature large $\rho'' V_B'' / \rho' V_B'$ to achieve a large $-\Delta(V)$, giving rise to an increased coupling coefficient and extended the existence range of LAW. Equation (S4) implies that $\Delta(V)$ should be negative provided that:

$$\frac{\rho''\varsigma}{\rho'\Omega'(V)}\left(\frac{V_B''}{V_B'}\right)^2 \tan(\beta h\zeta) > 1. \tag{S5}$$

Given the sandwiched materials $SiO_2$ thanks to its good insulation properties, we can attempt to extend the range to $V_B'' < V_{\mathrm{LAW}} < V_B'$ when satisfying the condition of equation (S5). For simplicity, we define the following relationships:

$$\begin{cases} \frac{\rho''}{\rho} = x, \frac{V_B''}{V_B} = y \\ \frac{\rho''}{\rho'} = a, \frac{V_B''}{V_B'} = b \end{cases}.$$

Considering the electrode thickness 0.06$\lambda$, we have $V_B < V_{\mathrm{LAW}} < V_B'' < V_B'$,i.e., $b < 1 < y$. $\Omega(V) + \Delta(V)$ is numerically solved to show monotonical decrease with $V$.

For $V_{\mathrm{LAW}} < V_B''$, it exists $\Omega(V_B'') + \Delta(V_B'') = \sqrt{1 - (V_B''/V_B)^2} + 0 \leq K^2$, which can be derived as

$$y \geq \sqrt{1 - K^2}. \tag{S6}$$

The relation show in Equation (S7) is naturally satisfied for $y > 1$ known from existing velocity relations.

For $V_B < V_{\mathrm{LAW}}$, it exists

$$\Omega(V_B) + \Delta(V_B) = xy^2\sqrt{1-\left(\frac{1}{y}\right)^2}\,\frac{1 + ab^2\frac{\sqrt{1-\left(\frac{1}{y}\right)^2}}{\sqrt{1-\left(\frac{b}{y}\right)^2}}\tanh\left(\beta h\sqrt{1-\left(\frac{1}{y}\right)^2}\right)}{ab^2\frac{\sqrt{1-\left(\frac{1}{y}\right)^2}}{\sqrt{1-\left(\frac{b}{y}\right)^2}} + \tanh\left(\beta h\sqrt{1-\left(\frac{1}{y}\right)^2}\right)} \geq K^2. \tag{S10}$$

From Equation (S10), the LAW can be confined within the range:

$$b \leq \left(\sqrt{\frac{y^2C}{a^2} + \frac{C^2}{4a^4}} - \frac{C}{2a^2}\right)^{1/2}, \tag{S11}$$

where $C$ is denoted as:

$$C = \frac{1}{y^2 - 1}\left( (\frac{xy\sqrt{y^2 - 1} - K^2 \tanh\left(\beta h\sqrt{1 - \left(\frac{1}{y}\right)^2}\right)}{K^2 - xy\sqrt{y^2 - 1}\tanh\left(\beta h\sqrt{1 - \left(\frac{1}{y}\right)^2}\right)}) \right)^2 . \qquad \text{(S12)}$$

## 2 Energy confinement comparison of LAW transducers with different cladding layers

Early investigations reveal that the power dissipation peak arises not only from Joule heating but also from SAW propagation losses, including bulk wave radiation. This necessitates energy confinement as the primary criterion for evaluating LAW configuration feasibility. As discussed in Fig. 2 of the main text, the very structure of a LAW transducer inherently requires an additional elastic medium atop a conventional TF-SAW design. While shear horizontal (SH) acoustic wave confinement has been validated for the $LiNbO_3$ on sapphire platform[3], energy confinement at the $LiNbO_3$/cladding interface remains critical to evaluate. To verify it, quasi-3D unit-cell frequency-domain simulations using COMSOL Multiphysics 6.0 are performed to compare α-Si and α-$SiO_2$ cladding layers. **Supplementary Fig. S1(a)** displays the simulated admittance curves for LAW transducers with 3 μm α-Si and α-$SiO_2$. For α-Si silicon cladding, as shown in **Supplementary Fig. S1a**, the targeted SH resonance mode achieves a high electromechanical coupling coefficient of 17.8% with minimal bulk wave radiation, confirming the validity of boundary engineering. Mode A, corresponding to the first mechanical resonance, exhibits strong acoustic energy confinement at the $LiNbO_3$/α-Si interface, characterized by rapid energy decay in the thickness direction. Mode B, representing the second mechanical resonance, displays quasi-Sezawa behavior by coupling the fundamental symmetrical (S0) and asymmetrical (A0) Lamb wave modes. In contrast, the α-$SiO_2$ cladding shown in **Supplementary Fig. S1a** introduces significant spurious responses beyond the first resonance. The main mode presents a small $k_t^2$ of 8.1% and a much smaller impedance ratio (40 dB) compared to that of α-Si. Mode C, associated with the first mechanical resonance, shows slow SH component decay across the amorphous silicon dioxide layer, while Modes D and E correspond to bulk shear horizontal wave radiation, indicating poor energy confinement of the LAW transducer with α-$SiO_2$ cladding layer. These findings agree well with our theoretical calculation for LAW transducers design, validating the methodology of boundary engineering for acoustic energy confinement.

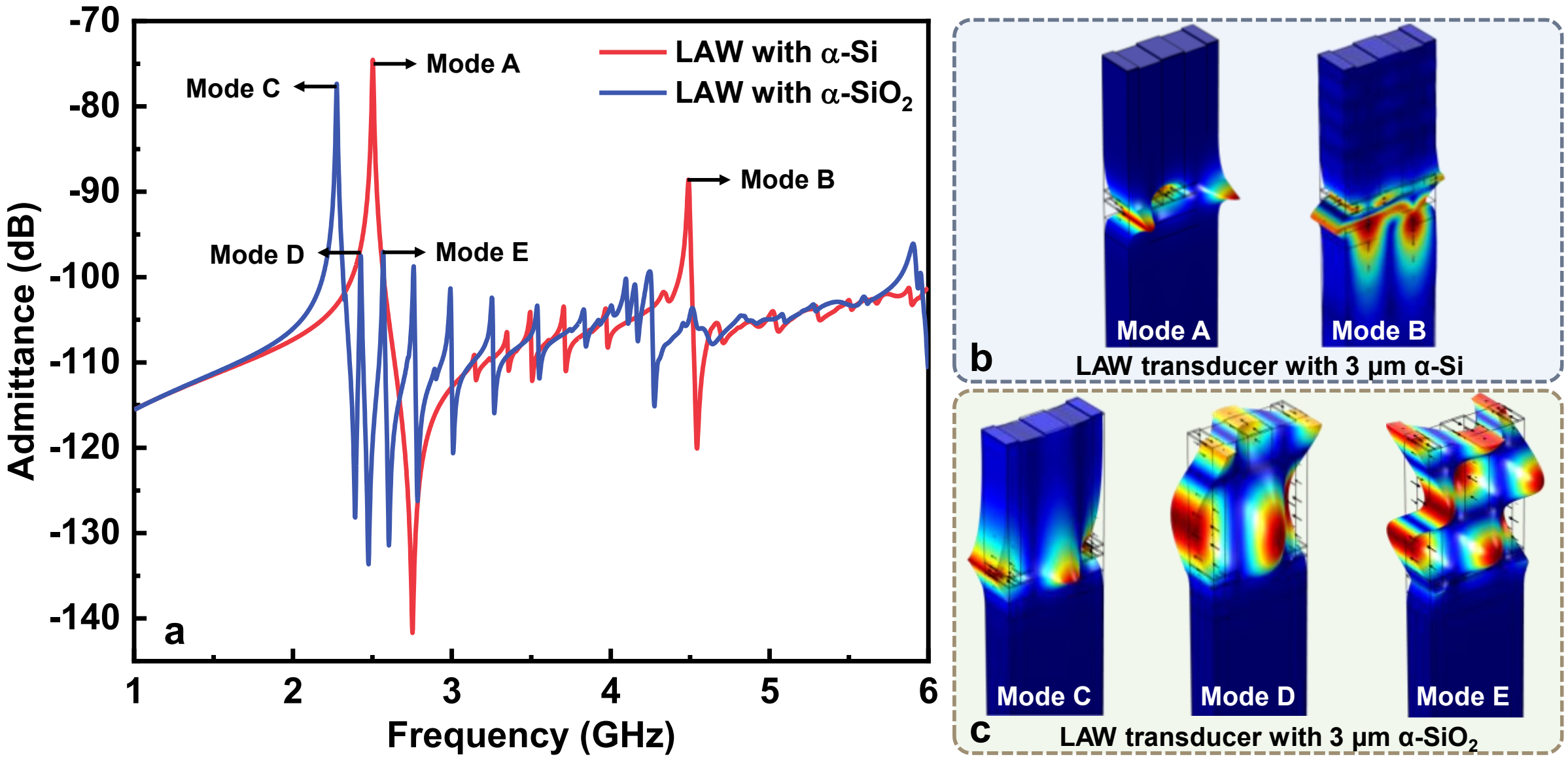

**Supplementary Fig. S1 | Energy confinement comparison of LAW transducers with different thick cladding layers.** **(a)** Simulated admittance curves of LAW transducers with thick α-Si and α-$SiO_2$ cladding layers. **(b)** Corresponding displacement mode shapes at $f_r$ of Mode A and Mode B on the LAW transducer with 3 μm α-Si cladding layer. **(c)** Corresponding displacement mode shapes at $f_r$ of Mode C-E on the LAW transducer with 3 μm α-$SiO_2$ cladding layer.

## 3 Measured TCF results of TF-SAW and $SiO_2$-overcoated TF-SAW devices

As shown in **Supplementary Fig. S2(a)**, the measured admittance curve of an Au/$LiNbO_3$/Sapphire TF-SAW transducer (λ = 1.2 μm) exhibits a downward shift with increasing temperature, confirming a negative temperature coefficient of frequency (TCF). As the temperature decreases from 25 °C to 5 °C, the 3-dB quality factor ($Q_p$) at parallel frequency ($f_p$) increases sharply from 98 to 314, marking a turnover point in dominant loss mechanisms. Additionally, spurious response near $f_p$ also deteriorates $Q_p$ at 25 °C. Continuously decreasing the temperature to -35 °C results in negligible $Q_p$ variation, indicating stabilized energy loss behavior. A quadratic polynomial is utilized to fit the extracted frequency shifts at different temperatures, which shows a first-order TCF of -64.30 ppm/°C and a second-order TCF of -9.672 ppb/°C$^2$, as shown in **Supplementary Fig. S2(b).** Compared to the uncoated TF-SAW transducer, $f_r$ of 500 nm $SiO_2$-overcoated TF-SAW transducer shifts from 1.941 GHz to 2.318 GHz but exhibits degraded performance, as illustrated in **Supplementary Figs. S2(c) and S2(d)**. The admittance ratio decreases from 58 dB to 41 dB, and the $k_t^2$ also decreases from 28% to 11.40%. Notably, the $SiO_2$ coating exacerbates the TCF, increasing the first-order and second-order TCF to -161.96 ppm/°C and -622.18 ppb/°C$^2$, respectively.

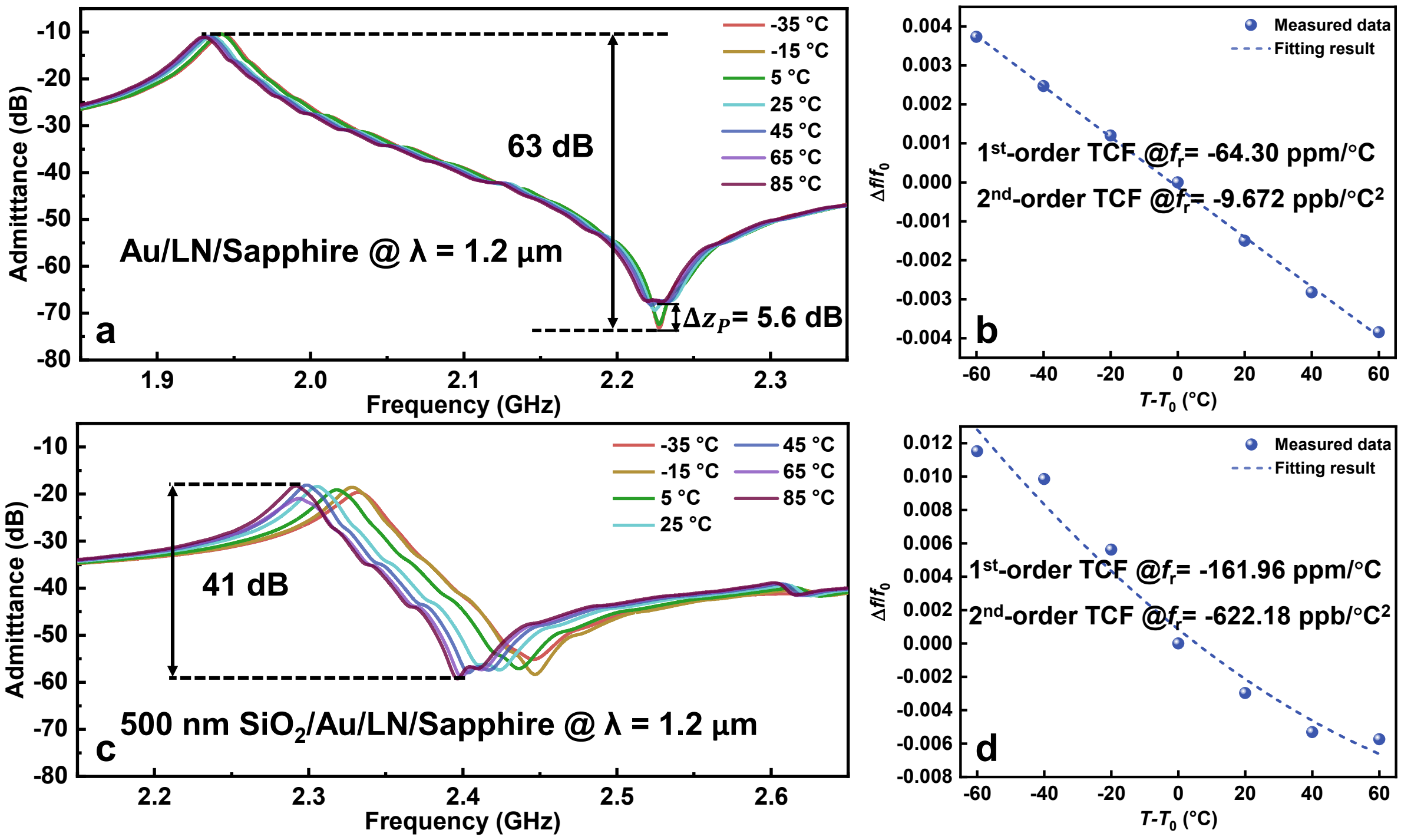

**Supplementary Fig. S2 | Temperature-dependent admittance curves of TF-SAW and $SiO_2$-overcoated TF-SAW transducers. (a)** Measured admittance curves of a TF-SAW transducer (λ = 1.2 μm) across a temperature of -35 °C to 85 °C. **(b)** Relative $f_r$ shift of the TF-SAW transducer (λ = 1.2 μm) under testing temperatures ranging from -35 °C to 85 °C, with 1$^{st}$ and 2$^{nd}$-order TCF fitting results. **(c)** Measured admittance curves of a 500 nm $SiO_2$-overcoated TF-SAW transducer (λ = 1.2 μm) across a temperature of -35 °C to 85 °C. **(d)** Relative $f_r$ shift of a 500 nm $SiO_2$-overcoated TF-SAW transducer (λ = 1.2 μm) under testing temperatures ranging from -35 °C to 85 °C, with 1$^{st}$ and 2$^{nd}$-order TCF fitting results.

## 4 Velocity saturation effect on the LAW platform

As shown in **Supplementary Fig. S3**, the resonant frequency of the fundamental layered SH0 mode increases with the thickness of the α-Si stress-manipulation layer, which modifies the effective Young's modulus and confines acoustic energy within the resonator cavity. Crucially, this frequency shift saturates at an α-Si thickness of approximately 0.6 μm, beyond which further increases have a negligible effect. This saturation behavior is elucidated by the simulated mode profile (inset of the figure for a 4-μm-thick α-Si layer), showing that the acoustic strain field is effectively confined at the IDT/$LiNbO_3$ interface and decays rapidly within the over-layer. As such, beyond the critical thickness, the stress-manipulation layer acts as a semi-infinite medium from the perspective of the guided mode, making the frequency independent of further thickness increases.

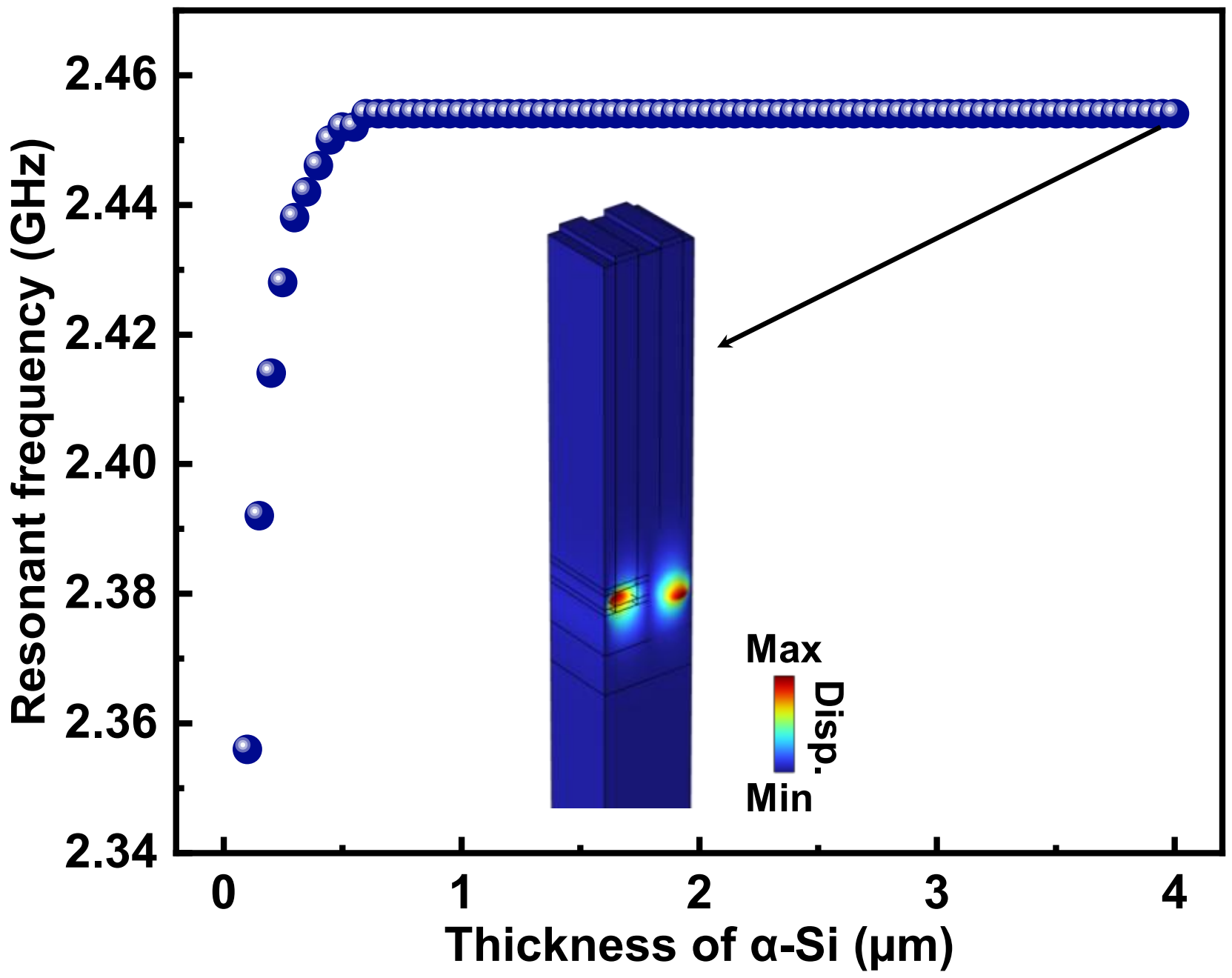

**Supplementary Fig. S3 | Resonant frequency dependence on the α-Si top-cladding thickness and corresponding mode profile at a thickness of 4 μm.**

## 5 EDS mapping of the as-fabricated LAW transducer

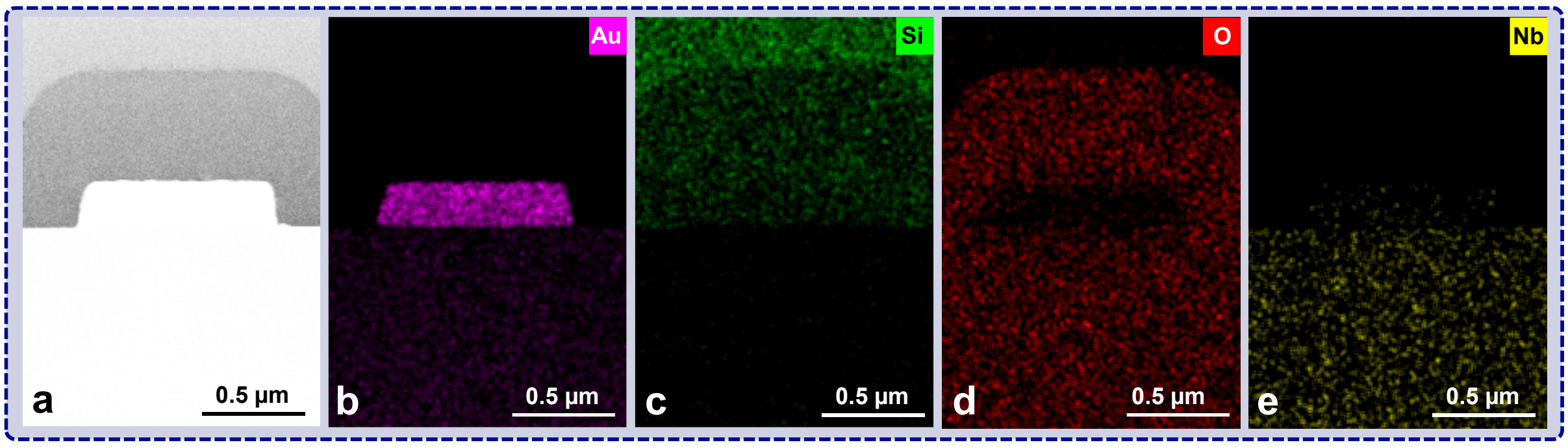

**Supplementary Fig. S4 | Energy dispersive spectroscopy (EDS) mapping of the as-fabricated LAW transducer. (a)** High-angle annular dark-field (HAADF) imaging STEM image illustrates the zoom-up view of LAW configuration having vertical stacks of α-Si/α-$SiO_2$/Au/$LiNbO_3$. Corresponding elemental mapping results for **(b)** Au, **(c)** Si, **(d)** O and **(e)** Nb, respectively. The EDS characterization well demonstrates the successful fabrication of the 3D-stacked LAW architecture.

## 6 Spurious modes suppression strategies for LAW transducers

The spurious modes observed in Figs. 4a and 4c are identified as two distinct types: transversal modes, arising from waveguiding in the aperture direction, and higher-order bulk waves, generated by scattering at vertical acoustic boundaries. The suppression of transversal modes has been extensively investigated in piezoelectric-on-insulator (POI) platforms. Established techniques, including piston-mode designs[4], tilted electrodes[5,6], and slowness curve modulation[7], as demonstrated in prior work, can effectively flatten the lateral velocity profile and mitigate these modes.

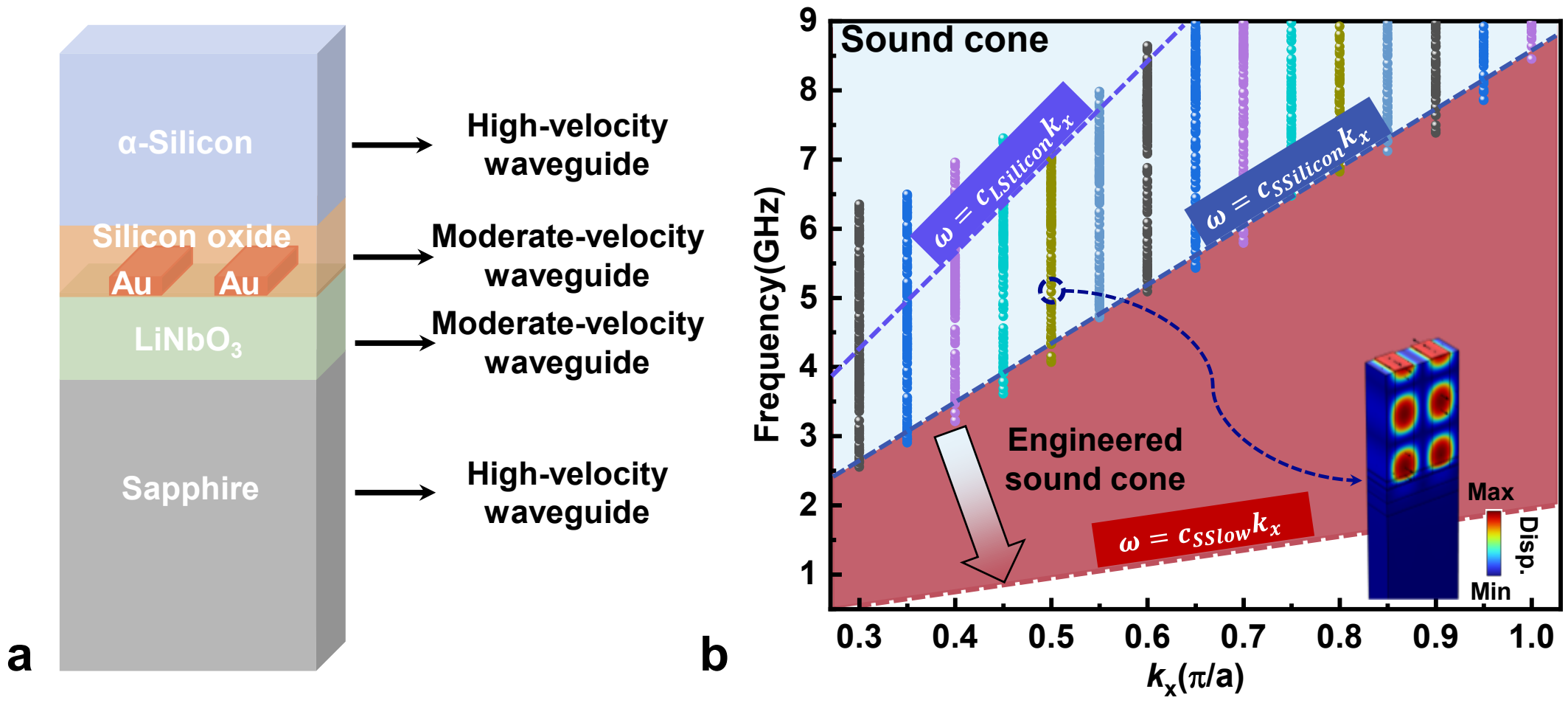

**Supplementary Fig. S5 | Acoustic waveguide design for spurious-mode suppression. (a)**, Schematic of the initial vertically stacked LAW transducer and its associated acoustic waveguides. **(b)**, Dispersion relation of an optimized LAW transducer design for suppressing bulk vibrational modes. The light-blue region denotes the initial sound cone for free-propagating bulk modes. The red-extended sound cone represents an engineered waveguide region with sufficiently low acoustic velocity, which redirects bulk modes away from the sandwiched resonator cavity.

To address the higher-order bulk waves, we propose and demonstrate a mechanical bandgap engineering strategy intrinsic to the layered architecture. The LAW structure can be conceptualized as a vertical stack of acoustic waveguides, as shown in **Supplementary Fig. S5a.** Given dispersion relationship for each mode, the target SH0 mode is confined within the $LiNbO_3$ layer, while parasitic bulk waves can propagate into the overlying $SiO_2$ and α-Si layers. To suppress those unwanted spurious waves, we engineer the sound cone in $\vec{k}$-space to guide them into an additional, uppermost slow-wave layer (**Supplementary Fig. S5b**). In this engineered sound cone, initial bulk modes vibrating in the upper $SiO_2$ and α-Si waveguides propagate into a slower wave layer, which sits on the top of the LAW structure (**Supplementary Fig. S6a**).

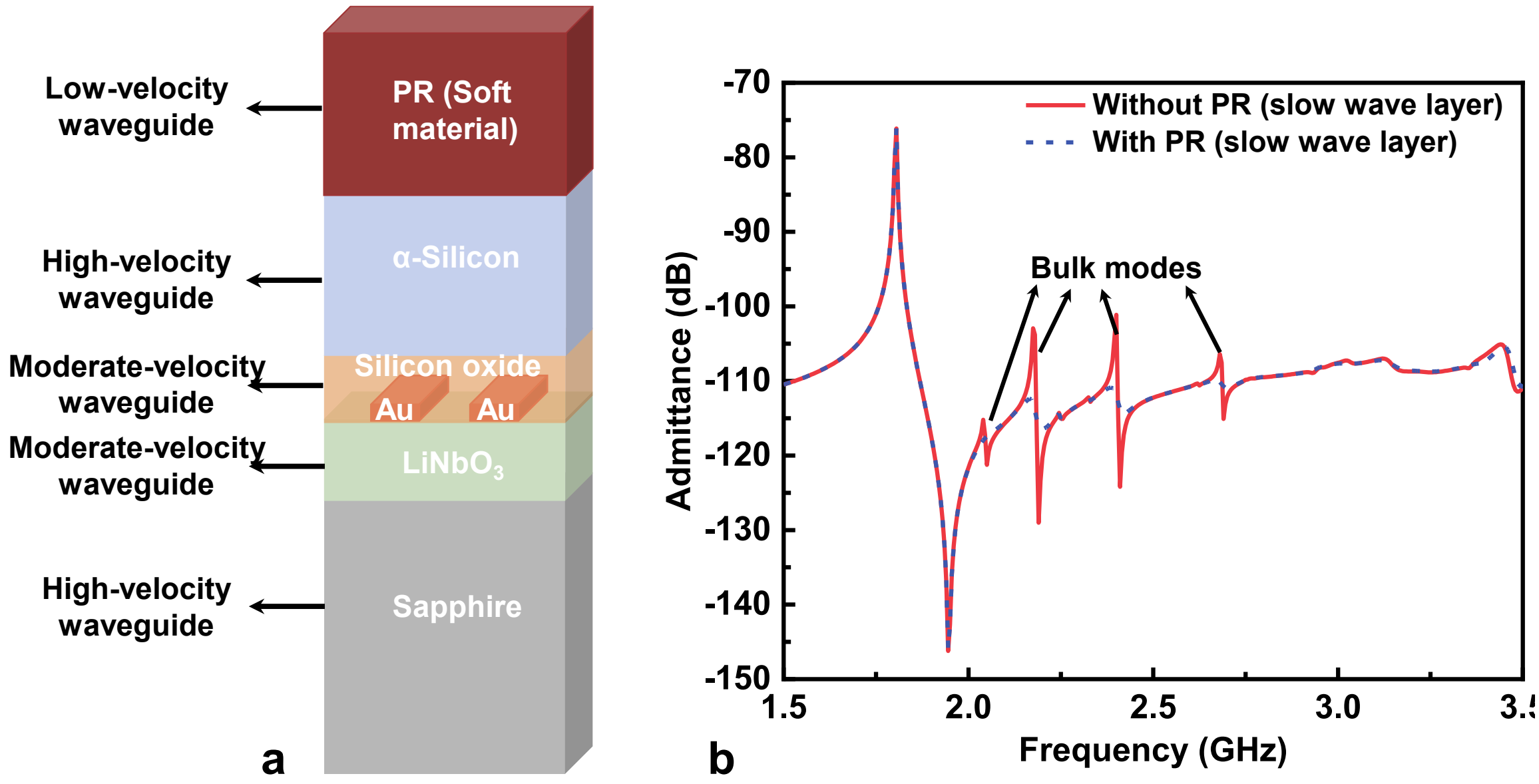

**Supplementary Fig. S6 | Simulated verification of bulk-mode suppression via a slow-wave photoresist (PR) waveguide. (a)**, Schematic of the engineered LAW transducer conceptualized as a vertical stack of distinct acoustic waveguides. **(b)**, Simulated admittance curves for the LAW transducer design with and without a slow-wave PR medium.

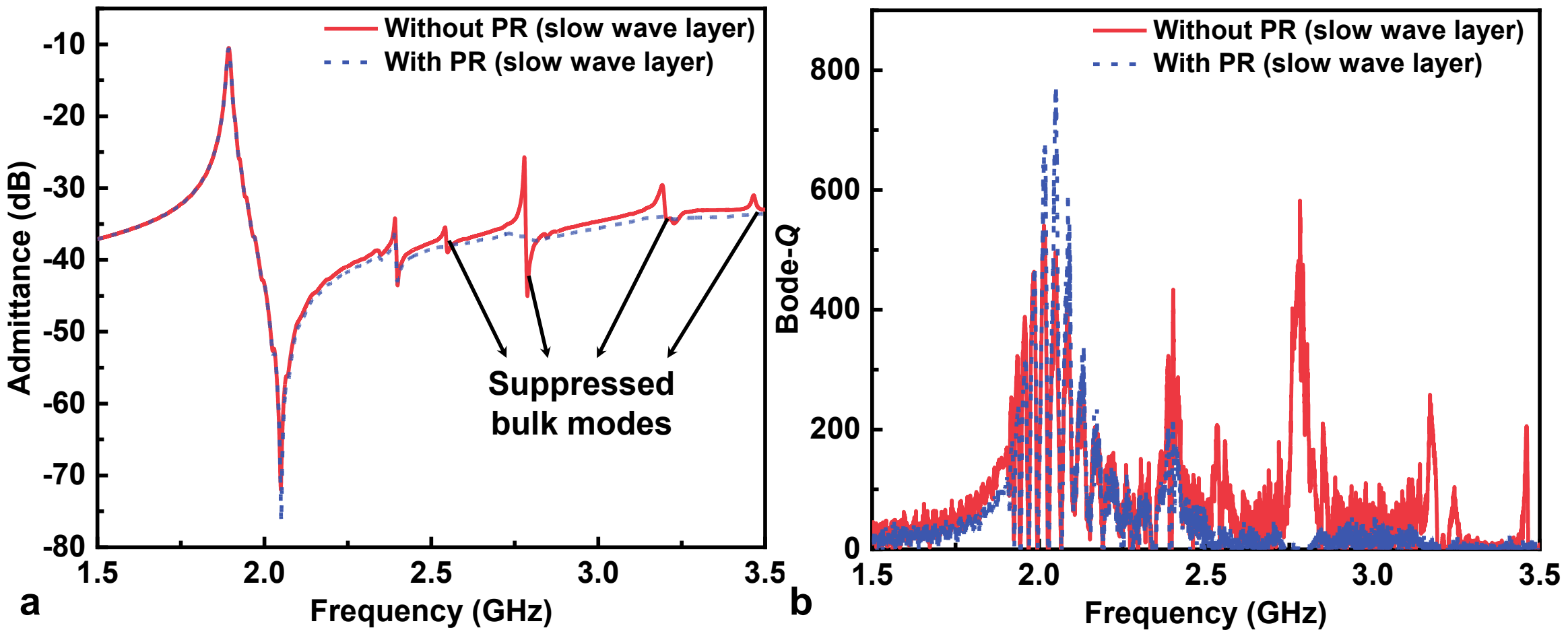

**Supplementary Fig. S7 | Experimental verification of spurious bulk-mode suppression. (a)**, Measured admittance and **(b)** extracted Bode-*Q* characteristics from one-port measurements, comparing the LAW transducer with (red) and without (blue) the slow-wave PR waveguide. The engineered mechanical boundary suppresses bulk modes, increasing the maximum Bode-*Q* of the targeted mode from 541 to 776.

Here, a photoresist (PR) layer serves as this slow-wave medium. Finite-element simulations confirm this principle: adding a PR waveguide significantly suppresses bulk-wave spurs without degrading the primary SH0 mode's electromechanical coupling ($k_t^2$) or quality factor (**Supplementary Fig. S6b**). We experimentally validated this by coating a 5.3-μm-thick PR layer on a LAW transducer (with probing pads exposed for electrical contact). The measured admittance (**Supplementary Fig. S7a**) shows effective bulk-mode suppression. the maximum Bode-$Q$ of the target mode increased from 541 to 776, while spurious modes were reduced to negligible levels (**Supplementary Fig. S7b**).

Wavelength-dependent measurements (**Supplementary Fig. S8**) further confirm the robustness of this mechanical boundary condition design. Key metrics, including admittance ratio (AR), Bode-$Q_{max}$, $k_t^2$, and the figure of merit (FoM = Bode-$Q_{max} \times k_t^2$), are summarized in **Supplementary Fig. S9.** The engineered LAW transducer with a slow-wave overlay shows significant performance enhancement over both traditional SAW and baseline LAW devices. Additionally, this suppression strategy does not compromise power-handling capability: the maximum von Mises stress at the critical metal/$LiNbO_3$ interface remains unchanged, confirming that the thermal-mechanical integrity of the core resonator is preserved.

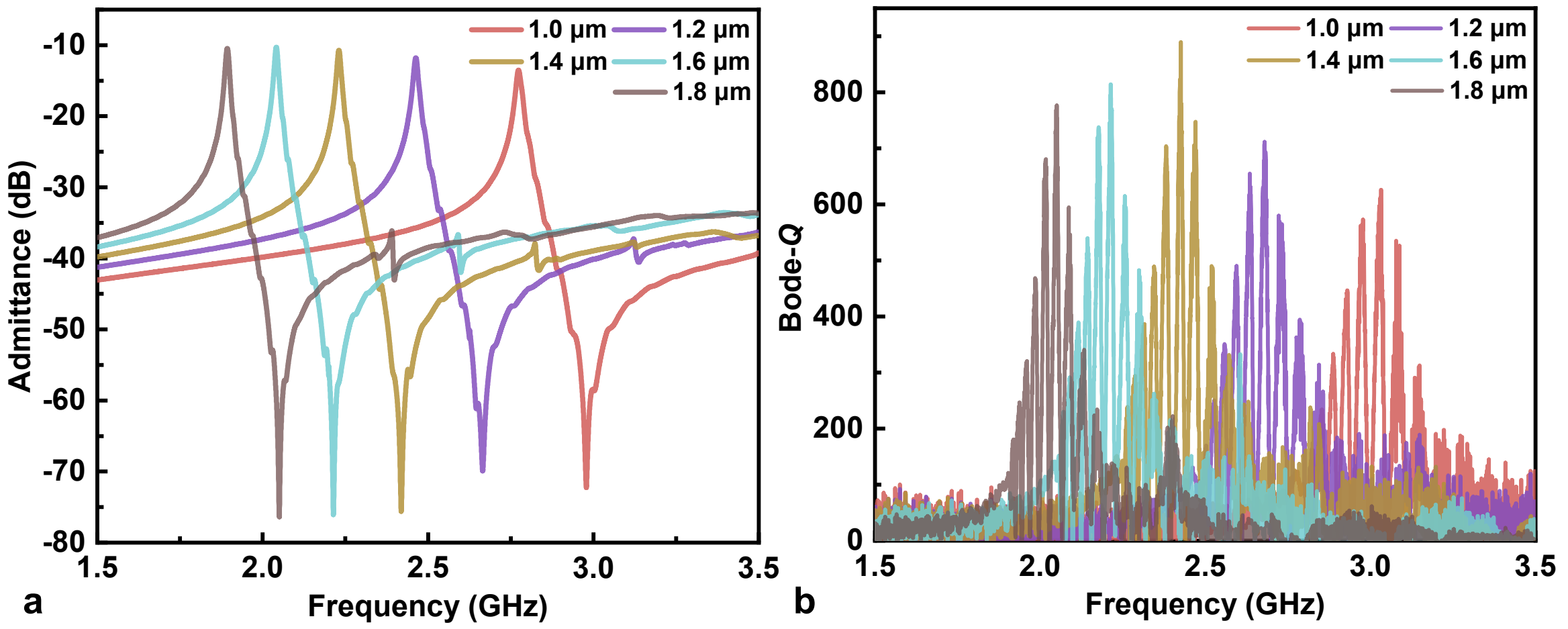

**Supplementary Fig. S8 | Wavelength-dependent performance of engineered LAW transducers with the slow-wave PR waveguide. (a)** Measured admittance curves and **(b)** extracted Bode-*Q* characteristics across different wavelengths, demonstrating the consistent performance enhancement benefiting from the spurious-mode suppression strategy.

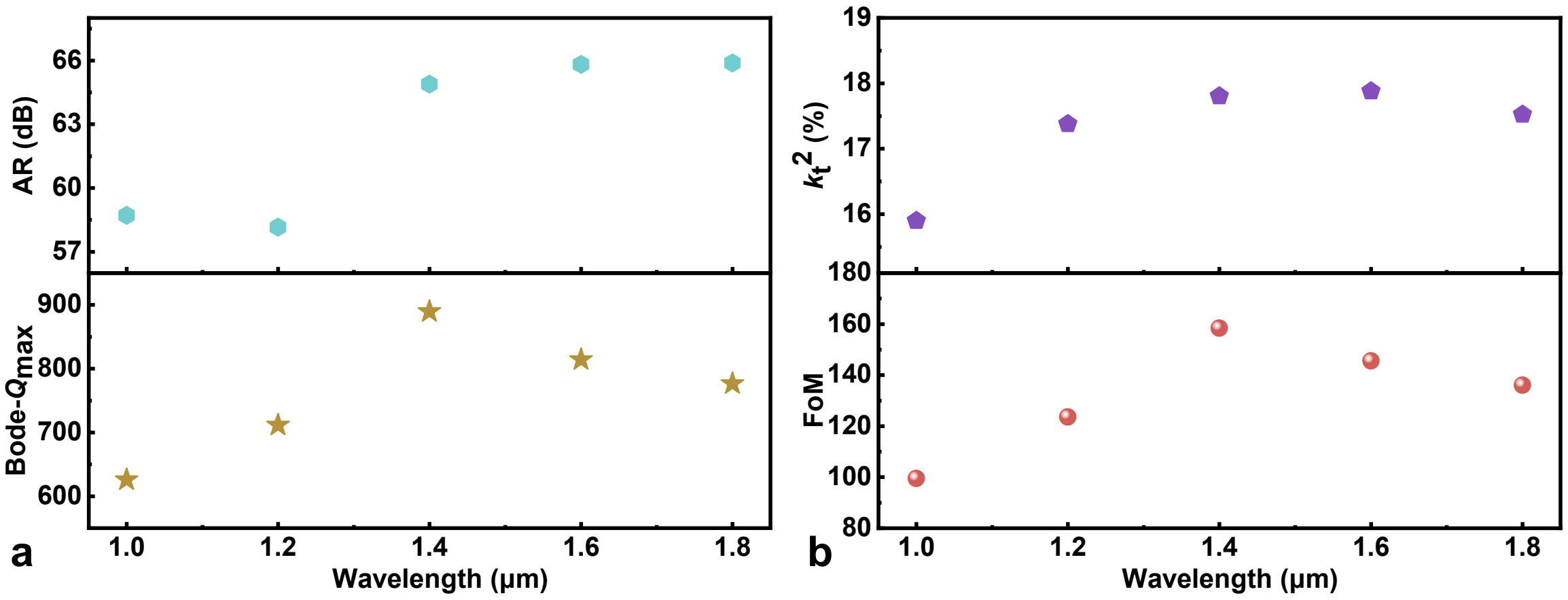

**Supplementary Fig. S9 | Wavelength-dependent performance metrics of the engineered LAW transducer.** Key metrics extracted for the device with the slow-wave PR layer across wavelengths, including **(a)** admittance ratio (AR) and maximum Bode-*Q* (Bode-$Q_{max}$), and (b) efficient electromechanical coupling coefficient ($k_t^2$) and figure of merit (FoM, defined as Bode-$Q_{max}$ × $k_t^2$).

# 7 Measurement results obtained from the baseline TF-SAW transducers

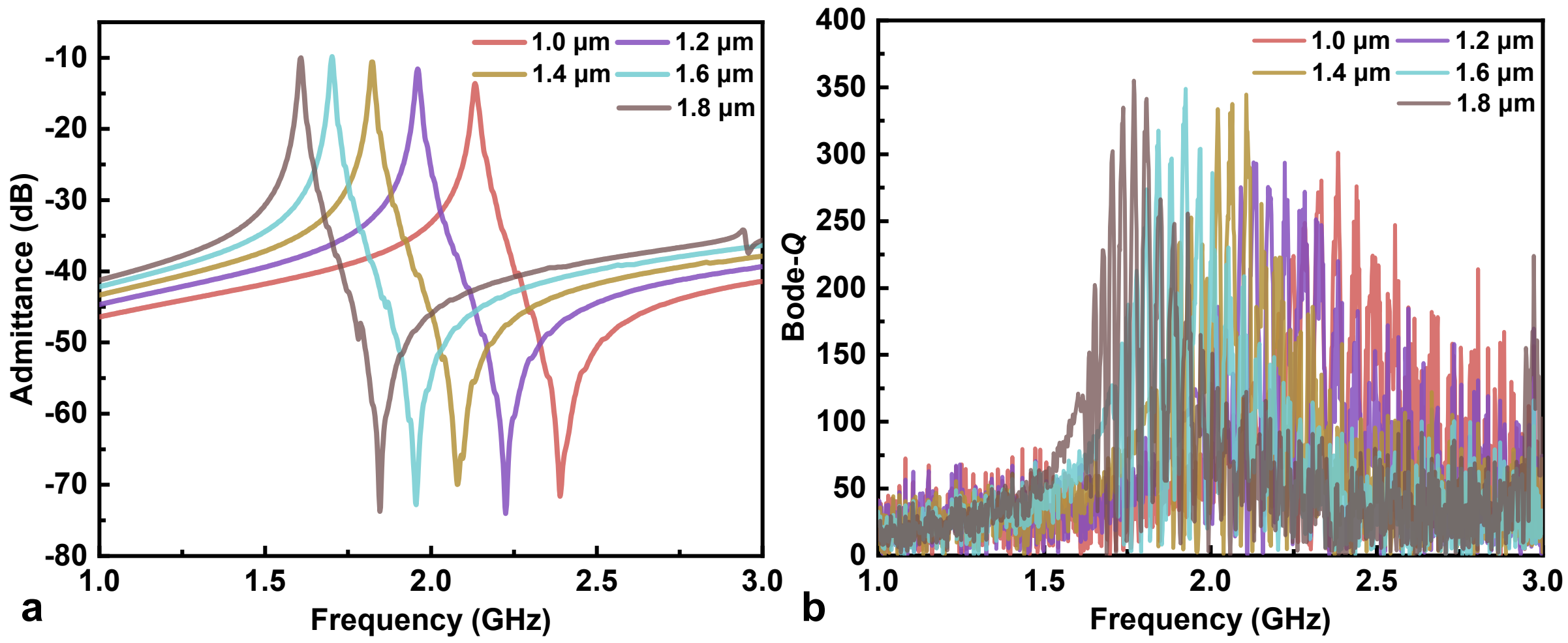

**Supplementary Fig. S10 | Baseline performance of the SAW transducer.** Wavelength dependence of the **(a)**, measured admittance curves and **(b)**, extracted Bode-*Q* characteristics for the TF-SAW transducer prior to the deposition of any upper cladding layers.

## 8 Performance comparison between TF-SAW, $SiO_2$-overcoated TF-SAW, and LAW transducers

**Supplementary Table S1. Summarized key metrics of transducers in three configurations**

| | TF-SAW transducer | $SiO_2$-overcoated TF-SAW transducer | LAW transducer |
|---|---|---|---|
| $f_s$ | 1.951 GHz | 2.205 GHz | 2.458 GHz |
| AR | 56.18 dB | 45.4 dB | 54.1 dB |
| $Q_r$ | 139.4 | 129.7 | 175.4 |
| $Q_a$ | 218.4 | 211.9 | 427.64 |
| $k_t^2$ | 23.73% | 12.47% | 14.67% |
| Bode $Q_{max}$ | 305 | 234 | 445 |
| FoM | 79.26 | 34.79 | 73.50 |
| $1^{st}$-order $TCF_r$ | -64.3 ppm/°C | -117.51 ppm/°C | -21.8 ppm/°C |
| $1^{st}$-order $TCF_a$ | -40.36 ppm/°C | -98.92 ppm/°C | -13 ppm/°C |

## 9 Advantages of LAW transducer in terms of temperature compensation effect

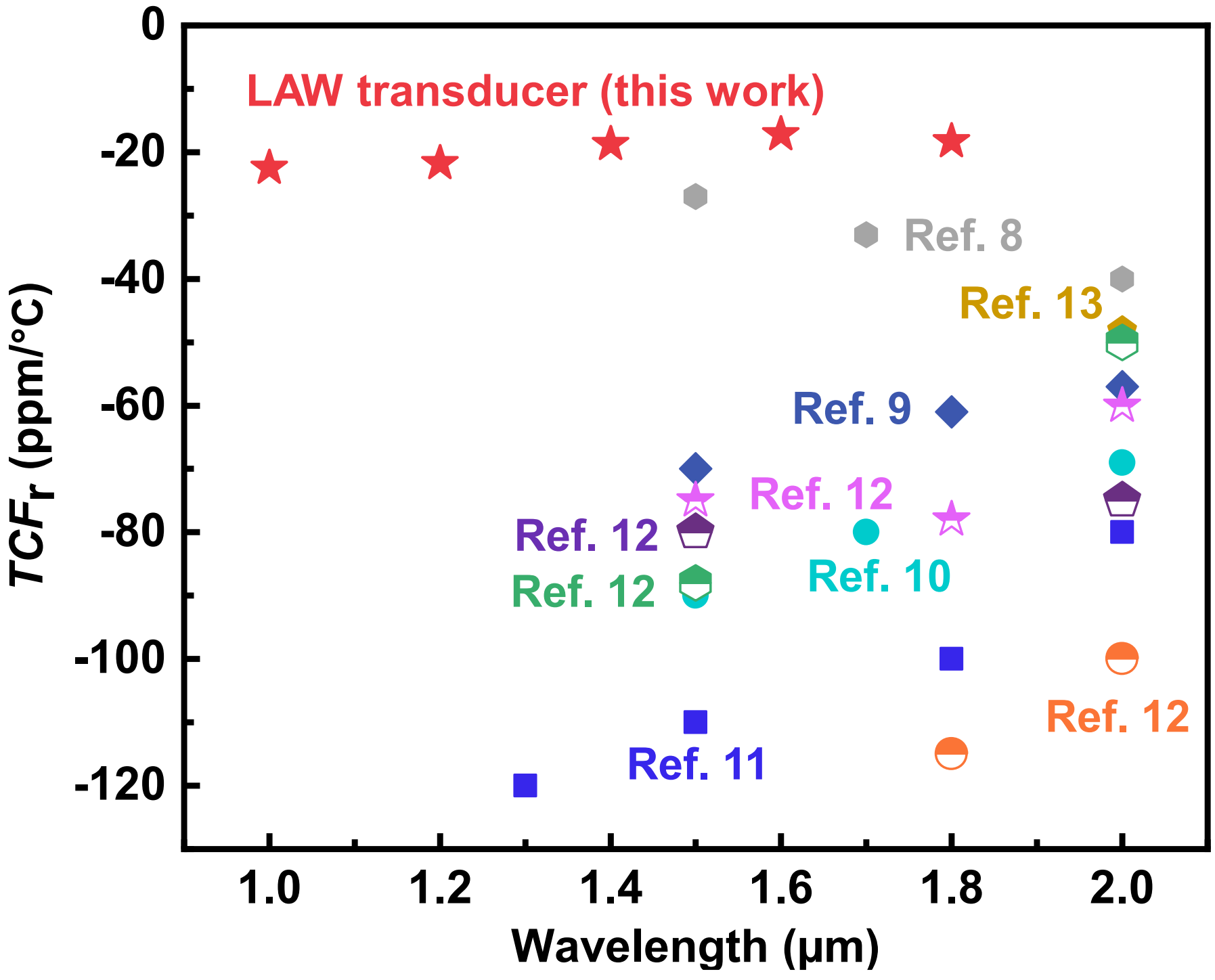

**Supplementary Fig. S11 | Advantages of LAW transducers in terms of temperature compensation effect.** The TCF at $f_r$ ($TCF_r$) of the SH-mode LAW transducers is compared with the advanced acoustic transducers across two device configurations: (1) $SiO_2$-overcoated bulk $LiNbO_3$ substrates[8] and (2) $LiNbO_3$-on-insulator (LNOI) TF-SAW platforms fabricated on silicon (Si)[9,10], quartz[11,12], and silicon carbide (SiC)[13] substrates. All devices share identical geometric parameters to ensure a fair comparison.

## 10 Power durability measurement setup

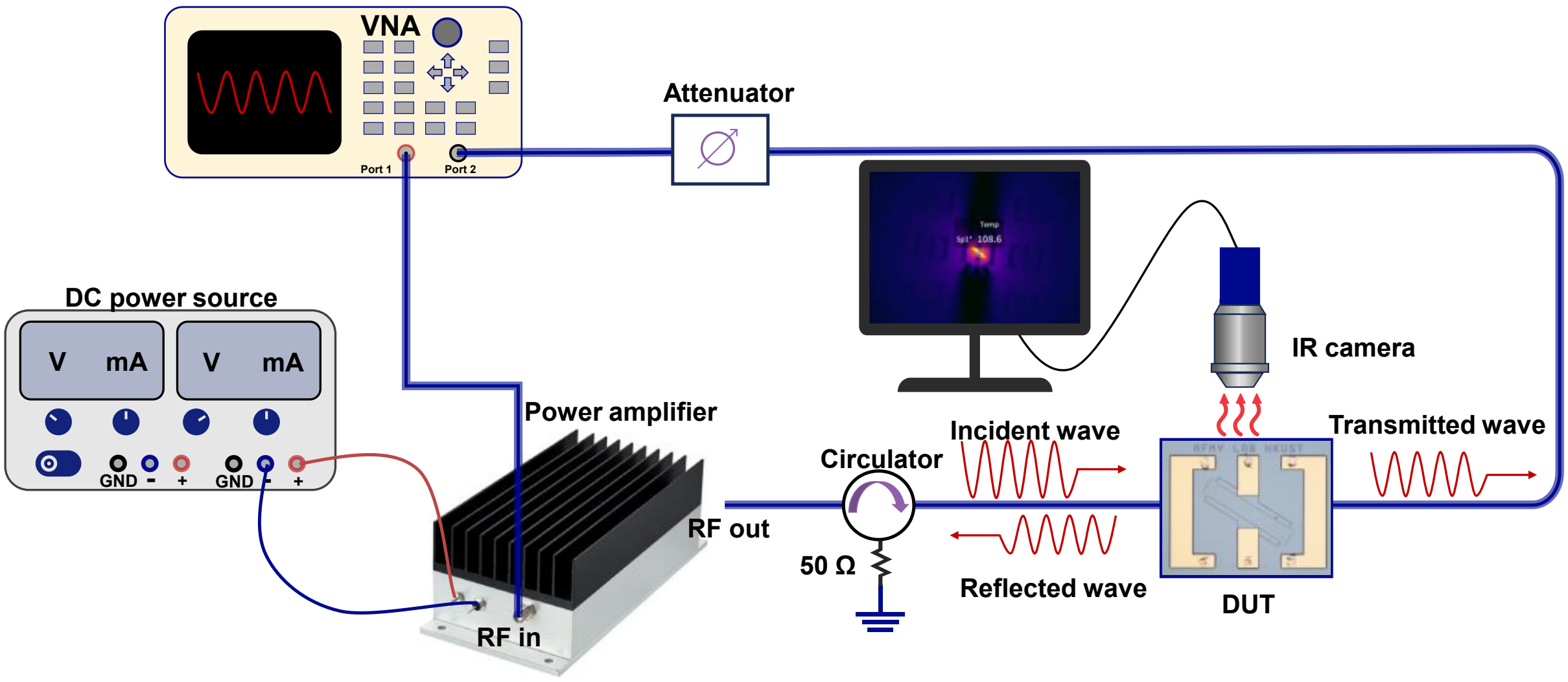

**Supplementary Fig. S12 | Power durability measurement setup.** Diagram of the experimental setup for the temperature mapping and the vectorial $S_{21}$ parameters measurements of the testing transducers under high-power loads. The inset figure in DUT refers to the microscopy image of a LAW transducer with a two-port configuration. A thermal profile of the two-port LAW transducer under an injected power density load of 45.61 dBm/mm$^2$ is also presented, indicating a temperature rise of 108.6 °C.

## 11 Power handling capability evaluation at filter-level

To substantiate our claim that power handling is a system-level property confounded by non-architectural factors, we performed a controlled, high-power study at the IHP-SAW filter level using identical constituent resonators and, critically, an identical filter layout. The three tested configurations, unpackaged (forward connection), unpackaged (reverse connection), and packaged (forward connection), differ only in their electrical port connection sequence and the presence/absence of encapsulation, while sharing the exact same physical layout design. We define "forward connection" as the configuration where incident power arrives first at the filter's input terminal, and "reverse connection" as the case where it arrives first at the output terminal. **Supplementary Fig. S13** presents the $S_{21}$ responses under increasing injected power. While all configurations remained stable up to 22.97 dBm, their failure points diverged sharply: 23.97 dBm (reverse), 26.97 dBm (forward), and 32.93 dBm (packaged). **Supplementary Fig. S14**, plots the corresponding minimum insertion loss (*IL*) versus injected power, revealing a maximum difference of 9.42 dB in failure threshold — a variation arising solely from connection and packaging, not from the acoustic resonators itself.

To investigate this further, we conducted continuous-wave tests on the forward-connected unpackaged filter at different frequencies within its passband (**Supplementary Fig. S15**). Thermal images captured by the infrared (IR) camera show that power dissipation is higher at the band edges (2.3942 GHz, 2.4842 GHz) than at the center (2.4515 GHz), with the spatial temperature profile, and thus the internal power flow distribution, shifting markedly with frequency. This directly visualizes the frequency-dependent, design-specific power routing within the composite acoustic-electromagnetic system, which also confirms that the physical locations and modes of resonator damage varied distinctly for each test frequency.

This combination of thermal, electrical, and mechanical evidence solidifies a key conclusion: the point of failure in a filter is not predetermined by the resonator alone but is a dynamic outcome of system-level variables such as power routing and packaging. Therefore, power handling measured at the filter level yields a metric that is intrinsically confounded by design-specific routing, connection, and encapsulation effects. Such a system-level result cannot serve as a direct or fair benchmark for comparing the inherent power-handling capability of different resonator technologies at the device-architecture level. Thus, to isolate and evaluate architectural innovation, we employ transducer-level benchmarking using an active-area-normalized injected power density, providing a more direct and accurate assessment of architectural innovation.

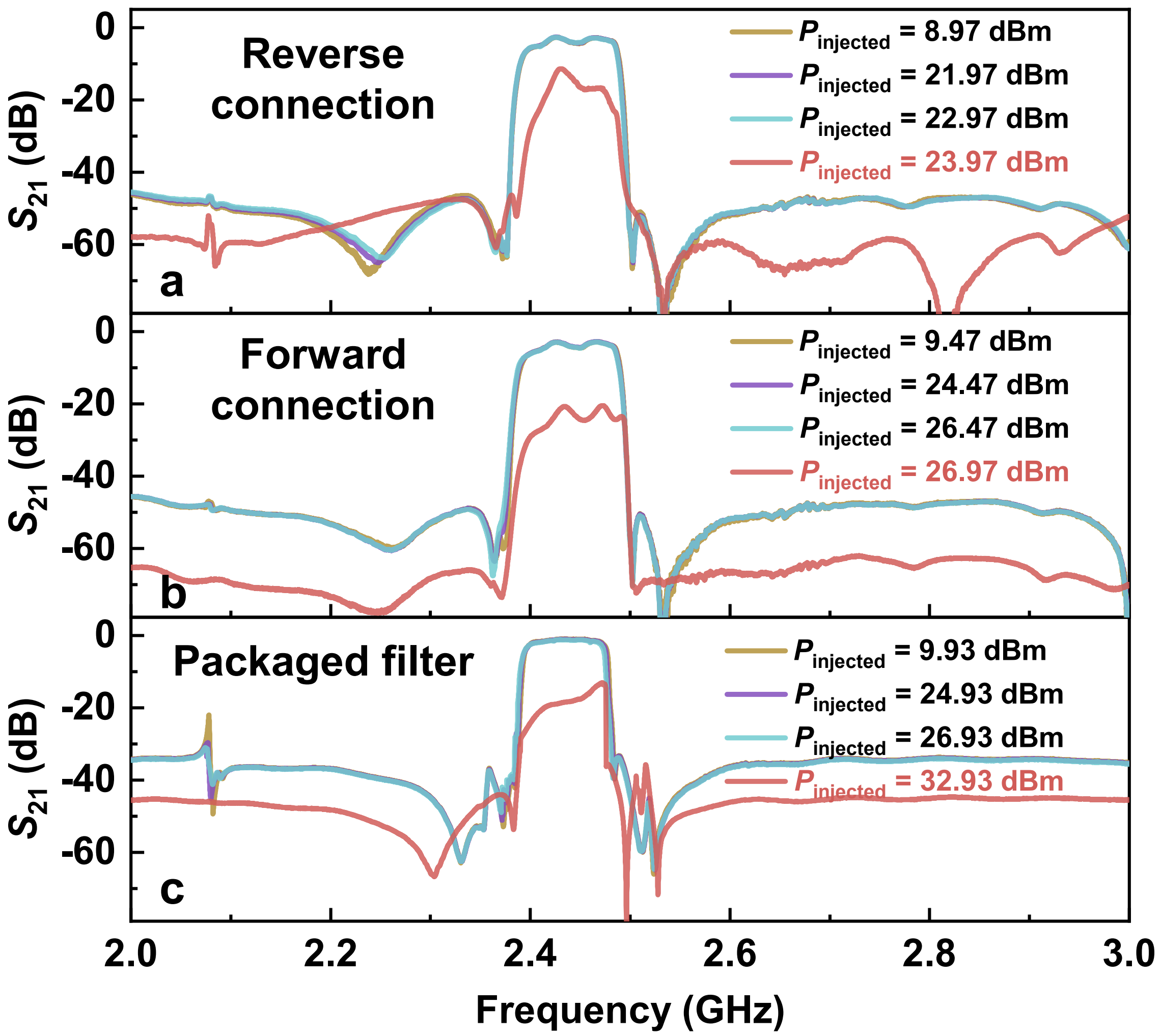

**Supplementary Fig. S13 | Power-handling comparison under varied connection and packaging configurations.** High-power $S_{21}$ responses for IHP-SAW filters with identical layout but different interfaces: **(a)** unpackaged with reverse connection, **(b)** unpackaged with forward connection, and (c) packaged filter with forward connection. The difference in failure points underlines that system-level power handling is also governed by electrical and packaging interfaces, not solely by the inherent resonator architecture itself.

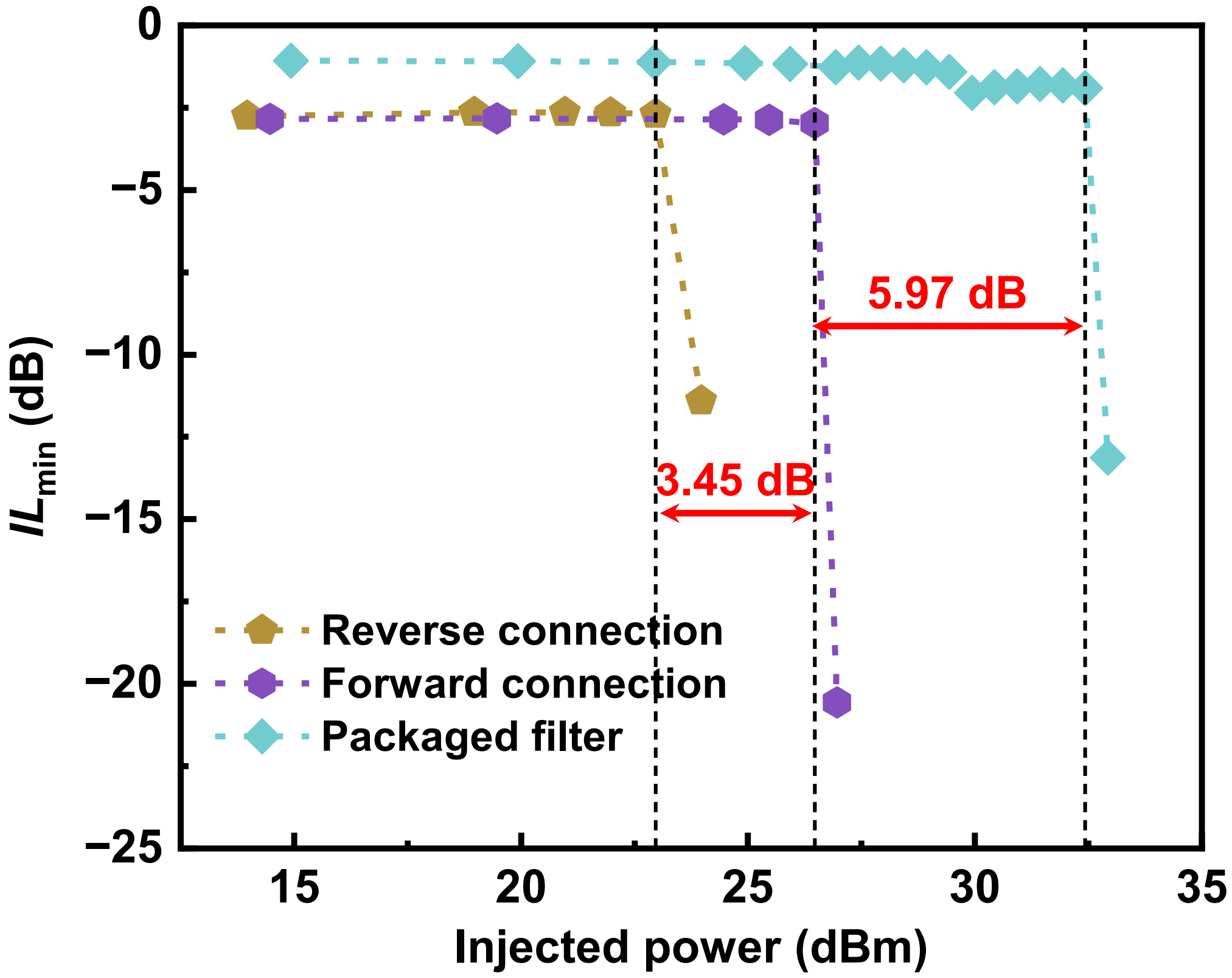

**Supplementary Fig. S14 | Failure threshold comparison across interface configurations.** Measured minimum insertion loss (*IL*) under high-power testing for the unpackaged (reverse/forward connection) and packaged IHP-SAW filters that share an identical layout. under high-power testing for the unpackaged (reverse/forward connection) and packaged IHP-SAW filters that share an identical layout.

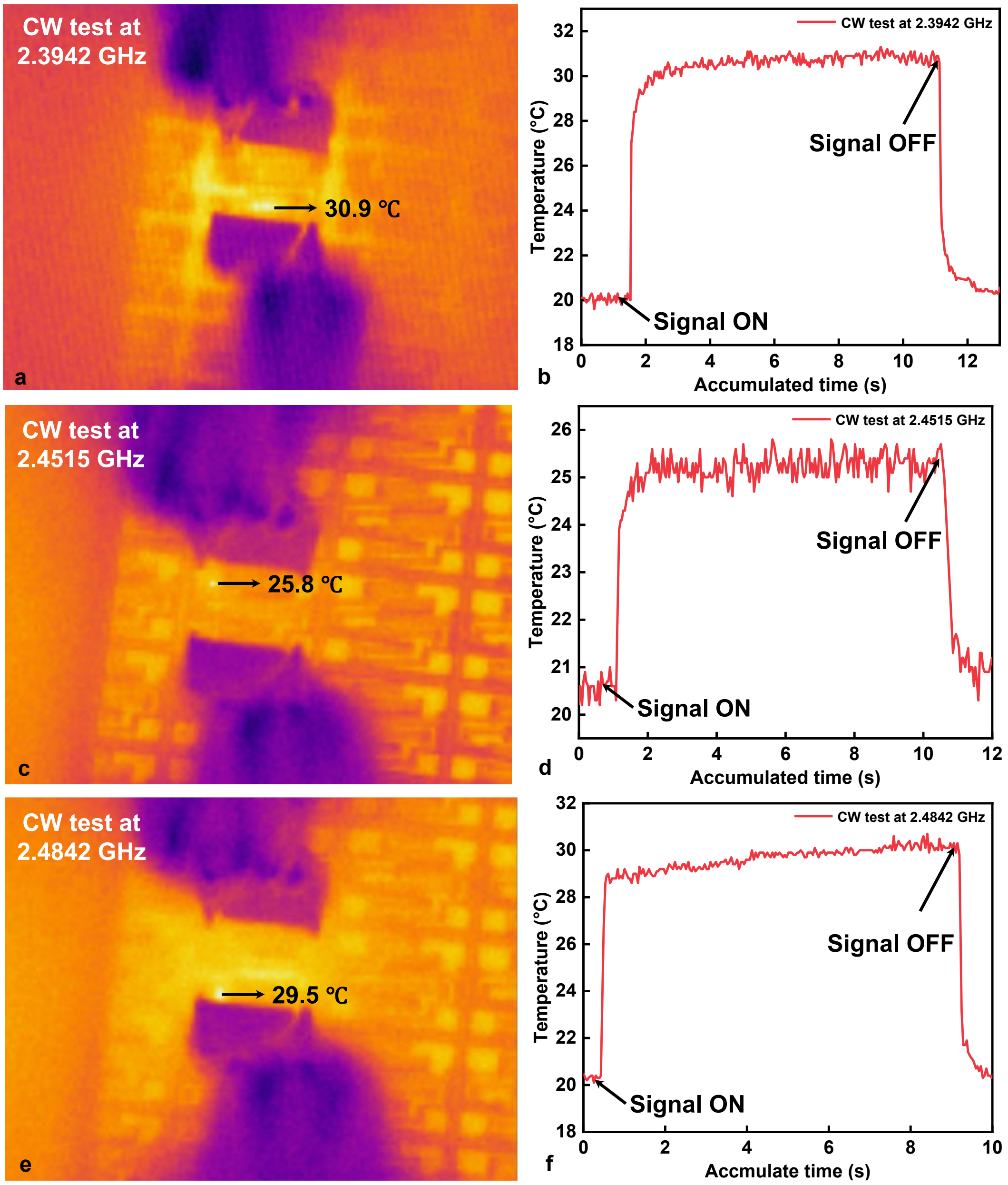

**Supplementary Fig. S15 | Frequency-dependent thermal analysis of an unpackaged IHP-SAW filter.** Thermal characterization under high-power continuous-wave (CW) excitation at distinct loading frequencies within the passband: **(a, b)** 2.3942 GHz (left band edge), **(c, d)** 2.4515 GHz (center), and **(e, f)** 2.4842 GHz (right band edge). Each frequency pair presents the infrared thermal image (left) and the corresponding transient temperature profile over time (right), revealing how power dissipation and hot-spot location shift with frequency due to system-level power-routing effects.

## 12 Frequency- and C0-dependent von Mises stress profiles

In principle, electromigration, which predominantly occurs at $f_r$, is another critical failure mode for acoustic wave devices. Therefore, to investigate the competing failure mechanisms in acoustic wave devices—specifically, electromigration (driven by current density) versus acoustomigration (driven by mechanical stress)—we performed finite element analysis (FEA) on a SAW transducer unit cell. The goal was to extract the frequency-dependent von Mises stress profile and identify the weakest failure point across the frequency domain. The extraction method is presented in **Supplementary Fig. S16**. A perfectly matched layer (PML) was applied at the substrate bottom to absorb leaked acoustic energy and eliminate spurious reflections. The simulated stress profile was obtained by sweeping the frequency and recording the maximum von Mises stress along the depth (from the IDT top surface into the substrate), which consistently peaked at the critical Au/$LiNbO_3$ interface. The results for devices with different static capacitances were normalized to their respective global maxima for direct comparison.

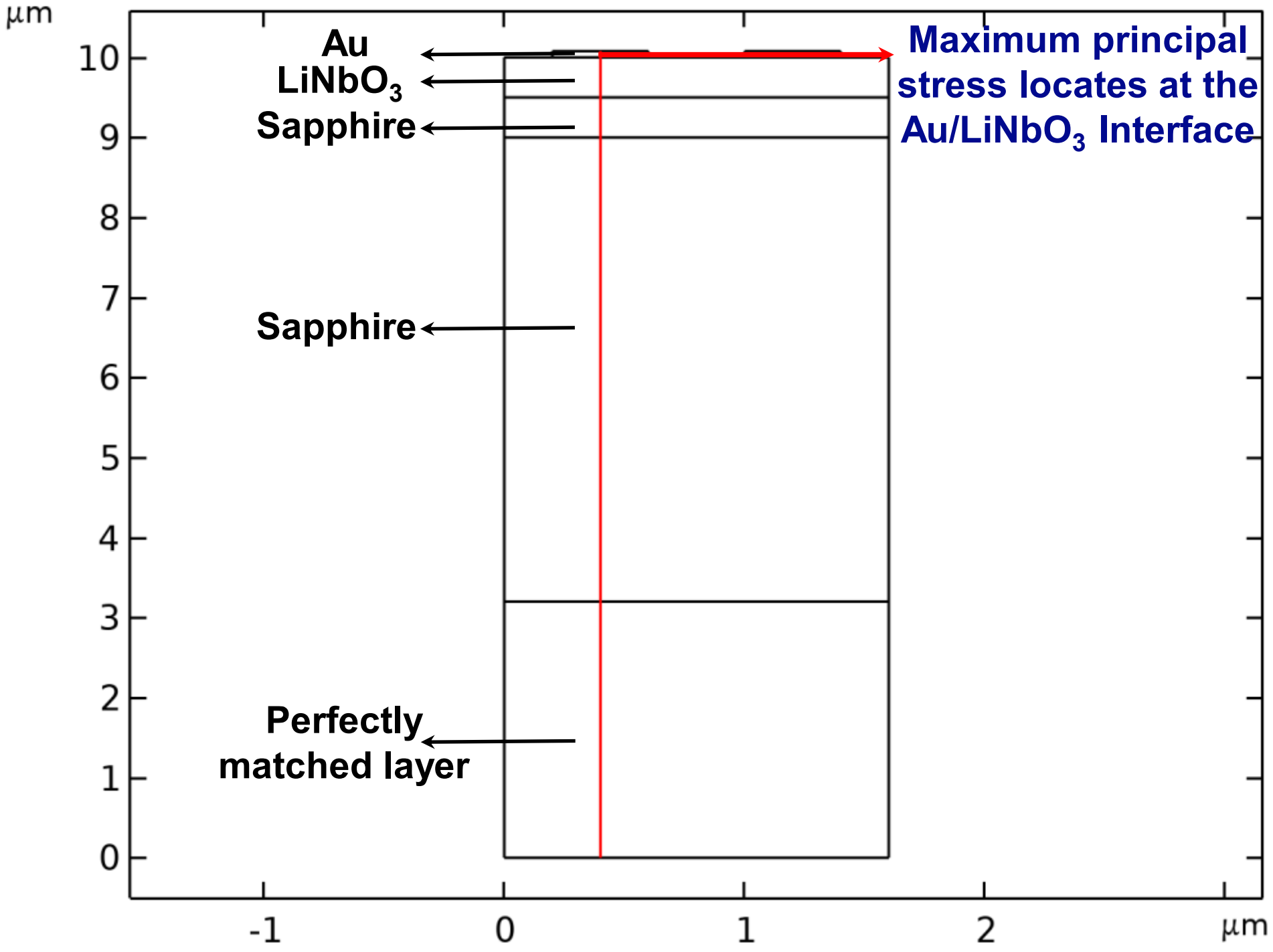

**Supplementary Fig. S16 | Method for extracting depth-dependent von Mises stress.** Schematic illustrating the finite-element-analysis (FEA) based procedure for evaluating the von Mises stress profile along the vertical axis, from the top surface of the interdigital transducer (IDT) layer down to the sapphire substrate.

The key finding is illustrated in **Supplementary Fig. S17**, where $f_r$ and $f_a$ are marked by dashed lines. At $f_r$, the transducer behaves as a near-short circuit. This condition maximizes current density, thereby promoting electromigration, but results in a minimal potential difference between IDT fingers, leading to negligible mechanical strain and thus minimal von Mises stress (thus minimizing acoustomigration). Conversely, as the load frequency increases towards $f_a$, the high impedance leads to a large potential difference, maximizing mechanical stress and acoustomigration risk. Therefore, the dominant failure mechanism and its corresponding "weakest failing point" in the frequency domain are not fixed; they shift between $f_r$ and $f_a$, depending on the device's electrical design (i.e., its static capacitance), as showcased in **Supplementary Fig. S17**.

To experimentally validate these simulation results, we performed comprehensive high-power characterizations on TF-SAW and LAW devices with different C0 values across the resonator band. **Supplementary Fig. S18a** presents a comparison of the injected power density thresholds for TF-SAW and LAW DUTs with C0 $\approx$ 520 fF, tested at three selected frequencies: $f_r$, the frequency of maximum dissipation ($f_{max}$), and $f_a$. The exact driving frequencies and their corresponding dissipation coefficients are indicated in the figure.

For the TF-SAW devices, the threshold injected power density at $f_{max}$ is 28.53 dBm/mm$^2$, representing the most failure-prone point. At $f_r$, the threshold is slightly higher at 29.15 dBm/mm$^2$, while at $f_a$ it increases to 31 dBm/mm$^2$ due to lower dissipation (**Supplementary Fig. S18a**). These results for the small-C0 device align with the expectation that failure susceptibility is highest near frequencies where acoustic energy is most strongly coupled. For the LAW device, the thresholds at $f_r$, $f_{max}$, and $f_a$ are 38.76, 38.46, and 41.63 dBm/mm$^2$, respectively. Although the improvement varies slightly across frequencies, the overall enhancement factor of approximately 11-fold for the LAW compared to the TF-SAW remains consistent with our earlier conclusions.

To further illustrate the difference in power handling robustness between the two architectures, **Supplementary Fig. S18b** and **S18c** show the admittance responses of the TF-SAW and LAW devices before and after high-power stress at their respective maximum-dissipation frequencies. After exposure to 28.53 dBm/mm$^2$, the TF-SAW device exhibits irreversible damage: spurious modes appear, the *Q*-factor drops significantly, the $k_t^2$ decreases, and the static capacitance C0 is also reduced (**Supplementary Fig. S18b**). In contrast, the LAW device stressed at 37.4 dBm/mm$^2$ shows only a minor frequency shift, while both $k_t^2$ and *Q* remain nearly unchanged (**Supplementary Fig. S18c**). Only when the power density is increased to 38.46 dBm/mm$^2$ does the LAW device begin to show degradation similar to that of the TF-SAW, yet its performance (e.g., $k_t^2$) is still superior to the damaged TF-SAW DUT.

We extended the same measurement for devices with C0 ≈ 5.5 pF. The frequency offset between $f_r$ and $f_{max}$ increases with C0: for C0 ≈ 520 fF, the offsets are 44 MHz for TF-SAW and 49 MHz for LAW; for C0 ≈ 5.5 pF, they become 121 MHz for TF-SAW and 108 MHz for LAW (**Supplementary Fig. S18d**). At the maximum dissipation frequency, the LAW again demonstrates a marked improvement: the critical power density for the TF-SAW is 28.19 dBm/mm$^2$, while the LAW withstands 38.75 dBm/mm$^2$ before showing any notable change, an enhancement factor of 11.38-fold. Post stress admittance comparisons confirm that at 28.19 dBm/mm$^2$ the TF-SAW is irreversibly damaged, with its admittance ratio drops to 23.8 dB and spurious modes appearing, whereas the LAW stressed at 38.75 dBm/mm$^2$ exhibits only a slight frequency shift with $k_t^2$ and *Q* essentially preserved. Only after exposure to 40.35 dBm/mm$^2$ does the LAW show similar degradation, with an admittance ratio of 25.4 dB and emerging spurious responses.

In a standard filter design, the static capacitance is typically tuned for 50-Ω impedance matching, which positions the peak power dissipation (and thus the most likely failure point) somewhere between $f_r$ and $f_a$. For the transducers used in our power-handling tests, the static capacitance was made to complement the spatial resolution of our infrared camera, enabling clear thermal imaging of the failure epicenter. This design choice thus shifts the point of peak stress (and thus the observed failure) to $f_a$. This controlled alignment further validates that the observed failures are primarily driven by acoustomigration at the designed stress maximum, rather than by electromigration.

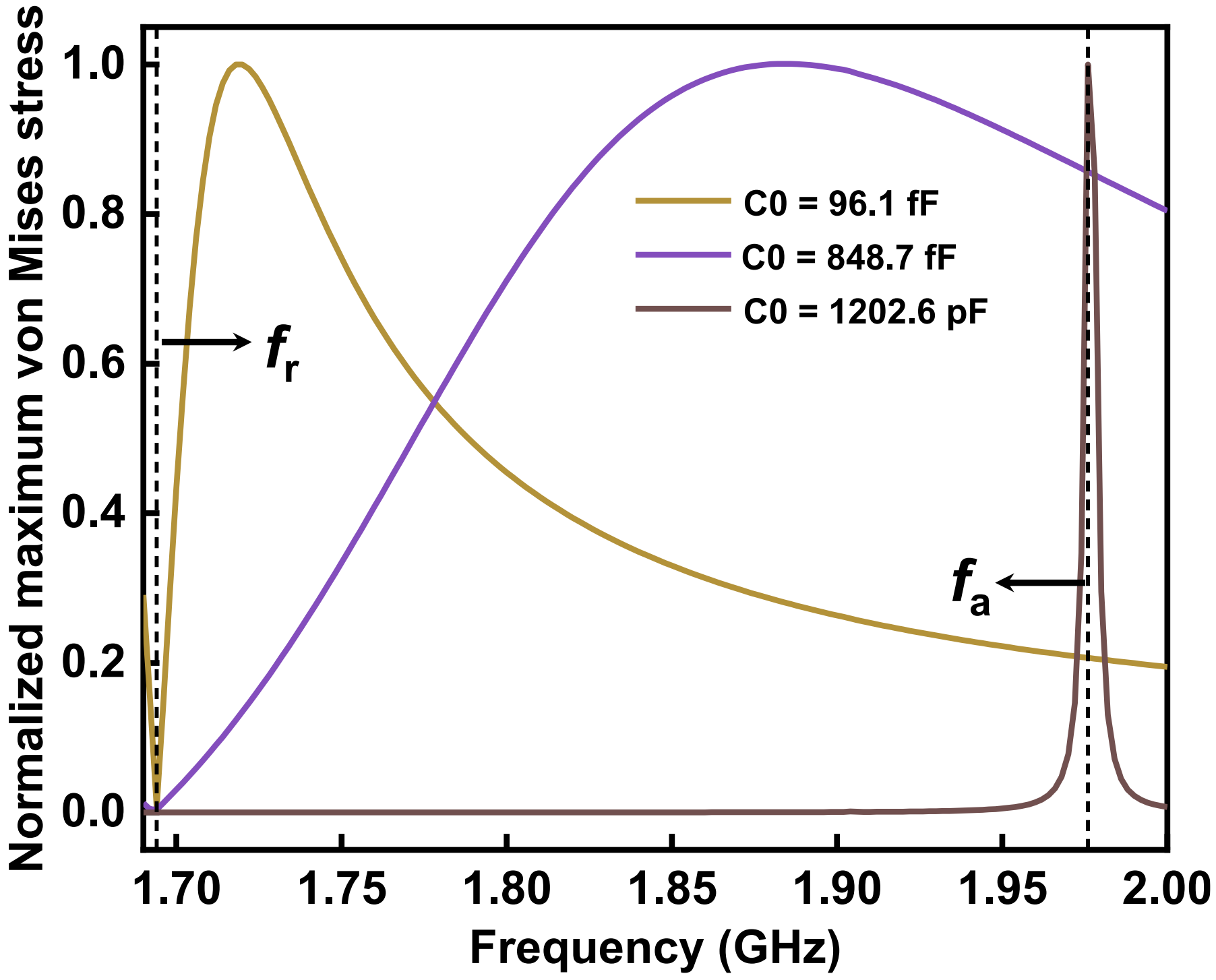

**Supplementary Fig. S17 | Simulated frequency- and static capacitance-dependent von Mises stress profiles.** The maximum stress is consistently located at the metal/$LiNbO_3$ interface. For each device, all profiles are normalized to the maximum von Mises stress in the depth direction extracted from the frequency sweep simulation.

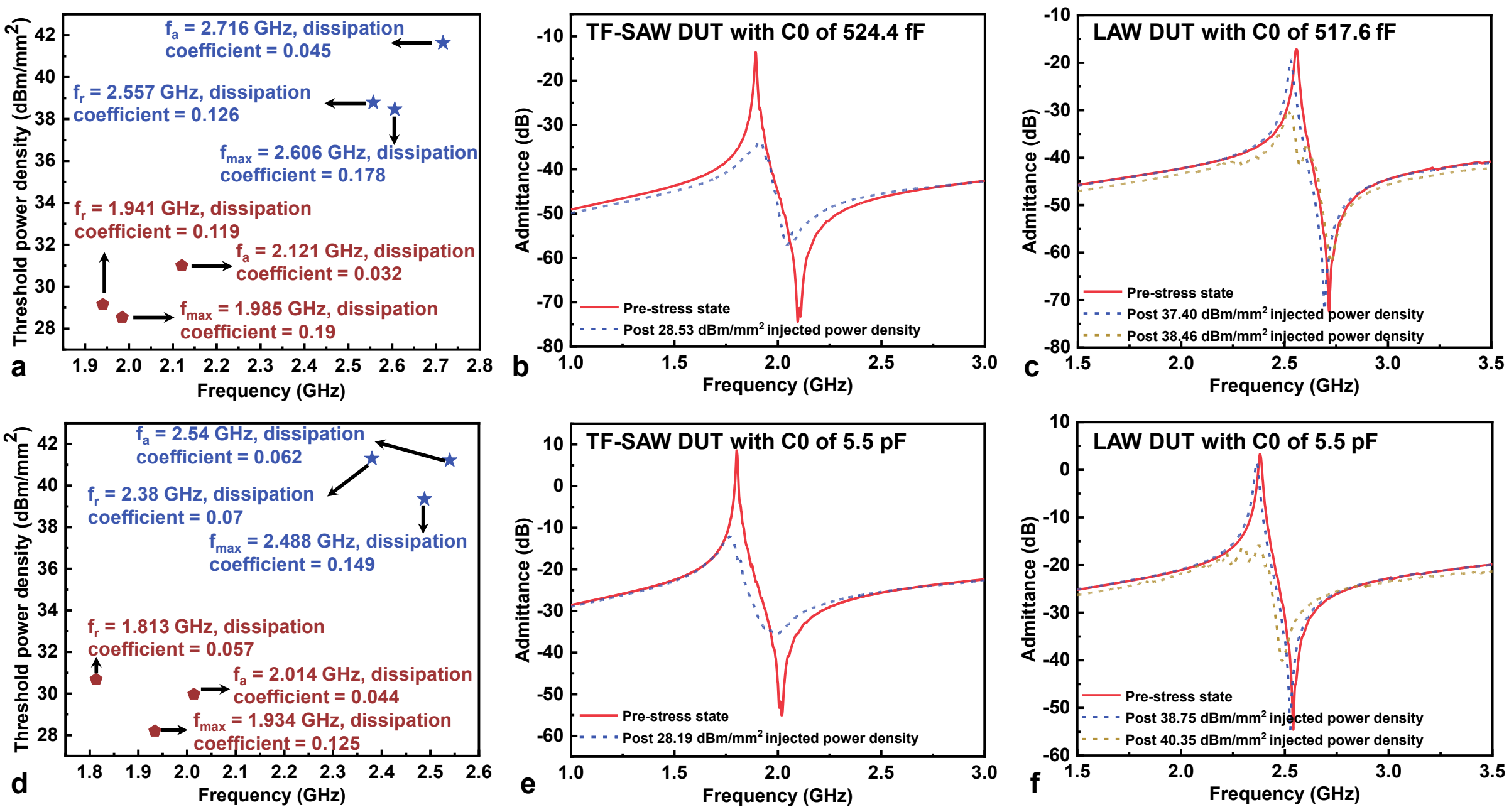

**Supplementary Fig. S18 | Power handling characterization of TF-SAW and LAW transducers with varying static capacitance. (a)**, Comparison of injected power density thresholds for TF-SAW and LAW DUTs with C0 ≈ 520 fF, measured at three selected frequencies across the resonator band. The driving frequencies and corresponding dissipation coefficients are indicated. **b-c**, Admittance responses before (pre-stress) and after (post-stress) high-power exposure for a TF-SAW DUT with C0 = 524.4fF **(b)** and a LAW DUT with C0 = 517.6 fF **(c)**. **(d)**, Same as a but for DUTs with C0 ≈ 5.5 pF. **e-f**, Corresponding pre- and post-stress admittance comparisons for a TF-SAW with C0 = 5.5 pF **(e)** and a LAW DUT with C0 = 5.5 pF **(f)**. The post-stress responses were obtained by small-signal frequency sweeps after the devices were subjected to continuous-wave high-power stress at the frequency corresponding to maximum dissipative absorption.

## 13 Measured *S*-parameter responses for TF-SAW and LAW transducers

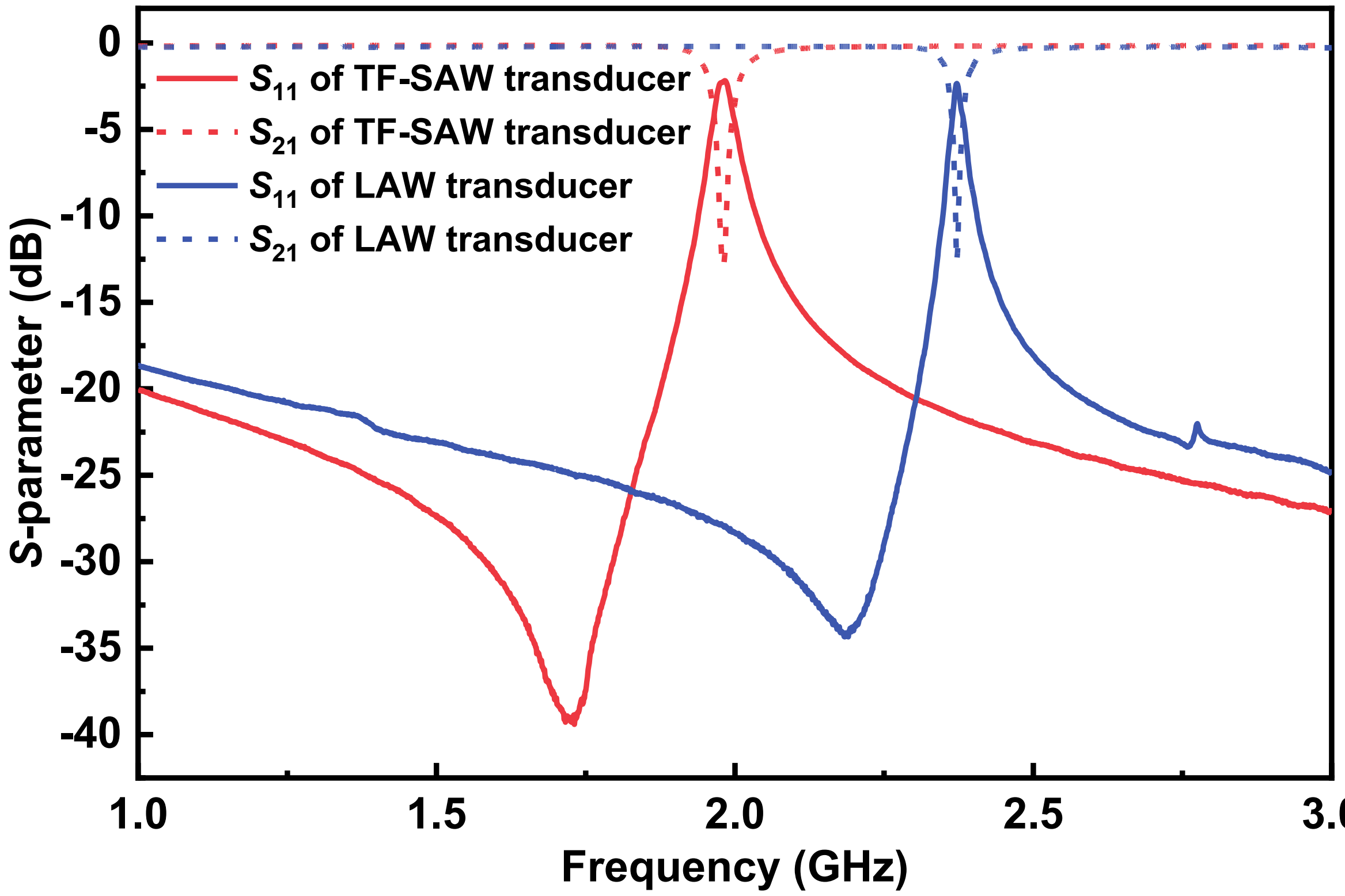

**Supplementary Fig. S19 | *S*-parameter responses for TF-SAW and LAW transducers measured under small-signal conditions.**

## 14 Dissipated power distribution of a typical LAW transducer vs. frequency

This section combines modified Butterworth-Van Dyke (mBVD) fitting and power flow simulation results for a typical LAW transducer to investigate the influence of different energy loss mechanisms and measurement port configurations on power dissipation profiles, with the aim of providing analysis for optimal driving frequency selection in high-power testing. **Supplementary Fig. S20** presents the admittance response of a typical LAW transducer and its mBVD fitting results. The parameters of each component in the mBVD model are summarized in the inset. Simulation results for the dissipated power within the LAW transducer in one-port and two-port configurations are illustrated in **Supplementary Fig. S20 (b)** and **Fig. S20 (c)**, respectively. The $f_r$ and $f_a$ are labelled by dashed lines. Note that different loss mechanisms dominate under different driving frequencies. The frequency-favoring dissipated power peak is located between $f_r$ and $f_a$. For the one-port configuration, peak power dissipation occurs near $f_r$, where the reflection coefficient is minimized. Otherwise, input power is strongly rejected with port impedance mismatch. At $f_r$, nearly 23.5% of input power is dissipated, dominated by ohmic loss in the electrodes due to peak current density. The peak of viscous losses (modeled by $R_m$) is also the strongest nearby $f_r$. While a similar frequency-dependent power dissipation trend is observed in the two-port configuration, the dissipation peak shifts toward $f_a$, with only 7.8% of input power absorbed by the LAW transducer. Notably, viscous losses constitute the largest proportion of total dissipation in the two-port case. At $f_a$, minimal current flow renders ohmic losses negligible for both configurations. Additionally, power dissipation in $R_0$ (dielectric loss) exhibits an inverse trend relative to ohmic losses. Far from the passband, dielectric and ohmic losses dominate, while viscous losses diminish to negligible levels. Under these conditions, the LAW transducer can be treated as a capacitor. These simulation results clarify that the contribution of different energy loss mechanisms to total energy dissipation varies from frequency to frequency[14–16], highlighting the importance of frequency selection for accessing the power handling capability of each testing transducer.

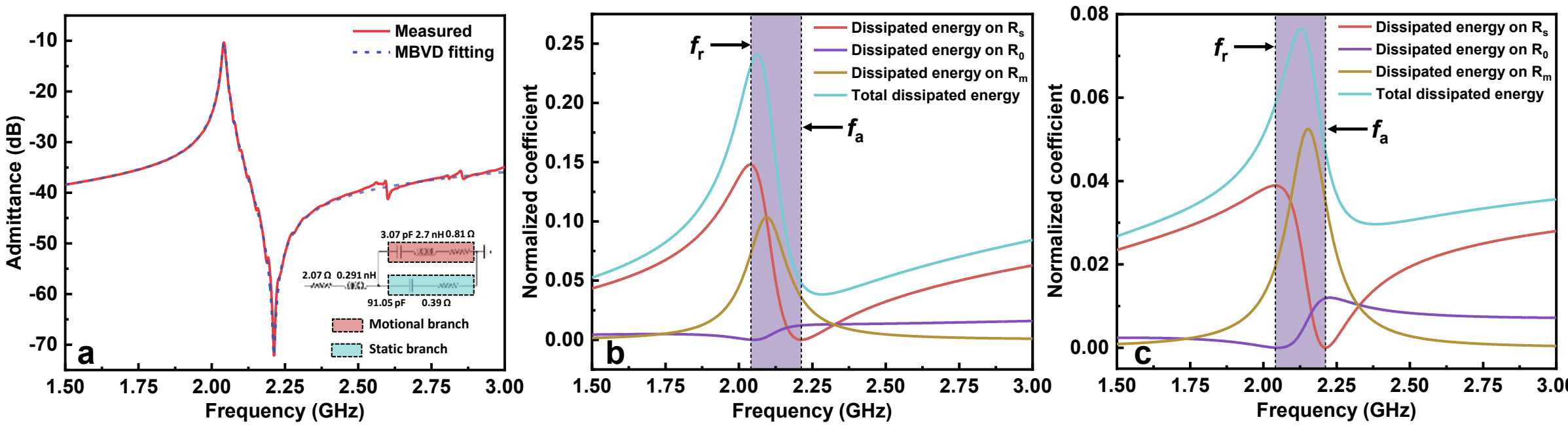

**Supplementary Fig. S20 | mBVD fitting and power dissipation simulation results. (a)** Measured and fitted admittance curves of a typical LAW transducer. The inset plot summarizes the parameters of each component in the mBVD model. Power dissipation distribution in the LAW transducer with **(b)** one-port configuration and **(c)** two-port configuration. $f_r$ and $f_a$ denote the resonant and anti-resonant frequency of the LAW transducer.

## 15 High-power test for SAW transducer at $f_r$

To investigate the dominant failure mechanism in acoustic transducers, we characterized the RF power budget and conducted high-power stress tests at the resonant frequency ($f_r$). **Supplementary Fig. S21** presents the measured power reflection, transmission, and dissipation for a SAW transducer across a 10-MHz band centered at $f_r$. Within this band, the power reflection coefficient is negligible (0.00017), and the transmission coefficient is high (0.9834), indicating minimal power rejection. The associated dissipation at $f_r$, while slightly elevated relative to adjacent bands, remains orders of magnitude lower than that near the anti-resonant frequency ($f_a$).

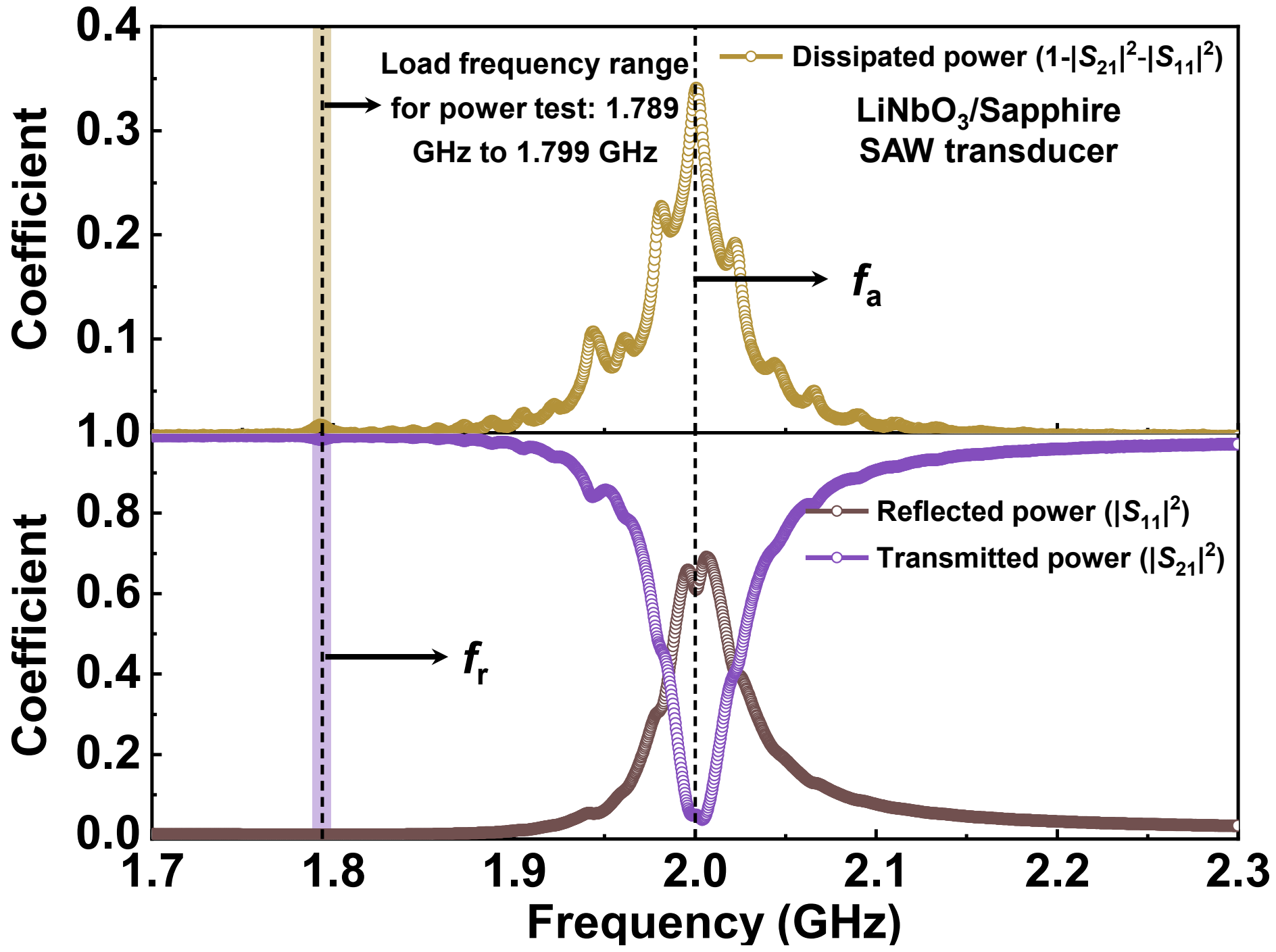

**Supplementary Fig. S21 | Power budget analysis of a SAW transducer. (a)** Measured dissipated, reflected, and transmitted power coefficients for the $LiNbO_3$/Sapphire TF-SAW transducer under a -15 dBm incident load. Frequency range for power test: 1.789–1.799 GHz.

Following the test protocol established in the main text, the device was then subjected to a high-power stress test with injected power density swept from 25.93 dBm/mm$^2$ to 45.93 dBm/mm$^2$ — exceeding the failure point of the LAW transducer reported in **Fig. 5c**. Notably, no appreciable steady-state temperature rise was detected via infrared thermography, even at these high-power levels. Post-stress low-power (–15 dBm) *S*-parameter measurements, performed after re-calibration, show no degradation in device performance (**Supplementary Fig. S22**), as confirmed by the almost unchanged frequency response.

The apparent robustness near $f_r$ for our large-C0 DUT (C0 = 1083 pF) is not due to poor matching, but rather to the current path distribution dictated by the mBVD model. At $f_r$, the static branch impedance is only ~0.14 Ω, while the motional branch impedance is ~0.6 Ω. Consequently, most of the RF current flows through the static branch, effectively bypassing the motional branch where acoustomigration would occur. In contrast, for a small-C0 DUT, the static branch impedance is much larger than that of the motional branch, forcing current through the motional branch and enabling acoustomigration-induced failure. This observation reinforces our central claim: the device becomes vulnerable when significant acoustic energy is present in the motional branch. The combination of the measurement results in Supplementary Information, Section 12 and Section 15, including minimal electrical dissipation at $f_r$, the absence of significant temperature-rise, and the preservation of device performance post-stress, provides strong, multi-faceted evidence that the predominant failure mechanism in these acoustic transducers is acoustomigration (driven by high mechanical stress) rather than electromigration (driven by current density and Joule heating).

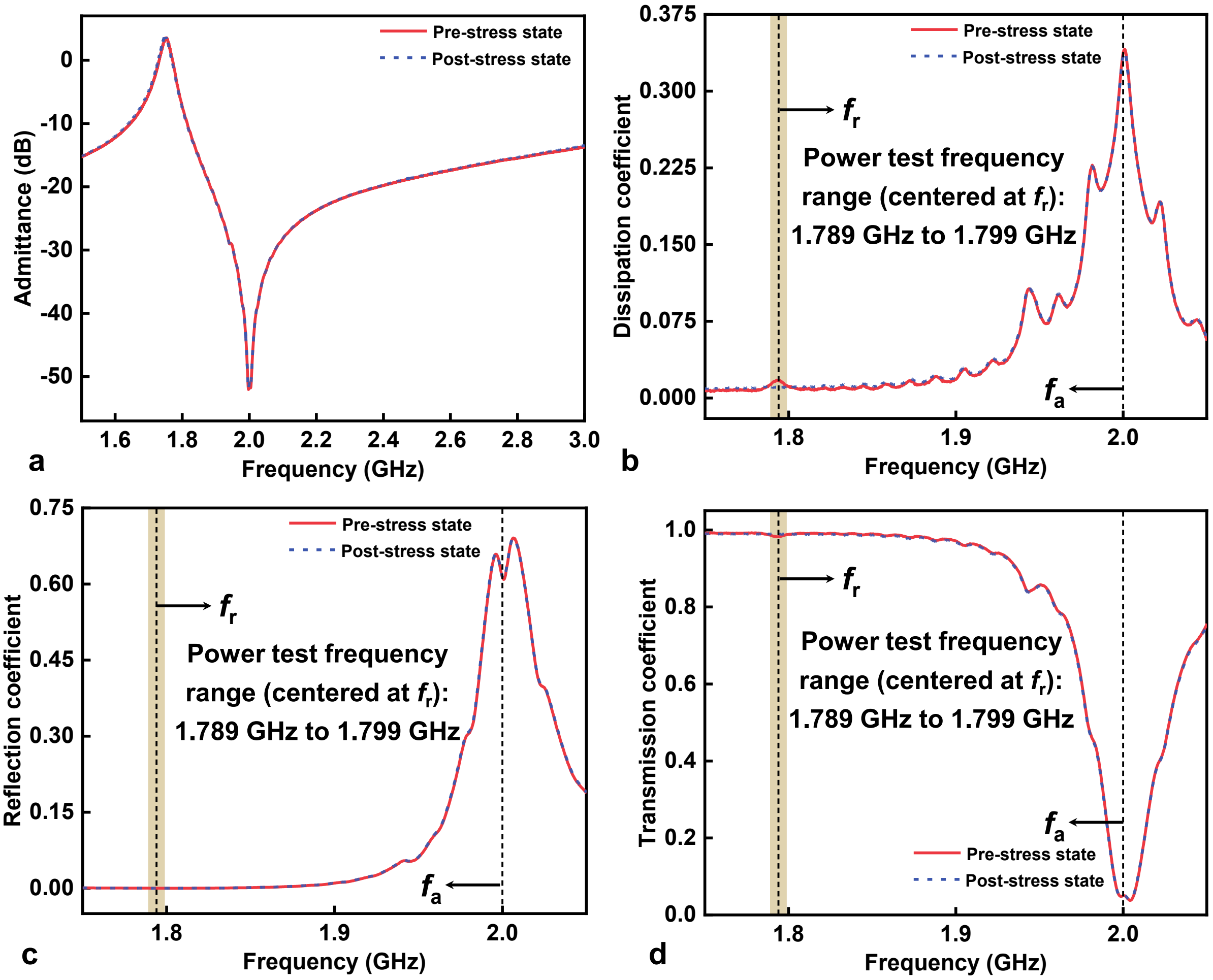

**Supplementary Fig. S22 | Pre- and post-stress characteristics of the SAW transducer.** Key parameters for the $LiNbO_3$/Sapphire TF-SAW DUT measured under a -15 dBm incident load: **(a)** Admittance curves, **(b)**, dissipation, **(c)**, reflection, and **(d)**, transmission coefficients for the LiNbO3/Sapphire TF-SAW DUT, measured under small-signal conditions (-15 dBm) before and after being subjected to high-power stress. The high-power stress was applied at a maximum injected power density of 45.93 dBm/mm2 over a narrow frequency range centered at the resonant frequency fr (1.789–1.799 GHz).

## 16 Thermal analysis of a LAW transducer under high RF loads

Based on the electrical modelling method described in **Supplementary Section 15**, power dissipation can be accurately calculated for each frequency, providing an estimate of the optimal frequency range required to heat the transducer effectively. To verify this methodology for power durability testing, thermal images were recorded during frequency sweeps at varying input power levels (input power is defined as the isolator's output power). The actual dissipated power depends on driving frequencies due to differences in reflection, dissipation, and transmission characteristics at each frequency, as shown in **Supplementary Fig. S23a**. The simulated power dissipation profile aligns closely with post-calculated experimental data. From the power dissipation curves, it can be observed that most RF power is absorbed by the LAW transducer at $f_a$. Following the analysis of frequency characteristics, IR radiation is recorded and calibrated using the emissivity calibration method detailed in the main text. **Supplementary Fig. S23b** indicates that the temperature rises scales with the calculated dissipated power. Despite the compensated TCF remaining low, the maximum temperature rise shifts to lower frequencies, which can be attributed to the non-zero TCF as temperature increases. Temperature rise measurements under varying input power loads further highlight the criticality of the tested frequency range: as input power increases, both temperature and dissipated power rise, driving $f_a$ downward, and these behaviors again amplify the dissipated power until device failure occurs[17,18].

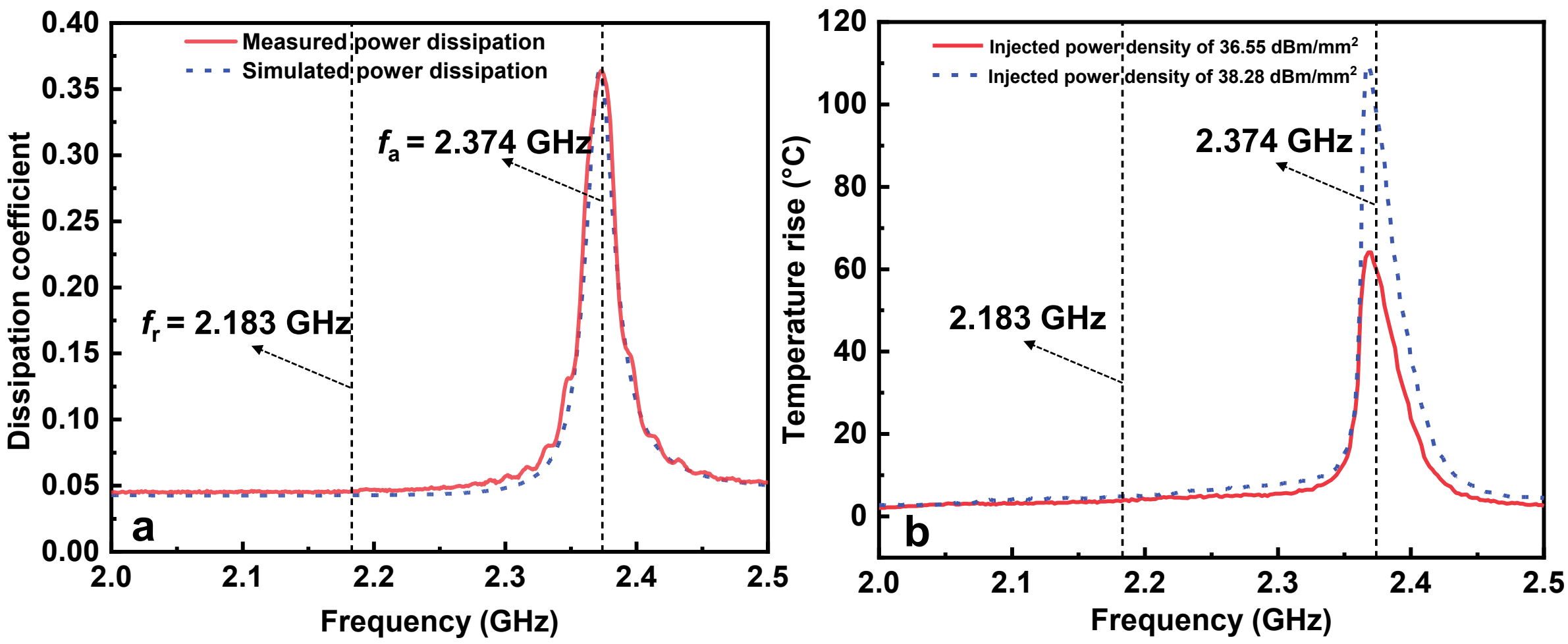

**Supplementary Fig. S23 | Power and thermal analysis of a LAW transducer under high RF load.** **(a)** Post-calculated and simulated power dissipation profiles in the LAW transducer. **(b)** Temperature curves of the LAW transducer during frequency sweeps at varying input power levels.

## 17 Performance comparisons of TF-SAW transducers before and after power tests

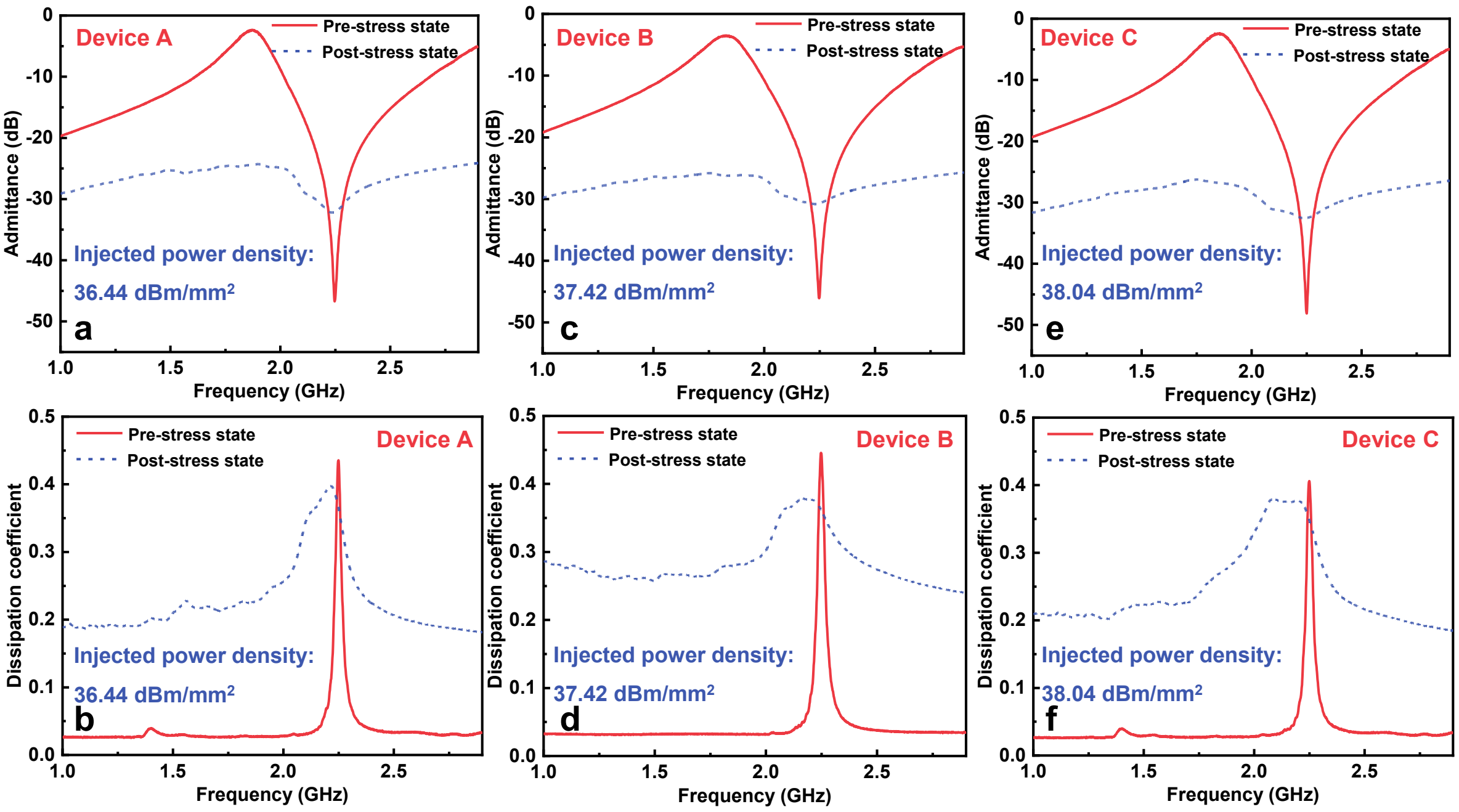

**Supplementary Fig. S24 | Performance comparison of TF-SAW transducers before and after high-power loads under different temperatures. (a)** Admittance curves and **(b)** power dissipation profiles of Device A before and after the high-power load test under -85 °C. **(c)** Admittance curves and **(d)** power dissipation profiles of Device B before and after the high-power load test under -45 °C. **(e)** Admittance curves and **(f)** power dissipation profiles of Device C before and after the high-power load test under -5 °C.

## 18 Performance comparisons of LAW transducers before and after power tests

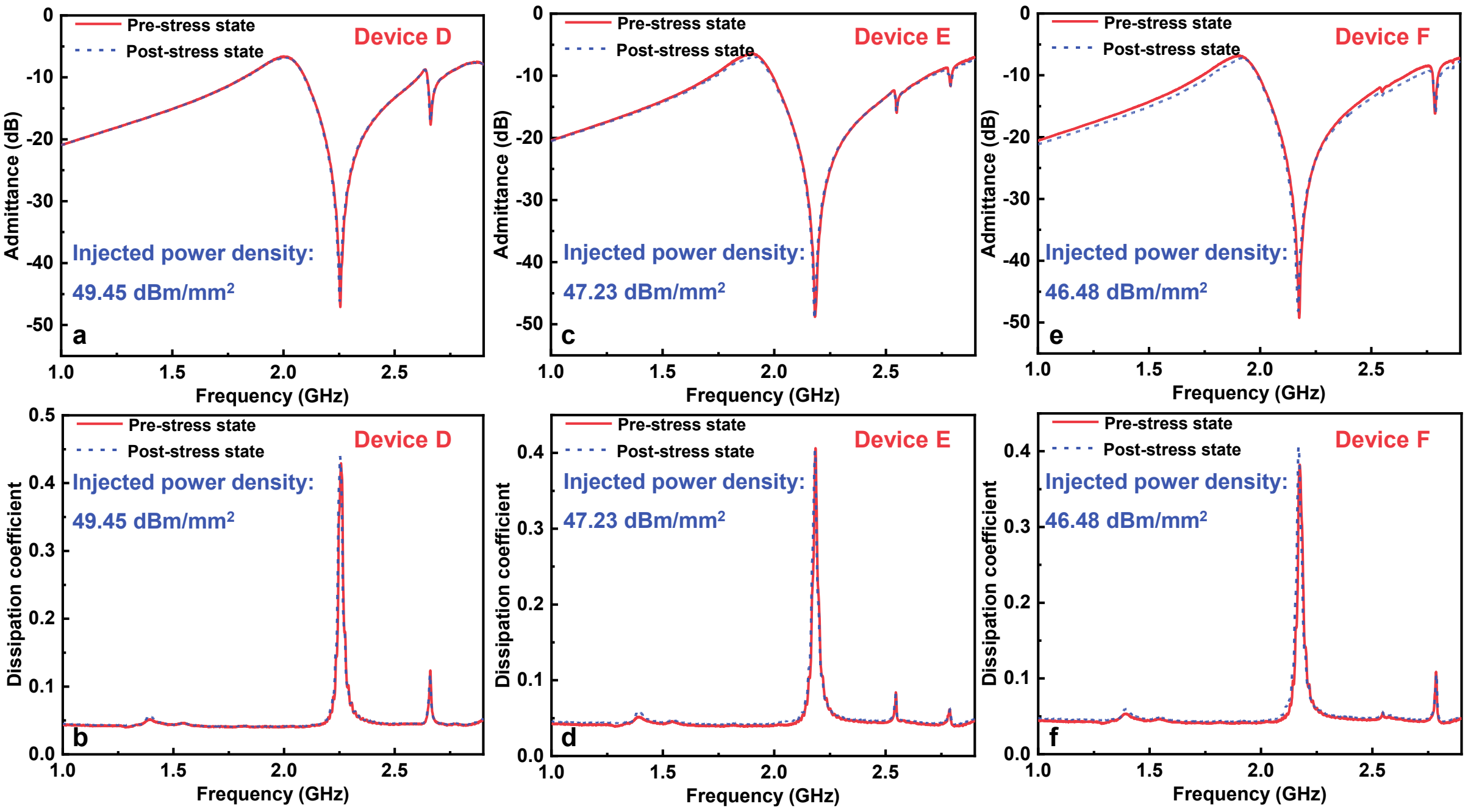

**Supplementary Fig. S25 | Performance comparison of LAW transducers before and after high-power loads under different temperatures. (a)** Admittance curves and **(b)** power dissipation profiles of Device D before and after the high-power load test under -85 °C. **(c)** Admittance curves and **(d)** power dissipation profiles of Device E before and after the high-power load test under -5 °C. **(e)** Admittance curves and **(f)** power dissipation profiles of Device F before and after the high-power load test under -80 °C.

## 19 Acoustomigration investigation on broken TF-SAW transducers (State B)

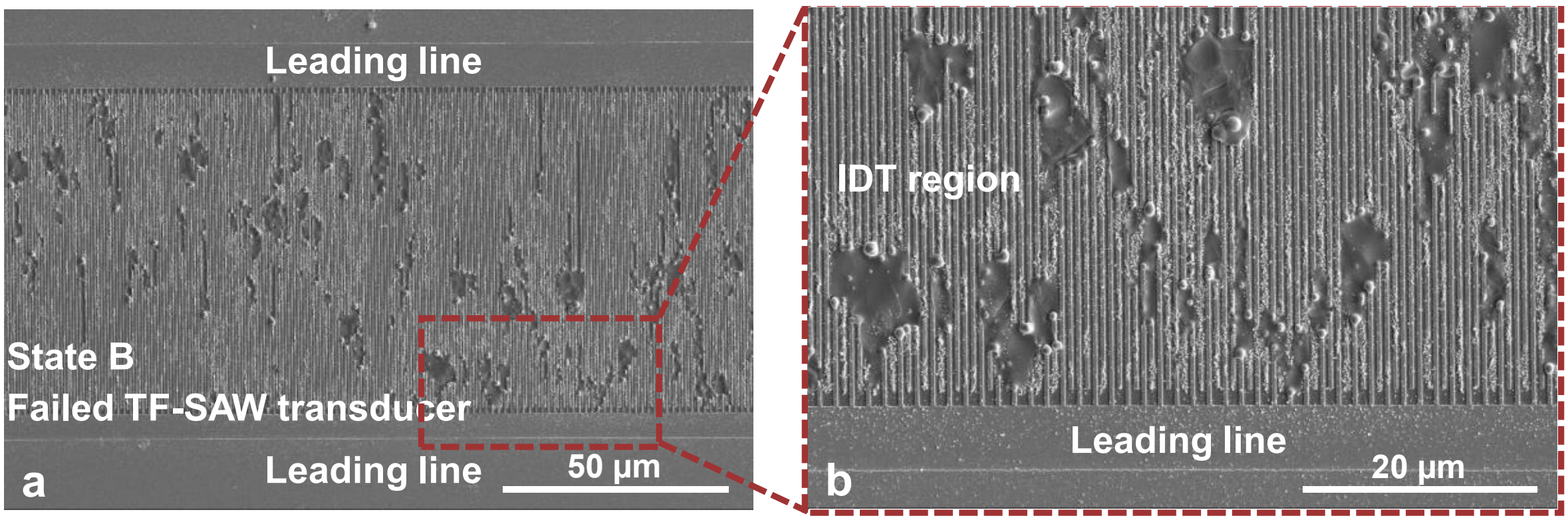

**Supplementary Fig. S26 | Post-failure analysis of a TF-SAW transducer (State B). (a)** Low-magnification top-view SEM image showing random and widespread damage across the entire transduction region. **(b)** High-magnification image of a selected area, revealing irregular electrode damage, including localized melting and metal balling, accompanied by numerous irreversible cracks propagating into the underlying $LiNbO_3$ substrate.

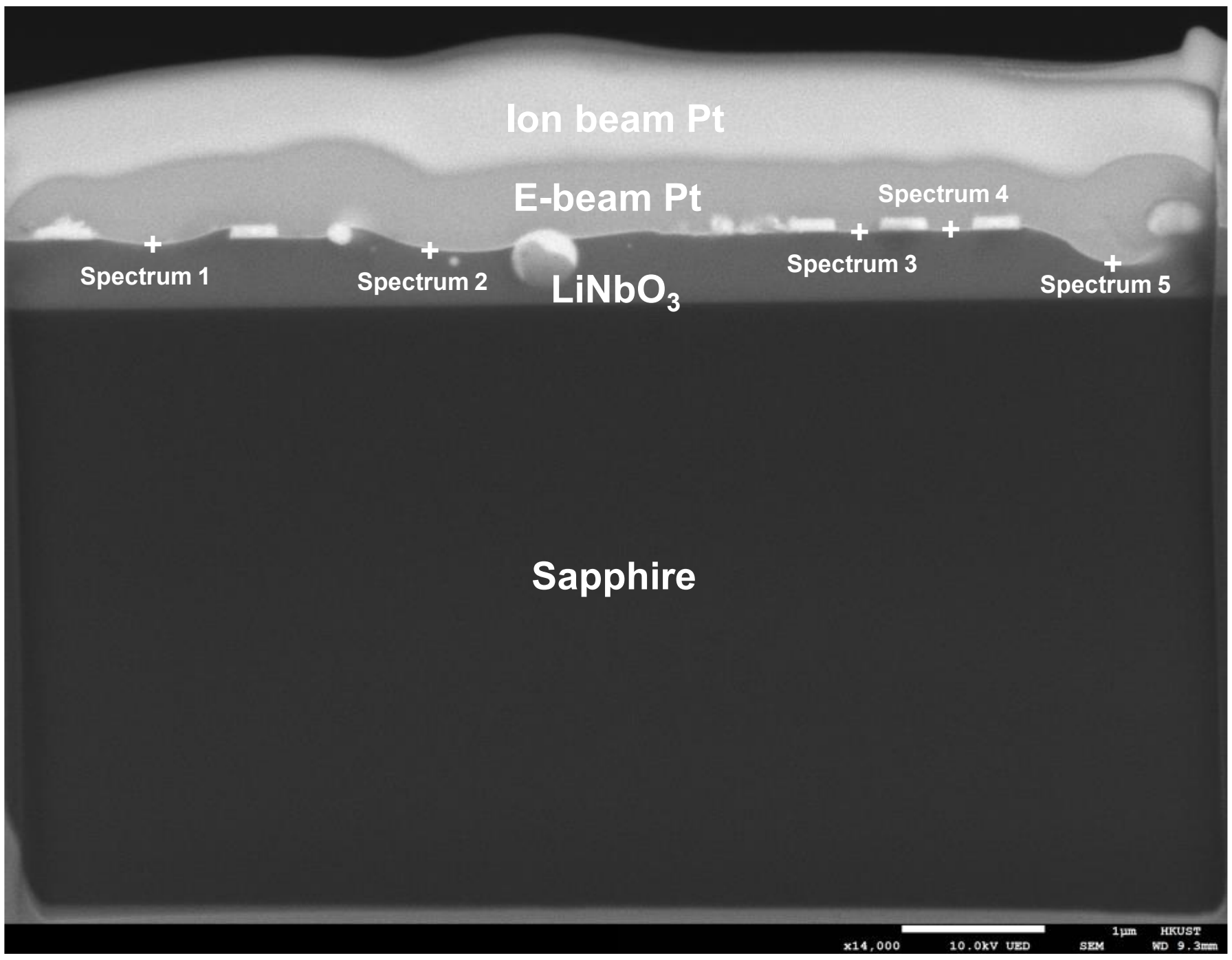

**Supplementary Fig. S27 | Acoustomigration investigation on a broken TF-SAW transducer (State B).** A cross-sectional scanning electron microscope (SEM) image of a broken TF-SAW transducer following high-power testing. The $LiNbO_3$ layer exhibits significant deformation, with a distinct continuous bright line observed at its surface. An energy-dispersive X-ray spectroscopy (EDS) analysis was performed at five labeled points along this feature, located at the Pt/$LiNbO_3$ interface, as marked in **Supplementary Fig. S11**. At the marked locations, gold is identified as the second most abundant element at these sites, even though it should not be present under small signal working conditions. This anomalous presence strongly indicates that acoustic-induced migration under high-power conditions

has caused gold to accumulate in these regions.

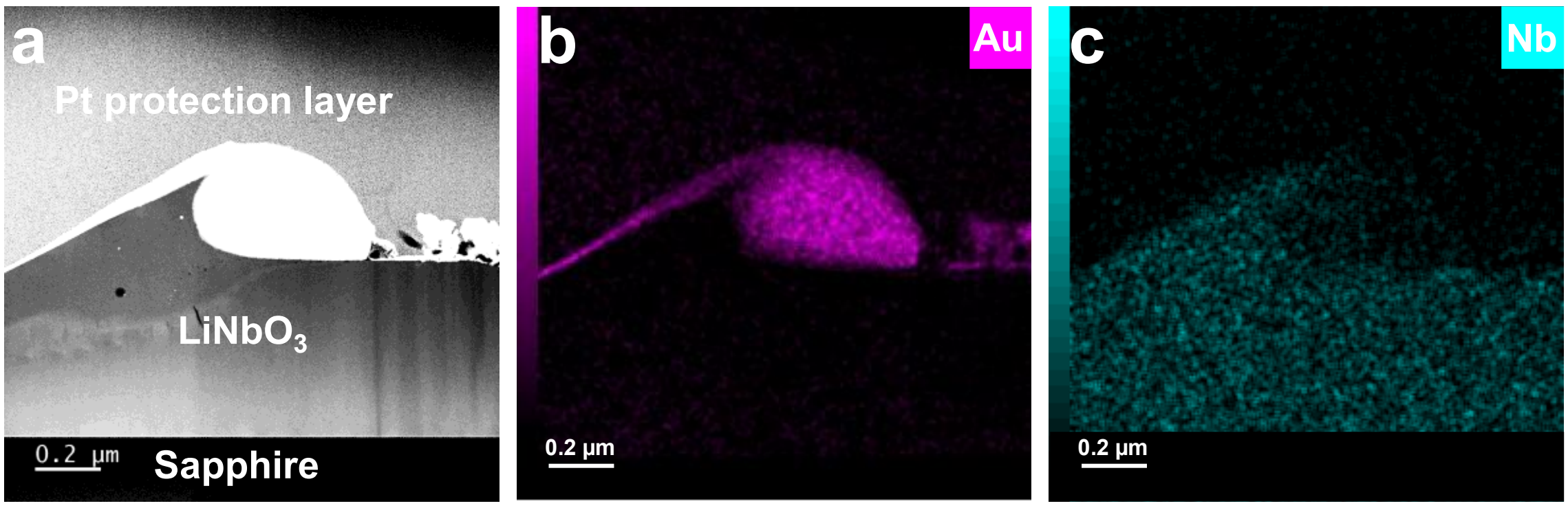

**Supplementary Fig. S28 | EDS mapping of broken TF-SAW transducers (State B). (a)** HAADF imaging STEM image illustrates the zoomed-up view of a failed TF-SAW transducer. Corresponding elemental mapping results for **(b)** Au and **(c)** Nb in **(a)**. The EDS mapping results confirm the severely deformed $LiNbO_3$ layer and IDTs under high-power loads.

**Supplementary Table S2. Elemental analysis at the five labelled sites of the failed TF-SAW transducer**

| | Spectrum 1 | Spectrum 2 | Spectrum 3 | Spectrum 4 | Spectrum 5 |
|---|---|---|---|---|---|
| **Pt (M series)** | 27.4 % | 27.8 % | 24.5 % | 26.4 % | 19.0 % |
| **Au (M series)** | 22.0 % | 20.4 % | 21.9 % | 17.8 % | 23.6 % |
| **Nb (L series)** | 11.2 % | 10.3 % | 14.1 % | 15.2 % | 19.8 % |
| **O (K series)** | 10.0 % | 11.5 % | 12.2 % | 13.4 % | 18.7 % |
| **C (K series)** | 15.6 % | 14.5 % | 12.7 % | 12.8 % | 11.6 % |
| **Cu (K series)** | 7.3 % | 7.7 % | 7.6 % | 6.6 % | 6.2 % |
| **Al (K series)** | 5.3 % | 6.5 % | 5.7 % | 6.3 % | 0 % |
| **Ni (K series)** | 0.8 % | 0.8 % | 0.6 % | 0.8 % | 0.6 % |
| **Cr (K series)** | 0 % | 0 % | 0 % | 0.4 % | 0.5 % |
| **Ga (K series)** | 0.4 % | 0.5 % | 0.7 % | 0.3 % | 0 % |

## 20 EDS mapping of a failed LAW transducer (State D)

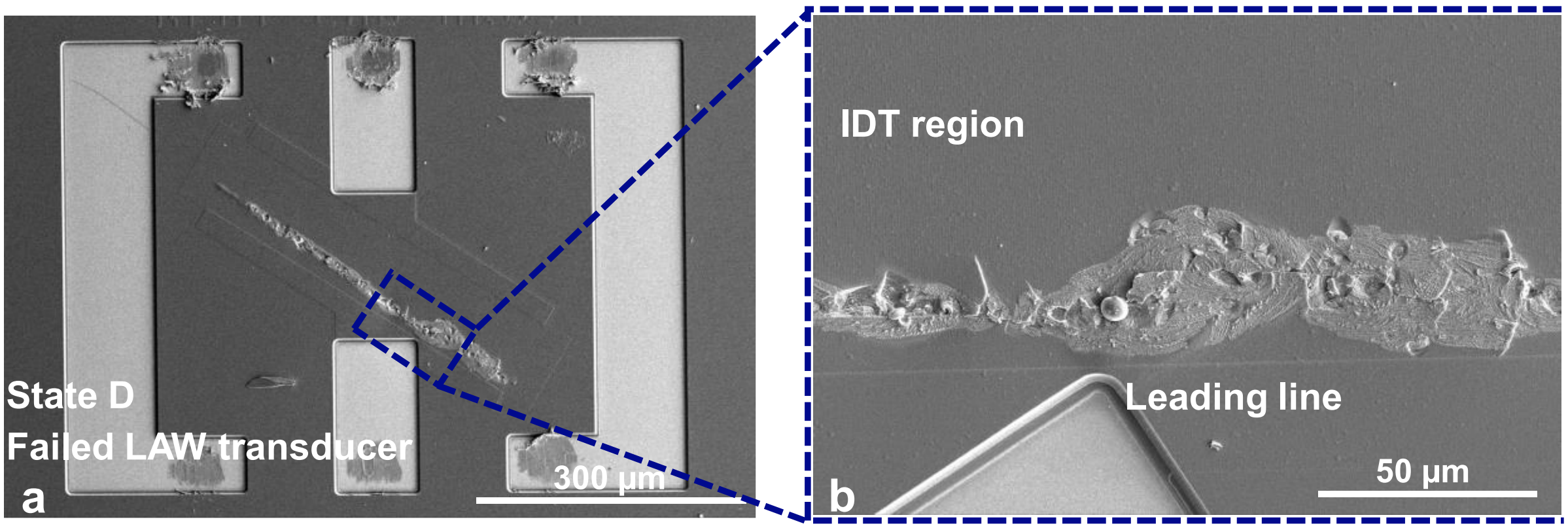

**Supplementary Fig. S29 | Post-failure analysis of a LAW transducer (State D). (a)** A zoomed-out view demonstrating that structural damage is highly localized for LAW DUT at State D, in contrast to the widespread failure observed in conventional SAW devices. **(b)** A magnified view of the failure region near the signal output terminal, showing a well-defined, near-parallel crack along the busline direction. The localized nature of the crack indicates that stress concentration, rather than global acoustic overload, governs the ultimate failure under high-power stress.

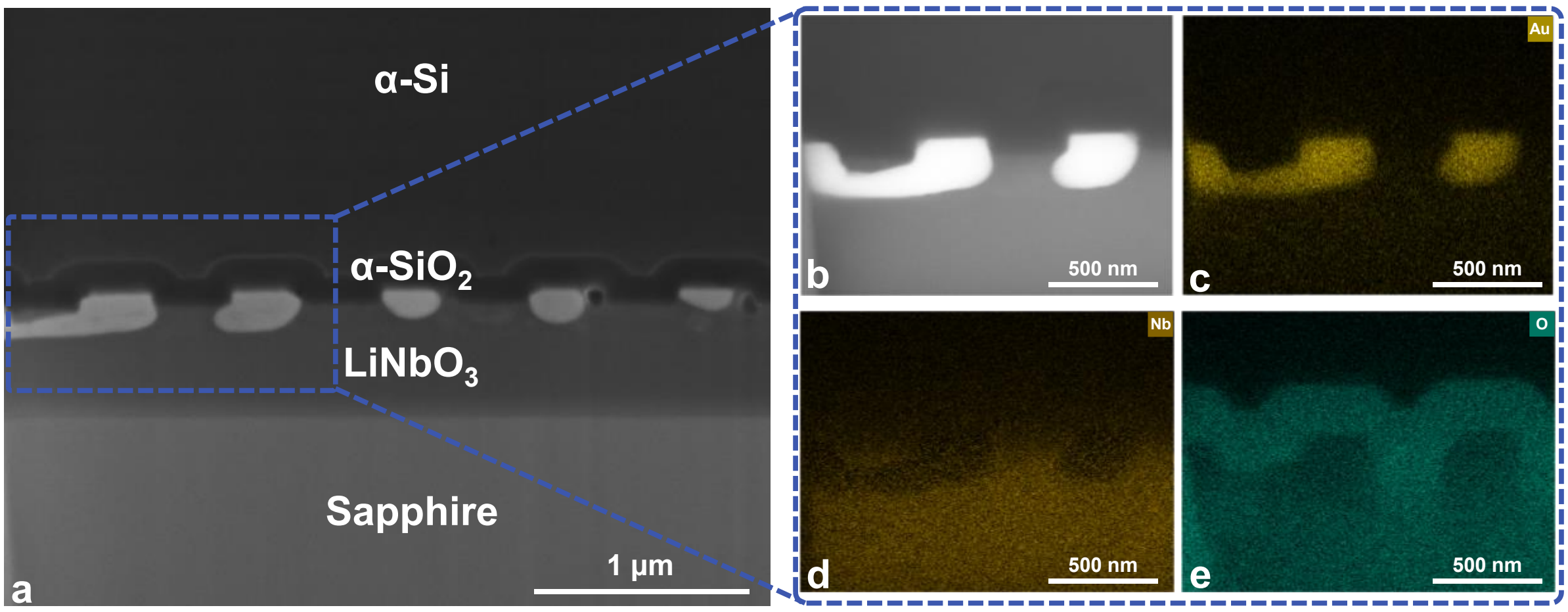

**Supplementary Fig. S30 | EDS mapping of broken LAW transducers (State D). (a)** Zoomed-out and (b) zoomed-in cross-sectional scanning electron microscope (SEM) image of a broken LAW transducer. Corresponding elemental mapping results for **(c)** Au, **(d)** Nb, and **(e)** O in **(b)**. No acoustomigration behavior can be observed for the LAW transducer after ultra-high power loads, as illustrated in **(a)**. The IDTs shown in **(a)** exhibit severe morphological deformation. Furthermore, localized Nb accumulation is observed above the interface of a pair of interconnected IDTs, suggesting stress-induced material redistribution during mechanical deformation. Although minor delamination occurs between the α-$SiO_2$ insulating layer and IDTs, the top cladding layers retain conformal coverage across most of the IDT array, highlighting their structural integrity under high-power operational conditions

## 21 Influence of residual stress of the silicon cladding layer on device performance

Residual stress control is of paramount importance in thick silicon cladding layer deposition, as large residual stress can cause irreversible device damage and modify the material properties of the adjacent layer[19]. Surface profiler KLA Tencor P-7 was utilized to measure the surface profile of a 4-inch silicon wafer before and after α-silicon deposition. Average residual stress in thick α-silicon can be calculated from the measured profilometry data.

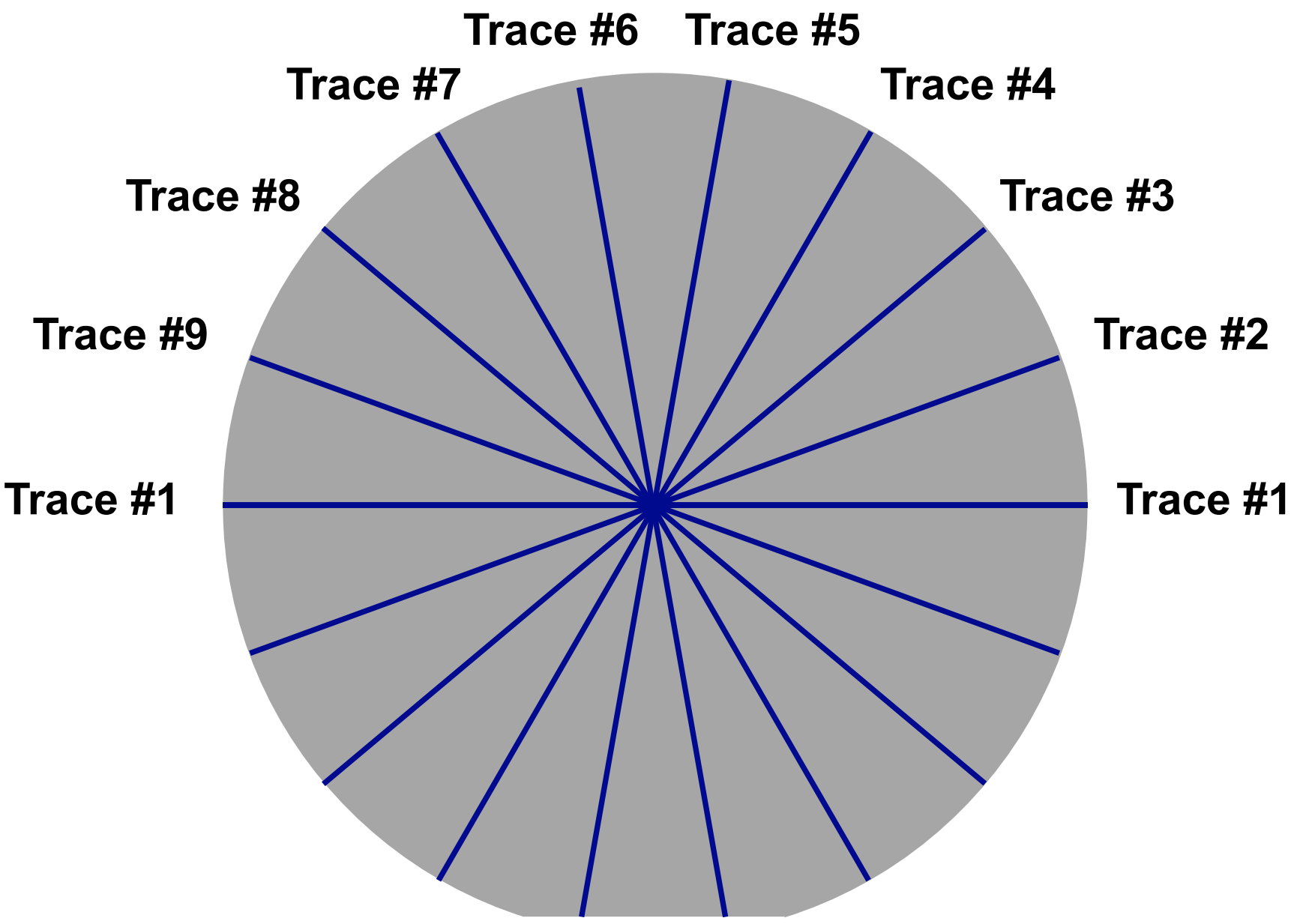

**Supplementary Fig. S31. | Methodology for residual stress extraction via wafer curvature measurement.** Schematic of the radius of curvature (ROC) measurement for a 4-inch wafer before and after α-Si deposition. Residual stress was extracted using the complete Stoney equation applied to nine uniformly distributed scans on a 4-inch wafer.

Profiles of 8-cm length and 5-μm sweeping resolution were obtained by rotating the 4-inch silicon (100) wafer and passing through the center, as shown in **Supplementary Fig. S31**. Subsequently, measured data was fitted by a fifth-order polynomial and utilized to calculate the radius ($R_{(x)}$) of curvature for each data point given by:

$$R_{(x)} = \frac{[1 + (dy/dx)^2]^{3/2}}{d^2y/dx^2} \tag{13}$$

Where $y$ is the height of difference trace and $x$ refers to the position. The residual stress ($\sigma_{(x)}$)

for each data point can be calculated by the Stoney equation[20]:

$$\sigma_{(x)} = \frac{1}{6R_{(x)}} \frac{E}{1-\nu} \frac{t_s^2}{t_f} \tag{14}$$

Where $E$ refers to the Young's modulus of the silicon (100) substrate, $\nu$ the Poisson's ratio for the silicon (100) substrate, $t_f$ and $t_s$ is the thickness of α-silicon thick film and the silicon (100) substrate. Finally, the average residual stress was determined by averaging the stress derived through the surface profiles.

**Supplementary Fig. S32a** presents the measured average radius of curvature (ROC) and its associated error bar for each of the nine traces, represented as a scatter plot with a fitted curve. The ROC is the primary experimental observable. Subsequently, the mean residual stress for each trace was calculated from its average ROC using Stoney's equation. **Supplementary Fig. S32b** shows the calculated average stress and its error bar for each trace, also as a scatter plot with a fit. Upon recalculating the weighted average after excluding two clear outliers — Trace #2 and Trace #8 (anomalously high ROC values, potentially from measurement near the wafer edge) — we obtain a refined average residual stress of 50.2 ± 6.9 MPa (with a corresponding average ROC of 337 ± 51 m).

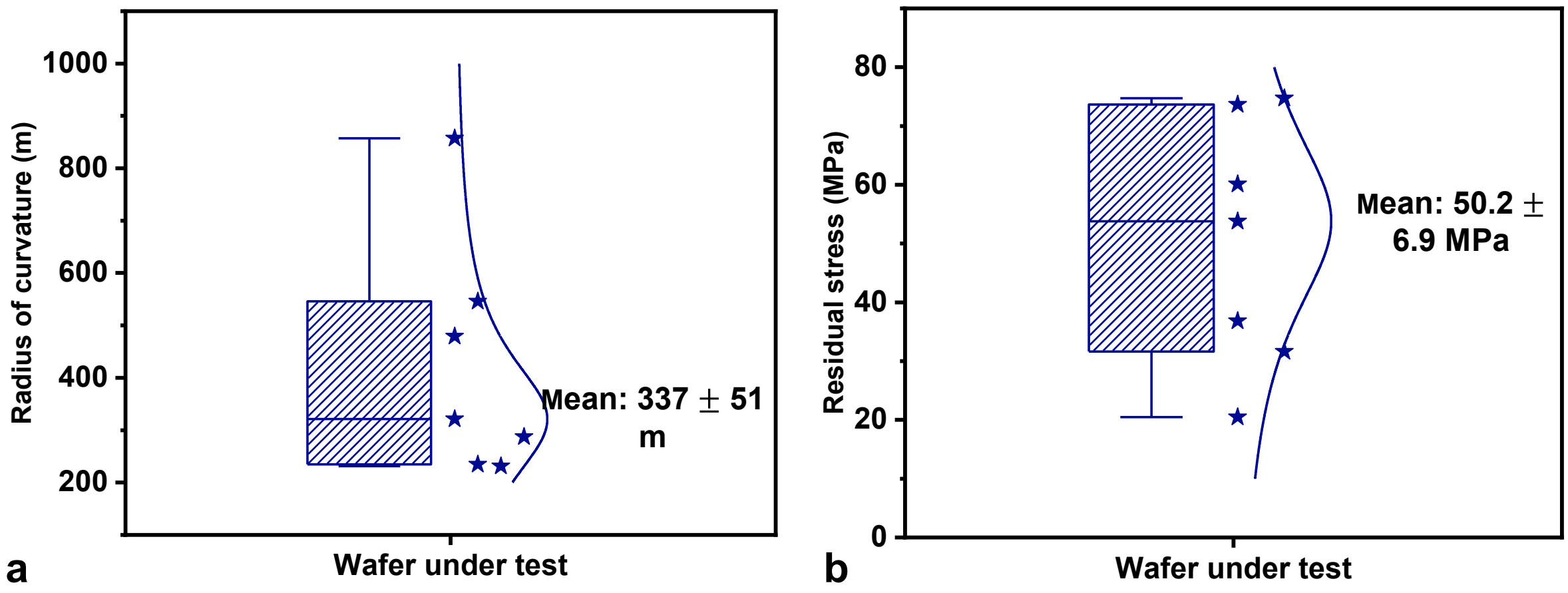

**Supplementary Fig. S32. | Statistical analysis of multi-trace curvature measurements.** Statistical distribution of the **(a)** ROC and **(b)** calculated residual stress for Traces #1 through #9, measured across the wafer. Data are presented as mean ± standard deviation, with average values of 337 ± 51 m (ROC) and 50.2 ± 6.9 MPa ( low tensile residual stress).

**Supplementary Fig. S33** shows measured admittance responses and Bode $Q$ curves on LAW transducers with varying residual stresses in the silicon cladding layer. Notably, the LAW transducer with high residual stress in the silicon cladding layer exhibits a small $k_t^2$ of 9.55%, AR of 40 dB, and Bode-$Q_{max}$ of 237. In contrast, the LAW transducer with near-zero residual stress in the silicon cladding layer significantly improves device performance, yielding a larger $k_t^2$ of 15.34%, AR of 55 dB, and Bode-$Q_{max}$ of 410.

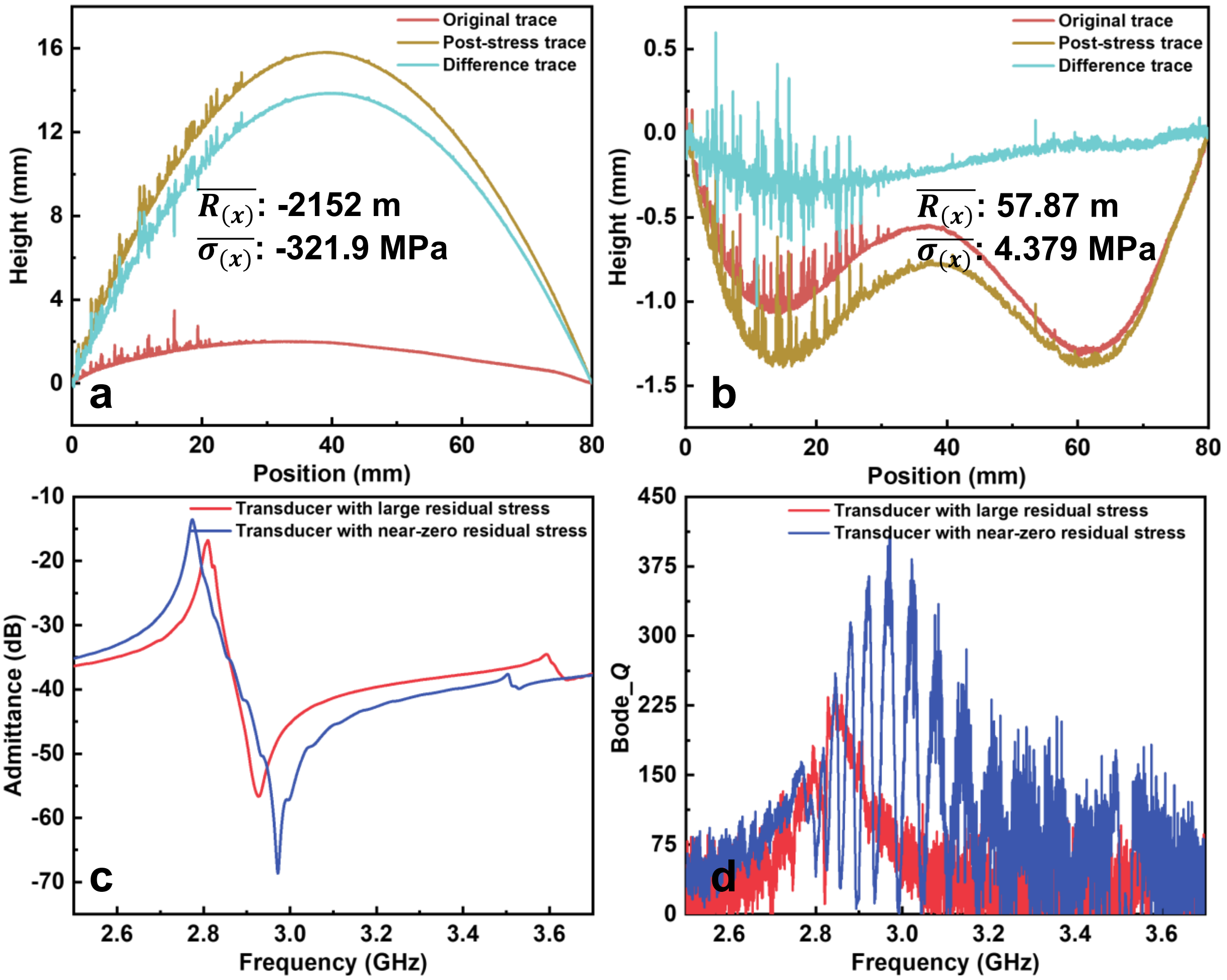

**Supplementary Fig. S33 | Influence of residual stress in silicon cladding layer on device performance.** Surface profiles of the α-silicon thick film under **(a)** high residual stress and **(b)** near-zero residual stress. The red, yellow, and cyan curves represent the pre-deposition profile, post-deposition profile, and differential profile (used to calculate residual stress in the silicon cladding layer), respectively. (c) Admittance curves and (d) Bode $Q$ curves comparing LAW transducers with varying residual stresses in the silicon cladding layer.

## 22 Calculation method of reflection coefficients for high-power tests

To validate the reflected coefficients extraction methodology, we compared the de-embedded $S_{21}$ (obtained using the power test setup in **Supplementary Fig. S10**) with $S_{21}$ measured directly in the VNA-only test loop (**Supplementary Fig. S34a**). The VNA output power in the power test setup was specifically set to match the load power delivered to the DUTs with that of the VNA-only test loop with calibrated reference plane. The close agreement between the de-embedded $S_{21}$ (via port-extension calibration) and the directly measured $S_{21}$ confirms the accurate de-embedding method. Building on mBVD fitting results under small-signal conditions, we further modified the mBVD model to fit the de-embedded $S_{21}$ under increasing load power by decreasing $f_r$ and $Q$, as indicated in **Supplementary Fig. S34b.** In summary, reflected and injected power coefficients derived via this mBVD fitting method perfectly match those calculated from direct measurements, validating the de-embedding methodology (**Supplementary Figs. S34c-S34d**).

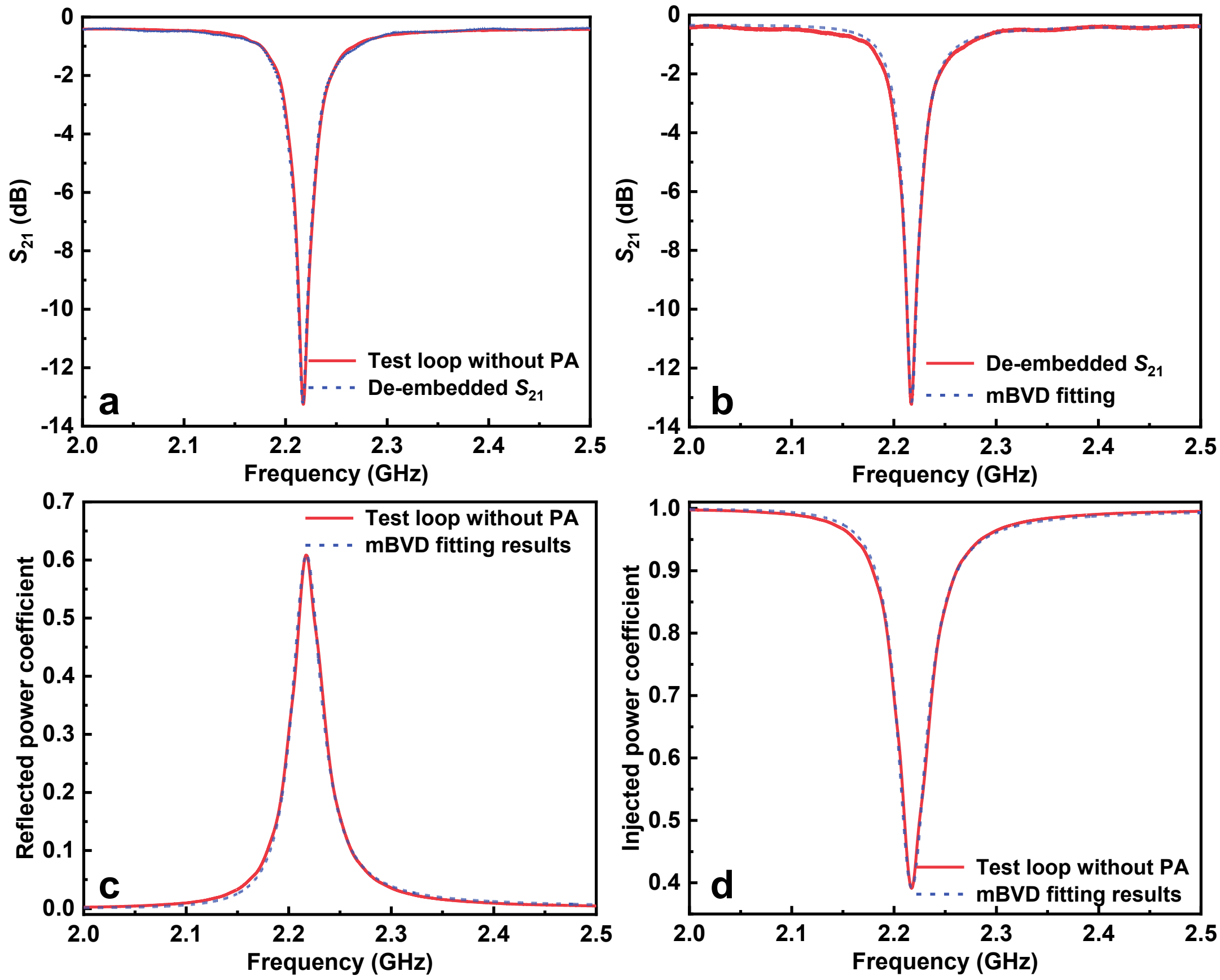

**Supplementary Fig. S34 | Extraction method of reflection coefficients for high-power tests. (a)**

Comparison between directly measured $S_{21}$ and de-embedded $S_{21}$ of a LAW transducer under a load power of 10 dBm. **(b)** mBVD fitted and de-embedded $S_{21}$ of the LAW transducer. **(c)** Reflection coefficient characteristics calculated from directly measured results and mBVD fitting results. **(d)** Injected power profiles obtained from directly measured results and mBVD fitting results.

## References

1. Wang, Y., Hashimoto, K., Omori, T. & Yamaguchi, M. Change in piezoelectric boundary acoustic wave characteristics with overlay and metal grating materials. *IEEE Trans. Ultrason. Ferroelectr. Freq. Control* **57**, 16–22 (2010).
2. Hashimoto, K. -y., Watanabe, Y., Akahane, M. & Yamaguchi, M. Analysis of acoustic properties of multi-layered structures by means of effective acoustic impedance matrix. In *Proc.* 1990 *IEEE Symposium on Ultrasonics* (IUS) 937–942 (IEEE, 1990).
3. Qian, F., Ho, T. F. & Yang, Y. Twist piezoelectric coupling properties to suppress spurious modes for lithium niobate thin-film acoustic devices. In *Proc. 2023 IEEE/MTT-S International Microwave Symposium (IMS)* 907–910 (IEEE, 2023).
4. Hashimoto, K. *et al.* Revisiting piston mode design for radio frequency surface acoustic wave resonators. In *Proc. 2022 IEEE MTT-S International Conference on Microwave Acoustics and Mechanics (IC-MAM)* 60–63 (IEEE, 2022).
5. Xu, H. *et al.* SAW filters on $LiNbO_3$/SiC heterostructure for 5G n77 and n78 band applications. *IEEE Trans. Ultrason. Ferroelectr. Freq. Control* **70**, 1157–1169 (2023).
6. Qian, F., Zheng, J., Xu, J. & Yang, Y. Heterogeneous interface-enhanced thin-film SAW devices using lithium niobate on Si. *IEEE Microw. Wirel. Technol. Lett.* **35**, 123–126 (2025).
7. Shen, J. *et al.* Suppressed transverse mode generation in TF-SAW resonators based on $LiTaO_3$/Sapphire. *IEEE Electron Device Lett.* **45**, 2241–2244 (2024).
8. Shen, J. *et al.* A low-loss wideband SAW filter with low drift using multilayered structure. *IEEE Electron Device Lett.* **43**, 1371–1374 (2022).
9. Su, R. *et al.* Wideband and low-loss surface acoustic wave filter based on 15° YX-$LiNbO_3$/$SiO_2$/Si structure. *IEEE Electron Device Lett.* **42**, 438–441 (2021).
10. Xu, H. *et al.* Large-range spurious mode elimination for wideband SAW filters on $LiNbO_3$/$SiO_2$/Si platform by $LiNbO_3$ cut angle modulation. *IEEE Trans. Ultrason. Ferroelectr. Freq. Control* **69**, 3117–3125 (2022).
11. Liu, P. *et al.* A spurious-free SAW resonator with near-zero TCF using $LiNbO_3$/$SiO_2$/quartz. *IEEE Electron Device Lett.* **44**, 1796–1799 (2023).
12. Xiao, B. *et al.* Anisotropy-matched LN/quartz heterostructure with inherent spurious mitigation for wideband SAW devices. *IEEE Trans. Microw. Theory Tech.* **73**, 10080–10094 (2025).
13. Zhang, S. *et al.* Surface acoustic wave devices using lithium niobate on silicon carbide. *IEEE Trans. Microw. Theory Tech.* **68**, 3653–3666 (2020).
14. Wen, Z. *et al.* A high power-handling laterally-excited bulk acoustic resonator with scattering vias in double-layer electrodes over +35 dBm. In *Proc. 2024 IEEE Ultrasonics, Ferroelectrics, and Frequency Control Joint Symposium (UFFC-JS)* 1–4 (2024). doi:10.1109/UFFC-JS60046.2024.10793761.
15. Sui, D. *et al.* Miniaturized A1 mode acoustic resonators and filters using inverted T-

shaped electrodes. *IEEE Trans. Microw. Theory Tech.* **73**, 10071–10079 (2025).

16. Fang, X. *et al.* Hybrid integration of dual-mode SAW resonators for high-power and wideband high-frequency filters. *IEEE Trans. Microw. Theory Tech.* **73**, 8490–8499 (2025).

17. Gonzalez-Rodriguez, M. *et al.* Method to measure reflection coefficient under CW high-power signals in SAW resonators. In *Proc. 2021 IEEE International Ultrasonics Symposium (IUS)* 1–4 (IEEE, 2021).

18. van der Wel, P. J., Wunnicke, O., de Bruijn, F. & Strijbos, R. C. Thermal behaviour and reliability of solidly mounted bulk acoustic wave duplexers under high power RF loads. In *Proc. 2009 IEEE International Reliability Physics Symposium (IRPS)* 557–561 (IEEE, 2009).

19. Pan, S., Memon, M. M., Wan, J., Wang, T. & Zhang, W. The influence of pressure on the TCF of AlN-based SAW pressure sensor. *IEEE Sens. J.* **22**, 3097–3104 (2022).

20. Stoney, G. G. & Parsons, C. A. The tension of metallic films deposited by electrolysis. *Proc. R. Soc. Lond. Ser. Contain. Pap. Math. Phys. Character* **82**, 172–175 (1997).